\documentclass[11pt]{article}

\usepackage[utf8]{inputenc} 
\usepackage[T1]{fontenc}    
\usepackage{hyperref}       
\usepackage{url}            
\usepackage{booktabs}       
\usepackage{amsfonts}       
\usepackage{nicefrac}       
\usepackage{microtype}      
\usepackage{xcolor}
\usepackage{mathtools}
\usepackage{algorithm}
\usepackage{algorithmic}
\usepackage{amssymb}
\usepackage{enumitem}
\usepackage{calc}
\usepackage{amsthm}
\usepackage{graphicx}
\usepackage{subcaption}
\usepackage{authblk}
\usepackage{wrapfig}
\usepackage{multirow}

\usepackage{float}

\usepackage{wrapfig}
\definecolor{dbblue}{RGB}{10,65,155}

\hypersetup{colorlinks,urlcolor=dbblue,citebordercolor=dbblue,linkbordercolor=dbblue,filecolor=dbblue,filebordercolor=dbblue,citecolor=dbblue,linkcolor=dbblue,colorlinks=true}

\usepackage[verbose=true,letterpaper]{geometry}

\renewcommand{\baselinestretch}{1.25}

\usepackage[numbers]{natbib}

\newtheorem{rem}{Remark}

\DeclareMathOperator{\sign}{sign}

\title{Limit Order Book Simulation and Trade Evaluation with $K$-Nearest-Neighbor Resampling\footnotetext{\scriptsize{\hskip-0.55cm Corresponding author: Michael Giegrich (\href{mailto:giegrich@maths.ox.ac.uk}{giegrich@maths.ox.ac.uk})}}} 

\author[a]{Michael Giegrich}
\affil[a]{Mathematical Institute, University of Oxford}
\affil[b]{Deutsche Bank, London}
\affil[c]{Department of Mathematics, Imperial College London}
\affil[d]{Oxford-Man Institute of Quantitative Finance, University of Oxford}
\author[b,c]{Roel Oomen}
\author[a,d]{Christoph Reisinger}

\begin{document}

\maketitle


\begin{abstract}
In this paper, we show how $K$-nearest neighbor ($K$-NN) resampling, an off-policy evaluation method proposed in \cite{giegrich2023k}, can be applied to simulate limit order book (LOB) markets and how it can be used to evaluate and calibrate trading strategies. Using historical LOB data, we demonstrate that our simulation method is capable of recreating realistic LOB dynamics and that synthetic trading within the simulation leads to a market impact in line with the corresponding literature. Compared to other statistical LOB simulation methods, our algorithm has theoretical convergence guarantees under general conditions, does not require optimization, is easy to implement and computationally efficient. Furthermore, we show that in a benchmark comparison our method outperforms a deep learning-based algorithm for several key statistics. In the context of a LOB with pro-rata type matching, we demonstrate how our algorithm can calibrate the size of limit orders for a liquidation strategy. Finally, we describe how $K$-NN resampling can be modified for choices of higher dimensional state spaces. 
\end{abstract}

\section{Introduction}

Central limit order books (LOBs) are the prevalent organisational mechanism for trading in a wide range of assets, from stocks and bonds to derivatives. A LOB is a centralized registry of traders' commitments to buy or sell a certain amount of an asset for a certain price, referred to as limit orders. Other traders may choose to execute a certain amount against those limit orders, by posting market orders. The appeal of LOBs is their ability to efficiently aggregate traders' preferences via limit and market orders. Due to their practical importance and the complex interaction patterns present in them, LOBs are a widely researched subject in academia and industry. From a practical perspective, an important question is to understand how well a trader's strategy performs within a limit order book and how the market reacts to this trader's strategy, noting that the placement of orders can have an adversarial effect (see, e.g.\ \cite{gatheral2010no}).

Direct testing of trading strategies within an actual market is inherently risky and can lead to sizeable losses or a disorderly market. On the other hand, large amounts of historical data for LOBs are available. In reinforcement learning, the subfield of off-policy evaluation deals with the problem of how to extract information about the performance of a strategy using such observational data. In this context, we proposed $K$-nearest neighbor (K-NN) resampling, a method that has theoretical convergence guarantees under general conditions, performs well in experiments in environments with intrinsic noise, is easy to implement and computationally efficient \citep{giegrich2023k}. The main idea of K-NN resampling is that similar states in a metric sense lead to similar transitions of the underlying random dynamical system. Given an initial state, the algorithm selects at random a new state from a list of nearest neighbors from the initial state. From this neighboring state, the historical transition from the data is used leading to a subsequent state. Chaining these random matchings together yields an approximate path for the underlying system and repeatedly restarting at the initial states leads to a collection of paths, i.e.\ a simulation. In this paper, we show how we can apply this idea for the simulation of LOBs and the evaluation of trading strategies within LOBs. 

The academic interest in the simulation of LOBs is well documented by the broad and recent literature. 
A current literature review of this subject matter \citep{jain2024limit} identifies four major approaches to LOB simulations: Point processes \citep{bouchaud2002statistical, cont2010stochastic, bacry2015hawkes, huang2015simulating, lu2018order, morariu2022state, jerome2022model}, agent based modelling \citep{paddrik2012agent, byrd2019abides, belcak2020fast, shi2023neural}, stochastic differential equations models \citep{cont2013price, korolev2015modeling,huang2017ergodicity, horst2017law, cont2021stochastic} and deep/statistical learning \citep{li2020generating, shi2021limit, lim2021intra, coletta2023conditional,cont2023limit,nagy2023generative, hultin2023generative}.
We note that the first three approaches usually require explicit modelling choices on either market dynamics or interactions of market participants. While this view allows for the direct choice of relevant market characteristics, it poses the risk of fundamental modelling biases in the market dynamics. For applications in actual markets, these approaches often require the additional non-trivial step of calibration to market data and may require a high level of granularity in the data.  

The algorithm we propose can be understood as a statistical learning method to order book simulation. In particular, our algorithm selects transitions between LOB snapshots using a non-parametric approach.  This is most similar to the scope and methodology proposed in \cite{cont2023limit}. The authors also sample transitions between LOB snapshots where the sampling distribution is given by a conditional generative adversarial neural network (CGAN). They show that their model is capable of reproducing stylized facts observed in the market and that the impact of trading within their simulator behaves as expected. Since the scope of our paper is partially similar, we reuse some of the stylized facts considered in this paper for testing our model's quality. Furthermore, we use the CGAN approach as a benchmark and see that our method outperforms it on key statistics for simulation quality. Beyond the scope of \cite{cont2023limit}, we also consider the evaluation and selection of trading strategies using limit orders. 

The other papers using deep learning for LOB simulation often have a slightly different focus. Namely, \citep{li2020generating, shi2021limit, lim2021intra, coletta2023conditional,nagy2023generative, hultin2023generative} considers event-by-event simulation within the LOB rather than the transition between LOB snapshots as in \cite{cont2023limit} and in this paper. We note that this may depend on the objective of the simulation. Event-by-event simulation are particularly useful for ultra-high frequency traders where a single event may cause the trader to react. Considering transitions between LOB snapshots on the other hand is often sufficient for practical applications in high to medium-high frequency trading. Note that our algorithm could be readily modified for event-by-event simulations -- we leave this investigation for future work.

Furthermore, all published statistical learning models that we are aware of use deep learning for generating simulations. In contrast, our method only uses nearest neighbor search. Not relying on deep learning models has several advantages. First, our algorithm does not need any optimization. In contrast, deep learning models require solving a highly non-convex optimization problem where optimization procedures often exhibit instability, as observed in \cite{cont2023limit}.   Furthermore, our method can be efficiently implemented using standard machine learning packages and purely relies on historical observations rather then sampling from a black box distribution. Finally, it does not require extensive hyperparameter tuning.

Since we apply our algorithm to a LOB with a pro-rata type matching mechanism, we also highlight \cite{belcak2020fast}, which is to our knowledge the only LOB simulator explicitly used for pro-rata markets. In contrast to our paper, the focus of their work is on the explanation of market phenomena by aggregation of agents rather than simulating a specific market or evaluating trading strategies. Furthermore, our method is applicable to different types of execution mechanisms and pro-rata type matching is only exploited for evaluating trading strategies with limit orders (see Section \ref{sec:polEval}). 
   
Our contributions in this paper can be summarized as follows:
\begin{enumerate}
	\item We provide a framework how K-NN resampling can be adapted for simulating LOBs and how to incorporate trades within these simulations.
	\item We describe how K-NN resampling can be used for evaluating trading strategies with both limit and market orders.
	\item We provide empirical evidence that our method is capable of realistically simulating LOB dynamics with and without trade interventions. In particular, we show that our method can outperform a strong deep learning-based benchmark on key statistics. 
	\item We demonstrate how K-NN resampling can be used to evaluate and choose trading strategies in a LOB with a pro-rata type matching mechanism. Specifically, we show how to calibrate the quantity of limit orders with our resampling algorithm. 
	\item We illustrate how K-NN resampling can effectively be used in a higher dimensional state space and how the algorithm can be combined with dimension reduction techniques.   
\end{enumerate}

We organize the paper as follows. Section \ref{sec:lob} introduces our notation and assumptions for LOBs. Section \ref{sec:alg} describes our algorithm and how it can be used for LOB simulations. Section \ref{sec:data} details the data set that we use to test the resampling algorithm. Section \ref{sec:results} explores the quality of the simulations, with and without trade interventions. In Section \ref{sec:polEval}, we demonstrate how trading strategies with limit orders can be evaluated and chosen based on our simulation approach. Section \ref{sec:state} considers a higher dimensional state space and usage of dimension reduction for the resampling. Section \ref{sec:conc} concludes the paper.

\paragraph*{Notation:}
$\mathbb{R}_{>0}$ and $\mathbb{N}_{>0}$ denotes the space of positive real and natural numbers, respectively. For $\delta\in\mathbb{R}_{>0}$, $\delta\mathbb{N}_{>0}$ denotes all the positive multiples of $\delta$.


\section{Limit order books}\label{sec:lob}

This section describes our notation and assumptions for LOBs as necessary for a precise definition of our algorithm. 

\subsection{Set-up}

Consider a trading time horizon $[0,T]$ where $T\in\mathbb{R}_{>0}$ and the tick size $\delta\in\mathbb{R}_{>0}$, then we can describe market orders,  limit orders and limit order cancellations as follows:
\begin{itemize}
		\item A limit order is a quadruple $x = (t, p, q, r)$ where $t\in [0,T]$ is the time the order is placed, $p\in\delta\mathbb{N}_{>0}$ is the price level, $q\in\mathbb{N}_{>0}$ is the quantity and $r\in\{\text{bid}, \text{ask}\}$ is the order side.
		\item A market order is a triple $y =(t, q, r)$ where $t\in [0,T]$ is the time the order is placed, $q\in\mathbb{N}$ is the quantity and $r\in\{\text{buy}, \text{sell}\}$ is the aggressor side of the order.
		\item  An order cancellation for a limit order $(t, p, \hat{q}, r)$ with remaining volume $\hat{q}>0$ and $t<T$ is a  5-tuple $c = (t,\hat{t}, p, \hat{q}, r) $ where $\hat{t}\in [t,T]$ is the time of cancellation.
\end{itemize}

The limit order book is the set of all active limit orders $L(t)$ at a given point in time $t\in [0,T]$. $L_p(t)$ denotes the set of all active limit orders at time $t\in [0,T]$ and price level $p\in\delta\mathbb{N}_{>0}$. We refer to $L_p(t)$ as LOB level. Note that each LOB level with active limit orders (i.e.\ $L_p(t)\neq \emptyset$) can be uniquely assigned to either order side at any point in time as otherwise LOB levels containing both bid and ask side would be executed against each other. This assignment is ordered and unique in the sense that there is a dividing price $p^*\in\mathbb{R}\setminus\delta\mathbb{N}$ for which all elements of $L_p(t)$ are bid orders, if $p<p^*$, and ask orders, if $p>p^*$. For our purposes, we assume that we can extend this unique assignment to all $L_p(t)$ even for empty LOB levels (i.e.\ $L_p(t)=\emptyset$)\footnote{In our practical implementation, we achieve this unique mapping for all $p$ by a rule-based assignment considering the last time a LOB level was active and its relative position in the LOB. Such an assignment is necessary to handle empty tick sizes in the algorithm.}. Furthermore, the volume $V_p(t)$ at a LOB level $L_p(t)$ is defined as $V_p(t) = \sum_{x\in L_p(t)} q_x$, if $L_p(t)\neq\emptyset$, and $V_p(t) = 0$, if $L_p(t)=\emptyset$.

\begin{figure}[h!]
	\centering
	\includegraphics[width=0.8\linewidth]{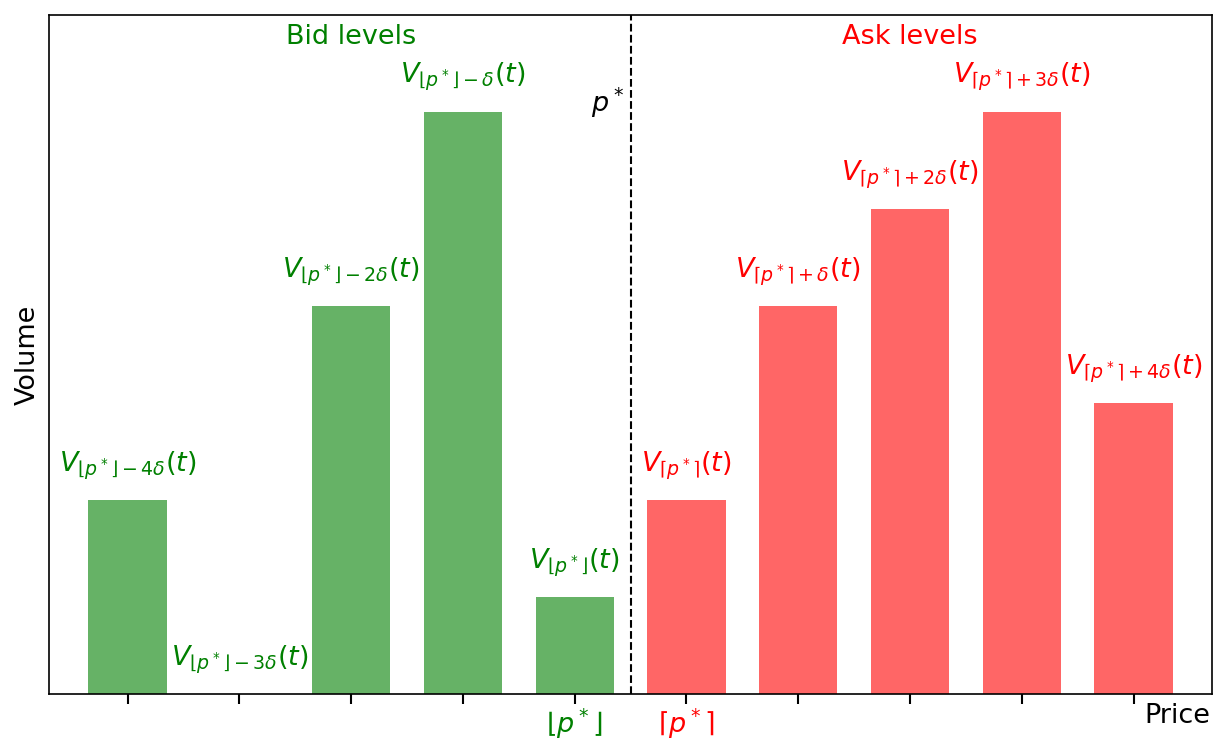}
	\caption{LOB snapshot $S^5_{p^*}(t)$ at time $t$ with dividing price $p^*$, $5$ levels on each market side and tick size $\delta$.}
	\label{fig:lobSnapshot}
\end{figure}

For a dividing price  $p^*\in\mathbb{R}\setminus\delta\mathbb{N}$, we denote the largest $p\in\delta\mathbb{N}$ for which $p<p^*$ by $\lfloor p^*\rfloor$ and define $\lceil p^*\rceil$ analogously.  Thus, $\lfloor p^*\rfloor$ is the best bid price and $\lceil p^*\rceil$ the best ask price. This notation allows us to define a centered LOB snapshot with $l$ levels. Let $l\in\mathbb{N}_{>0}$ and $p^*\in\mathbb{R}\setminus\delta\mathbb{N}$ a dividing price with $\lfloor p^*\rfloor>l\delta$, then the LOB snapshot with $l$ levels is defined as
\begin{equation*}
	S^l_{p^*}(t)= (V_{\lfloor p^*\rfloor-(l-1)\delta}(t),\dots,V_{\lfloor p^*\rfloor}(t), V_{\lceil p^*\rceil}(t),\dots,V_{\lceil p^*\rceil+(l-1)\delta}(t)).
\end{equation*}
The first $l$ entries correspond to the bid levels while the last $l$ entries are for the ask values. Figure \ref{fig:lobSnapshot} shows an example for an LOB snapshot.

\begin{figure}[h!]
	\centering
	\includegraphics[width=0.8\linewidth]{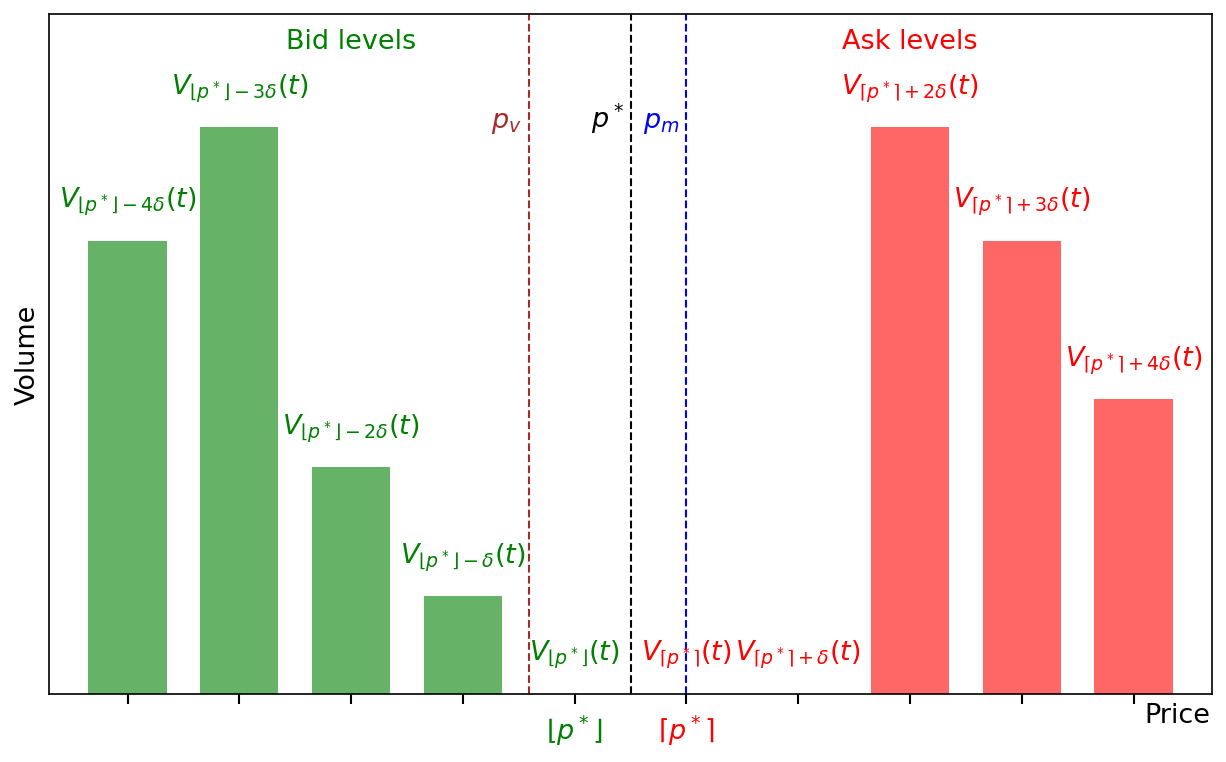}
	\caption{ LOB snapshot $S^3_{p^*}(t)$ at time $t$ with dividing price $p^*$, mid-price $p_m$ and weighted mid-price $p_v$.}
	\label{fig:lobPrices}
\end{figure}

Choosing the smallest $k_b,k_a\in\mathbb{N}$ such that $V_{\lfloor p^*\rfloor-(k_b-1)\delta}(t)>0$ and $V_{\lceil p^*\rceil+(k_a-1)\delta}(t)>0$, the (unweighted) mid-price is given by 
\begin{equation*}
	p_m(t) = ((\lfloor p^*\rfloor-(k_b-1)\delta) + (\lceil p^*\rceil+(k_a-1)\delta))/2. 
\end{equation*}
The weighted mid-price is then given by  
\begin{equation*}
	p_w(t) =\frac{(\lfloor p^*\rfloor+(k_b-1)\delta)V_{\lfloor p^*\rfloor-(k_b-1)\delta}(t) + (\lceil p^*\rceil-(k_a-1)\delta)V_{\lceil p^*\rceil+(k_a-1)\delta}(t)}{V_{\lfloor p^*\rfloor-(k_b-1)\delta}(t) + V_{\lceil p^*\rceil+(k_a-1)\delta}(t)}. 
\end{equation*}
Figure \ref{fig:lobPrices} demonstrates for an example LOB snapshot how prices can differ for different price definitions. The dividing price is seen as given (see footnote) while the mid-price and the weighted mid-price depend on the current shape of the order book.

Furthermore, we define the return for either the mid-price or the weighted price as the $\log$-return, i.e.\
\begin{equation*}
	\ln p_{\cdot}(t)-\ln p_{\cdot}(\tilde{t})
\end{equation*}
where $\tilde{t}<t$. Finally, we define analogous to \cite{cartea2018enhancing} the order book imbalance (OBI) as 
\begin{equation*}
	\rho(t)=\frac{V_{\lfloor p^*\rfloor}(t)- V_{\lceil p^*\rceil}(t)
	}{V_{\lfloor p^*\rfloor}(t)+ V_{\lceil p^*\rceil}(t)} \in[-1,1]
\end{equation*}
if $V_{\lfloor p^*\rfloor}(t)+ V_{\lceil p^*\rceil}(t)>0$ and $\rho(t)=0$ else.

\subsection{Assumptions on limit order book dynamics}
Limit order books are usually seen as evolving continuously over time, where market orders and limit orders may arrive at any time. The arriving orders are often modelled as random but reactive to the current state of the LO (see e.g.\ \cite{cont2010stochastic, rambaldi2017role}). We focus on a discretization that allows us to consider high to medium-high frequency trade decisions. In particular, we assume that the LOB evolves as a discrete random dynamical system transitioning from one LOB state-price pair $(Z_t, p_t^*)$ to the next, $(Z_{t+\Delta t}, p_{t+\Delta t}^*)$, where $\Delta t\in\mathbb{R}_{>0}$ is the time increment and $Z_t$ the state representation for the LOB\footnote{ We note that the state representation can be chosen in many different ways and is not limited to LOB snapshots. We discuss possible choices of the state below.}. Using transition functions $f_Z$, $f_p$ and a random noise $\varepsilon$, a single LOB transition is described by
\begin{equation*}
	(Z_{t+\Delta t}, p_{t+\Delta t}^*) = (f_Z(Z_t, \varepsilon), f_p(Z_t, p_{t}^*, \varepsilon)).
\end{equation*} 
For an initial state-price pair $(Z_{0}, p_{0}^*)$, the LOB dynamics are then defined by recursively applying the transition functions, i.e.\ 
\begin{equation*}
	(Z_{(k+1)\Delta t}, p_{(k+1)\Delta t}^*) = (f_Z(Z_{k\Delta t}, \varepsilon_{k\Delta t}), f_p(Z_{k\Delta t}, p_{k\Delta t}^*, \varepsilon_{k\Delta t}))
\end{equation*}  
where we assume that the noise is identically and independently distributed (iid) for all $s\in\mathbb{N}$. 

We choose to model the price explicitly in order to facilitate the assignment of empty tick sizes in the LOB snapshots and we assume that the price level does not have an impact on the shape of the subsequent order book. Furthermore, the iid assumption on the system noise implies that the system is Markovian. Random dynamical system formulations for LOBs are common in the literature and can, for example, be found in \cite{cont2010stochastic, cont2013price, abergel2013mathematical, horst2017law}.

Note that the LOB state can be chosen in potentially many different ways depending on the investigation. We will focus on a Markovian interpretation of the LOB snapshots where we use 
\begin{equation*}
Z_t = S^l_{p_t^*}(t)
\end{equation*}
for some LOB depth $l$. Other potential choices of $Z_t$ can include a statistic on the history of LOB snapshots to account for non-Markovian behavior of order books \cite{sirignano2019universal} or information on seasonalities (e.g.\ distance to fixing times) and other external events impacting the order book (e.g.\ trades in a derivative order book).

\subsection{Trading agent in the limit order book}
We describe how an additional trading agent can be added to the random dynamical system for the purpose of interacting with the LOB or evaluating trading strategies. We assume that for a trader their trading decision is based on the current state of the LOB and personal information. This personal information can contain a wide variety of variables. For example, it can include information on the trader's inventory and their risk profile, but it also could include alpha signals or information on other active traders in the market. Usually, we assume that the information contains the current LOB position $P(t)$ of the trader. A trader's LOB position is the set of all active limit orders placed by the trader. 

We formalize the interaction of a trading agent with the LOB by defining the trading strategy of a trader. Let $u$ be a function mapping from an LOB state $Z_t$ and trader information $I_t$ to a triple consisting of a market order $y$, a set of cancellations of outstanding limit orders $\{c_j\}_{j=1}^m$ and a set of new limit orders $\{x_i\}_{i=1}^n$, i.e.\ 
\begin{equation*}
	u(Z_t,I_t)=( \{c_j\}_{j=1}^m, y,\{x_i\}_{i=1}^n ).
\end{equation*}
To limit the possibility of self-fulfilment, we assume the following hierarchy in the execution in our algorithm. First, a trader cancels some of their outstanding orders according to $\{c_j\}_{j=1}^m$, then they execute their market order if the market order quantity is larger than zero and finally the new limit orders $\{x_i\}_{i=1}^n$ are placed.

\subsection{Interaction with the limit order book}
In the context of the random dynamical system formulation of the LOB dynamics, we make the assumption that we can incorporate the immediate effect of a trader's action on the LOB state. For this purpose, we define functions for limit orders and market orders that map onto a new LOB state.

For limit orders, we only consider the case where the order does not cross the spread\footnote{The general case where a limit order can cross the spread can be thought of as a straightforward combination of the immediate impact of a market order and a limit order where the mid-price shifts accordingly.}. For a $k\in \mathbb{N}$ with $k<l$, consider the limit order $x=(t, \lfloor p^*\rfloor-k\delta, q, \text{bid})$ and the LOB state-price pair $(S^l_{p_t^*}(t), p_t^*)$ at the same point in time, then the function 
\begin{equation*}
	g_L((S^l_{p_t^*}(t), p_t^*),x) = (\tilde{S}^l_{p_t^*}(t), p_t^*)
\end{equation*}
where 
\begin{equation*}
\tilde{S}^l_{p_t^*}(t) = (\dots,V_{\lfloor p^*\rfloor-(k+1)\delta}(t),V_{\lfloor p^*\rfloor-k\delta}(t)+q,V_{\lfloor p^*\rfloor-(k-1)\delta}(t),\dots)
\end{equation*}
incorporates bid side limit orders. The extension to the ask side is analogous. The third plot in Figure \ref{fig:loInteract} shows how the volume of a buy limit order and a sell limit order is placed in a centered LOB snapshot. Furthermore, we analogously define the function $g_C((S^l_{p_t^*}(t), p_t^*), c)$ where instead of adding the cancelled volume is subtracted from the appropriate place in the centered LOB snapshot (see first plot in Figure \ref{fig:loInteract}).

Consider the market order $y =(t, q, \text{buy})$ and a LOB state-price pair $(S^l_{p_t^*}(t), p_t^*)$ where $\sum_{i=0}^{k}V_{\lceil p^*\rceil+i\delta}(t)>q>\sum_{i=0}^{k-1}V_{\lceil p^*\rceil+i\delta}(t)$ for $k\in\mathbb{N}$ with $k\leq l-1$. Then,
\begin{equation*}
	g_M((S^l_{p_t^*}(t), p_t^*),y) = (\tilde{S}^l_{p_t^*}(t), p_t^*)
\end{equation*}
where 
\begin{equation*}
	\tilde{S}^l_{p_t^*}(t) = (\dots,V_{\lfloor p^*\rfloor}(t),0,\dots,0,\sum_{i=0}^{k}V_{\lceil p^*\rceil+i\delta}(t)-q,V_{\lceil p^*\rceil+(k+1)\delta}(t),\dots)
\end{equation*}
is used to account for the immediate effect of a buy market order on the LOB state. We visualize the placement of a buy market order in the second plot of Figure \ref{fig:loInteract}. A sell order can be treated analogously. For a market order, we can further directly observe the trade revenue. Using the buy order and the order book from above the trader's revenue is
\begin{equation*}
	-\left(\lceil p^*\rceil+k\delta\right)\left(\sum_{i=0}^{k}V_{\lceil p^*\rceil+i\delta}(t)-q\right) - \sum_{i=0}^{k-1}(\lceil p^*\rceil+i\delta)V_{\lceil p^*\rceil+i\delta}(t).
\end{equation*}
A sell order changes the sign of the trade revenue accordingly. 

\subsection{Limit order execution mechanisms}
There are multiple different matching mechanisms used in different markets that decide which limit order is executed, and to what extent, for an incoming market order. 
The order book mechanism producing our data is CME's
Allocation algorithm that is most prominently used for trading money market futures on the CME and is a variation of pro-rata matching. 

In pro-rata matching, orders are matched according to their relative size compared to the entire volume at a price level. For instance, for a fixed price level $p$, let $V_p$ be the volume in the order book and consider an arriving market order of size $q_M<V_p$. The executed amount of a limit order of size $q_L$ is given by 
\begin{equation*}
	\left\lfloor \frac{q_Mq_L}{V_p}\right\rfloor,
\end{equation*}
where $\lfloor\cdot\rfloor$ rounds to the next smaller integer value\footnote{The amount of the market order not executed pro-rata due to the rounding error is usually executed with a different matching algorithm as residual. In this paper we neglect residual execution although explicit modelling would be possible.}. Note that in this case only three values need to be known to determine the executed amount of a limit order. In particular, this allows us to observe the executed amount of limit orders by a trading agent by keeping track of the volume at the time market orders arrive and, thus, also allows us to calculate trade revenues of a limit order strategy.  Pure pro-rata mechanisms are for example used for Eurex equity options and CME agricultural options. 

In a variation of this, the Allocation algorithm prioritizes the execution of the first limit order placed at a more aggressive level over the limit orders arriving at the same level at a later point in time. An arriving market order is first executed against the aggressing limit order, if present. The remaining volume of the market order is executed against the non-aggressing limit orders on a pro-rata basis. For instance, for a fixed price level $p$, let $V_p$ be the volume in the order book, $q_A$ the size of the aggressing limit order and consider an arriving market order of size $q_A<q_M<V_p$. The executed amount of a non-agressing limit order of size $q_L$ is then given by 
\begin{equation*}
	\left\lfloor \frac{(q_M-q_A)q_L}{V_p-q_A}\right\rfloor.
\end{equation*}
While for the Allocation algorithm it is slightly more complicated to determine the executed amount of a non-aggressing limit order, it only requires tracking one additional variable. Thus, it is still feasible to calculate trade revenues for limit order strategies.
\footnote{Another popular matching mechanism is the first-in-first-out (FIFO) algorithm. 
The queue structure of FIFO matching makes following one particular limit order a relatively high dimensional problem as one needs to track all other limit orders that are currently ahead in the queue. 
}

\section{Algorithm}\label{sec:alg}

In this section, we explain how K-NN resampling can be applied to LOB simulation and trade evaluation, making use of the fact that we can observe the direct impact of a trading agent's decision on the LOB. We explain first how K-NN resampling as in  \cite{giegrich2023k} can be used for simulating the dynamics of LOBs with or without interaction before extending the algorithm for the evaluation of trading strategies.

\begin{algorithm}
	\renewcommand{\thealgorithm}{KNNR}
	\caption{ $K$-nearest neighbor resampling for LOB simulation}\label{alg:matchingSim}
	\hspace*{\algorithmicindent} \textbf{Input:}  Data set $\mathcal{D}$; trading strategy $u(\cdot,\cdot)$; order update functions $g_C,g_M,g_L$; nearest neighbors parameter $K$; path length $T_n$; number of resampled trajectories $N$ 
	\begin{algorithmic}[1]
		\STATE{$\text{paths}\gets[\cdot]$}
		\FOR{$r$ from $1$ to $N$}
		\STATE{$\text{path}\gets[\cdot]$}
		\STATE{Sample an initial state-price pair $(\widehat{S^l_{p_0}}(0),\widehat{p_0})$ from $\mathcal{D}$ and append path with $(\widehat{S^l_{p_0}}(0),\widehat{p_0})$}
		\FOR{$s$ from $0$ to $T_n-1$ }
		\STATE{Get agent's orders from the trading strategy $u(\widehat{S^l_{p_s}}(s),s)=(\{c_j\}_{j=1}^m, y,\{x_i\}_{i=1}^n )$}
		\STATE{Apply agent's orders to current LOB snapshot, i.e.\ 
			\begin{equation*}
				(\widetilde{S^l_{p_s}}(s),\widetilde{p_s}) = \prod_{j=1}^n g_L\circ g_M\circ \prod_{i=1}^mg_C\left(\widehat{S^l_{p_s}}(s),\widehat{p_s},(\{c_j\}_{j=1}^m, y,\{x_i\}_{i=1}^n )\right).
		\end{equation*}}	
		\STATE{Randomly choose $k$ from $1,\dots,K$}
		\STATE{Find the $k$-nearest neighbor of $\widetilde{S^l_{p_s}}(s)$ in $\mathcal{D}$ and denote it by $\overline{S^l_{p_{t_j}^*}}(t_j)$} 
		\STATE{Update price $\widehat{p_{s+1}} \gets \widehat{p_s} + (\overline{p_{t_j+\Delta t}^*}-\overline{p_{t_j}^*})$}
		\STATE{Update state $\widehat{S^l_{p_{s+1}}}(s+1) \gets\overline{S^l_{p_{t_j+\Delta t}^*}}(t_j+\Delta t)$}
		\STATE{ Append path with $(\widehat{S^l_{p_{s+1}}}(s+1),\widehat{p_{s+1}})$}
		\ENDFOR
		\STATE {Append paths with path}
		\ENDFOR
	\end{algorithmic}
	\hspace*{\algorithmicindent} \textbf{Output:} paths
\end{algorithm}

\subsection{$K$-NN resampling for LOB simulation with a trading agent}

\begin{figure}
	\begin{subfigure}{.5\linewidth}
		\centering
		\includegraphics[width=.9\linewidth]{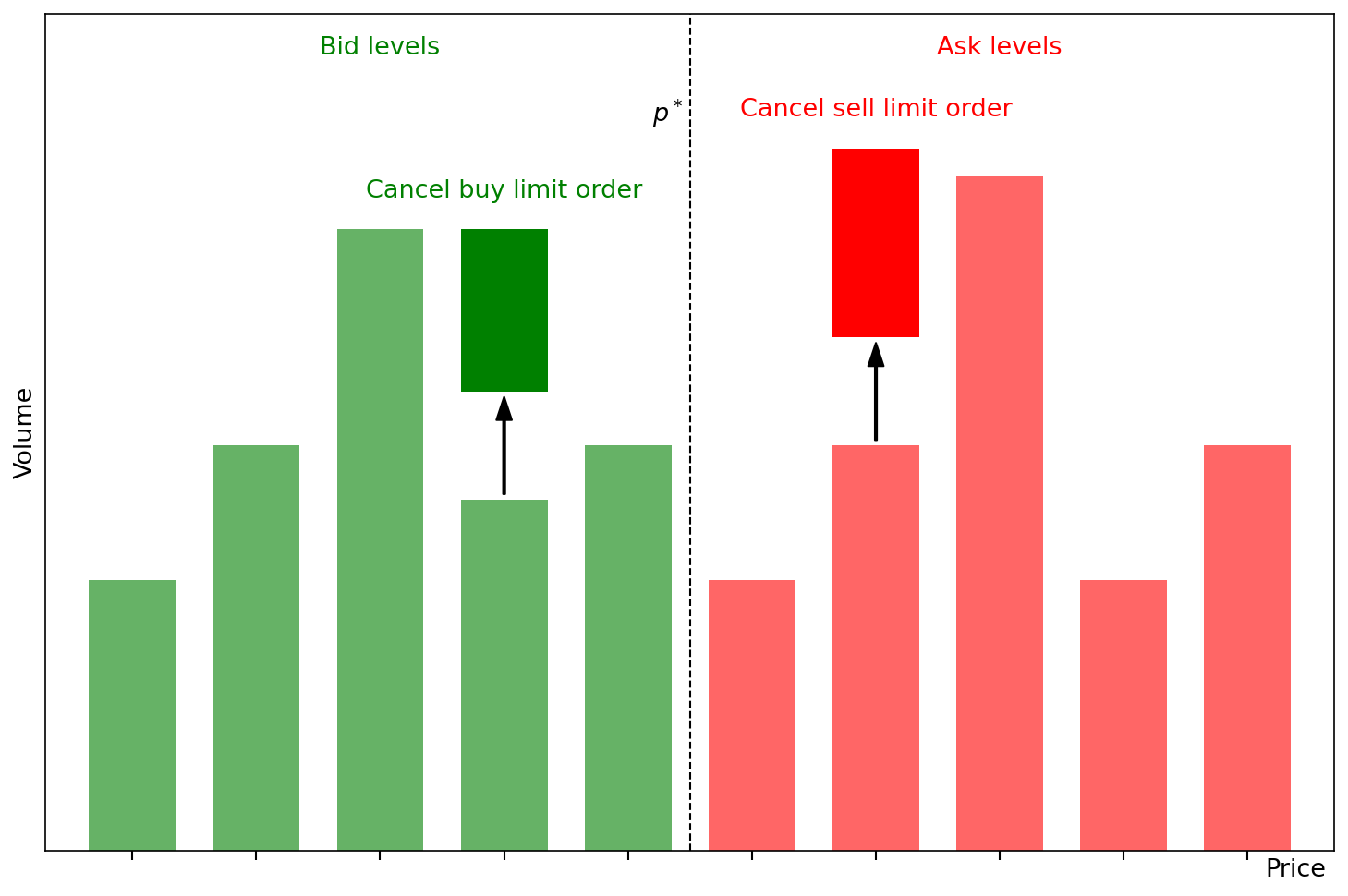}
		\label{fig:loCancel}
	\end{subfigure}%
	\begin{subfigure}{.5\linewidth}
		\centering
		\includegraphics[width=.9\linewidth]{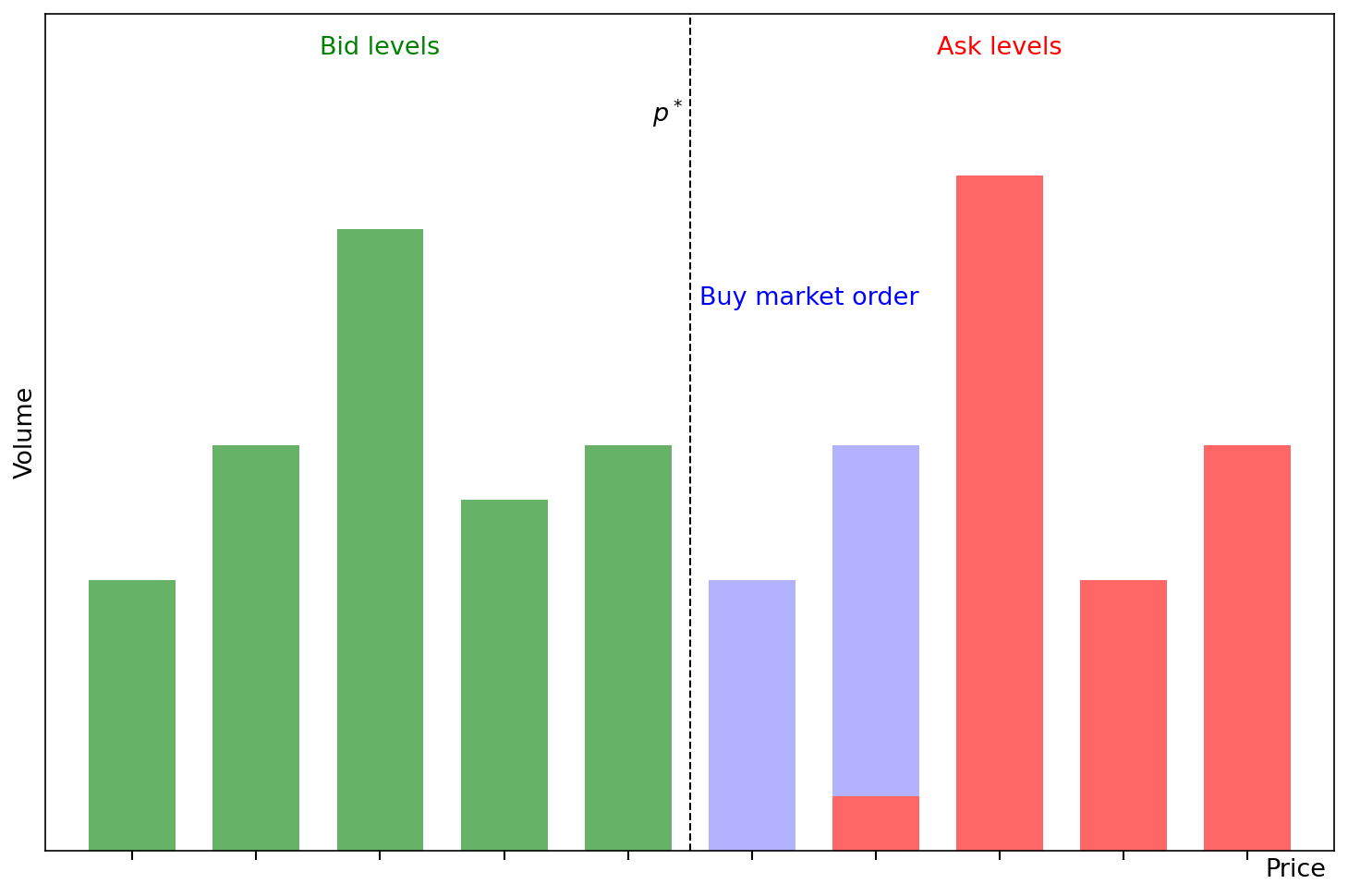}
		\label{fig:moPlacement}
	\end{subfigure}\\[1ex]
	\begin{subfigure}{\linewidth}
		\centering
		\includegraphics[width=.45\linewidth]{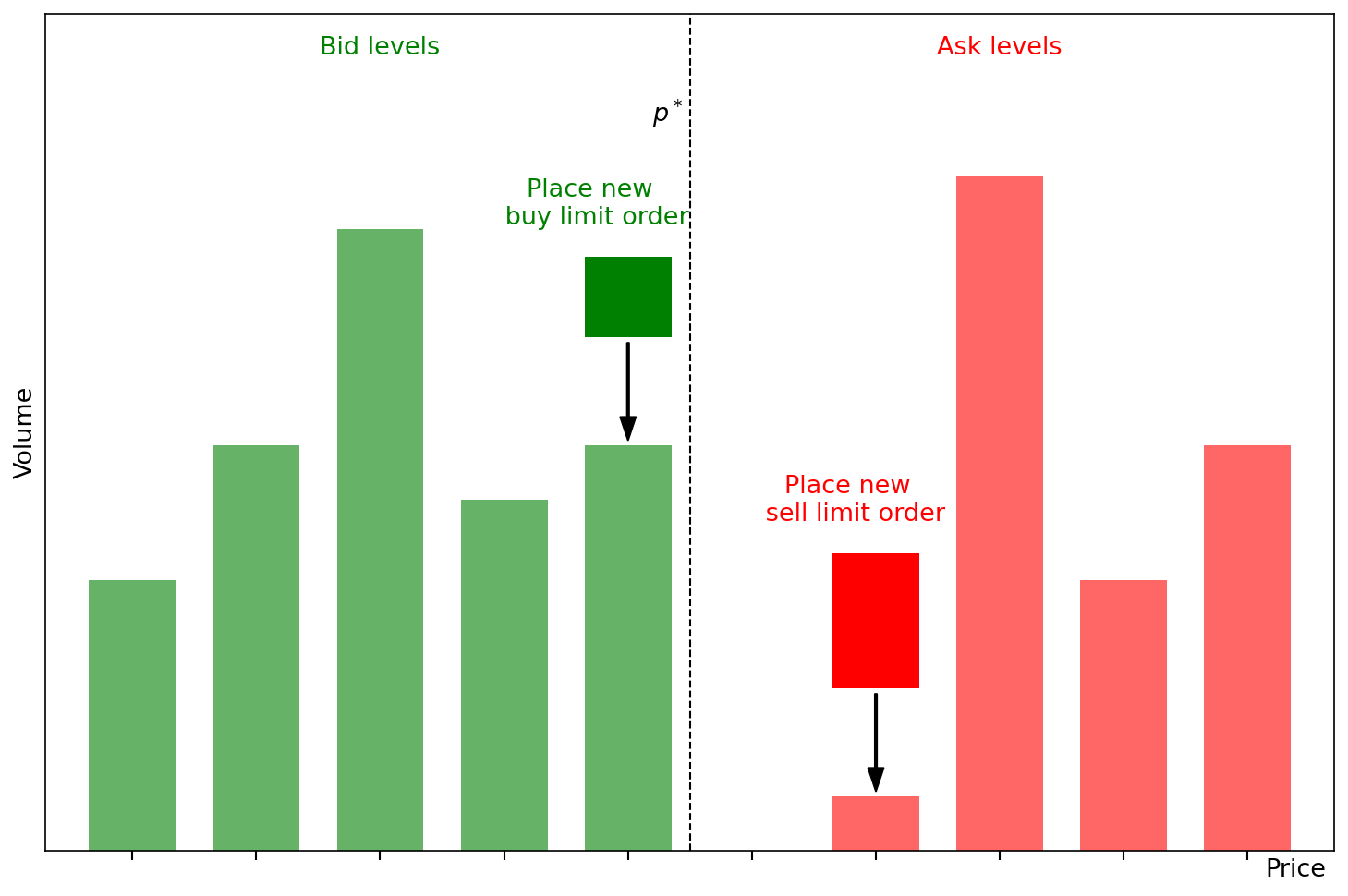}
		\label{fig:loPlacement}
	\end{subfigure}
	\caption{This figure gives an example for the incorporation of actions from a trading agent into the LOB (cf. line $7$ in Algorithm \ref{alg:matchingSim}). In the first plot, the trader cancels a buy limit order and a sell limit order. Then, in the second plot, a buy market order is placed depleting some of the liquidity in the LOB. In the third plot, the trading agent then places a new buy limit order and a new sell limit order in the LOB.}
	\label{fig:loInteract}
\end{figure}

We give a full description of $K$-NN resampling for LOB simulation in Algorithm \ref{alg:matchingSim}. 
We assume that we have access to a data set containing LOB transitions $\mathcal{D}=\{(S^l_{p_{t_i}^*}({t_i}), p_{t_i,i}^*,S^l_{p_{t_i+\Delta t,i}^*}({t_i+\Delta t}), p_{t_i+\Delta t,i}^*)\}_{i=1}^n$ and a trading agent with trading strategy $u(S^l_{p_{s}^*}({s}),s)$ which takes an LOB snapshot $S^l_{p_{s}^*}({s})$ and a time $s\in[0,T]$ as input. The trading strategy then gives us the cancellations, the market order and the limit orders of the trading agent. We choose uniformly randomly an initial state $(\widehat{S^l_{p_0}}(0),\widehat{p_0})=(S^l_{p_{t_i}^*}({t_i}), p_{t_i,i}^*)$. We then pass the state information to the trading strategy to get the trader's action $(\{c_j\}_{j=1}^m, y,\{x_i\}_{i=1}^n ) = u(\widehat{S^l_{p_0}}(0),0)$ . The trader's actions are then incorporated using the functions $g_C(\cdot), g_M(\cdot)$ and $g_L(\cdot)$ leading to the modified state $\widetilde{S^l_{p_0}}(0)$ and price  $\widetilde{p_0}$ corresponding to line $7$ in Algorithm \ref{alg:matchingSim}. We visualize an example for the steps of line $7$ in Algorithm \ref{alg:matchingSim} in Figure \ref{fig:loInteract}. We choose uniformly at random one of the $K$ nearest neighbors of $\widetilde{S^l_{p_0}}(0)$ and denote it by $\overline{S^l_{p_{t_j}^*}}(t_j)$. We then assign $\widehat{S^l_{p_1}}(1) = \overline{S^l_{p_{t_j+\Delta t}^*}}(t_j+\Delta t)$ and calculate the new price $\widehat{p_1} = \widehat{p_0} + (\overline{p_{t_j+\Delta t}^*}-\overline{p_{t_j}^*})$\footnote{Note that this is feasible since we assumed that the price level has no impact on the state transition.}. Repeating this procedure $T_n$-times for $T_n\in\mathbb{N}_{>0}$ gives us one simulated path. We can repeat the path simulation and use the saved paths for our analysis.  

Algorithm \ref{alg:matchingSim} is a so-called `lazy learning' algorithm as it does not require any optimization. This is in contrast to generative deep learning models, where training can be unstable. However, the method still provides a large variability in path generation as for any initial state the algorithm is able to build up to $K^{T_n}$ different possible paths which are all equally likely to occur. Furthermore, Algorithm \ref{alg:matchingSim} can be parallelized and efficiently implemented with a tree-based nearest neighbor search using standard machine learning libraries. Also it only requires the tuning of one hyper-parameter -- the number of nearest neighbors. These considerations make Algorithm \ref{alg:matchingSim} easy to implement and to run on any given data set. We refer the reader to \cite{giegrich2023k} for more details on implementation and computational considerations. 

\begin{rem}\leavevmode
	\begin{enumerate}
		\item For simplicity, we assume that the LOB state is given by centered LOB snapshots. For incorporating a trader's action in a more general LOB state we can introduce a mapping that uses the previous LOB state and the modified LOB snapshot to update the LOB state.
		\item We assume a trading strategy for the trading agent in this section that does not require the explicit modelling of the trader's information process. For the evaluation of trading strategies it may be necessary to do this, for example, by keeping track of the trader's position.
		\item If a simulation without trade intervention is desired, one can simply skip lines 6 and 7 of Algorithm \ref{alg:matchingSim}. 
		\item In our experiments, we choose the nearest neighbor parameter $K=20$ heuristically, which is common practice in machine learning, i.e. \cite{hastie2009elements}. We note that most theoretical results on the choice of $K$  only give growth rates in dependence of the data set size, i.e. \cite{gyorfi2002distribution}, rather then prescribing a specific $K$ for a fixed data set size. An additional consideration, beyond the classical bias-variance trade-off in nearest neighbor regressions is that  the parameter $K$ in Algorithm \ref{alg:matchingSim} also determines the number of possible paths that can be generated and may significantly impact the computational costs of the algorithm.  
		\item As metric for the nearest neighbor search, we use Euclidean distances to keep the implementation as simple as possible. Other choices could be a weighted Euclidean norm putting more emphasis on order book levels closer to the dividing price or an unbalanced Wasserstein metric (e.g.\ \cite{sejourne2021unbalanced}) emphasising a mass interpretation of the order book. To us, the appropriate notion of a metric in a LOB space is an open research question and is beyond the scope of this current work.  
	\end{enumerate}
\end{rem}
\subsection{K-NN resampling for trading strategy evaluation}
Algorithm \ref{alg:matchingSim} cannot only be used for LOB simulations but with slight adjustments also for the evaluation of trading strategies. In particular, the trader's information process may need to be modelled explicitly to keep track of the trader's current position, inventory and trading revenues. The agent's trading strategy may depend on this additional information. To gather the information on position, inventory and trading revenues, we record all the trades executed against either limit orders or market orders entered by the trader. For market orders, this is feasible for any common execution mechanism. For limit orders, keeping track is directly possible in LOBs using either pro-rata or CME's Allocation execution mechanism. We do this by replaying the trades that occurred between the LOB snapshots in the historical data set and using CME's Allocation mechanism for attribution as explained in the previous section.  For FIFO LOBs, one could potentially use a parametric or non-parametric model for the execution of limit orders. 

\section{Data set}\label{sec:data}
\begin{figure}
	\begin{subfigure}{.5\linewidth}
		\centering
		\includegraphics[width=.9\linewidth]{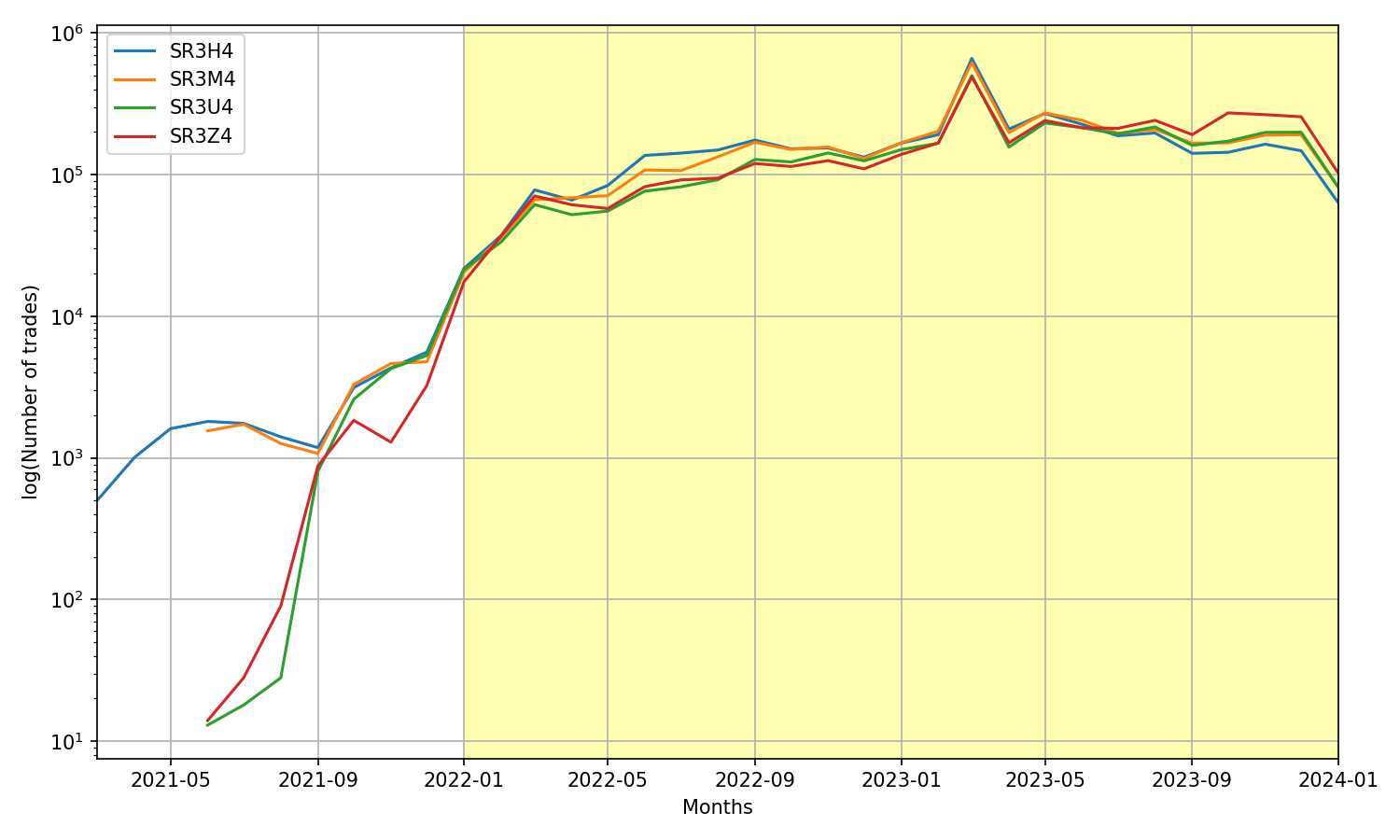}
		\label{fig:tradeCount}
	\end{subfigure}%
	\begin{subfigure}{.5\linewidth}
		\centering
		\includegraphics[width=.9\linewidth]{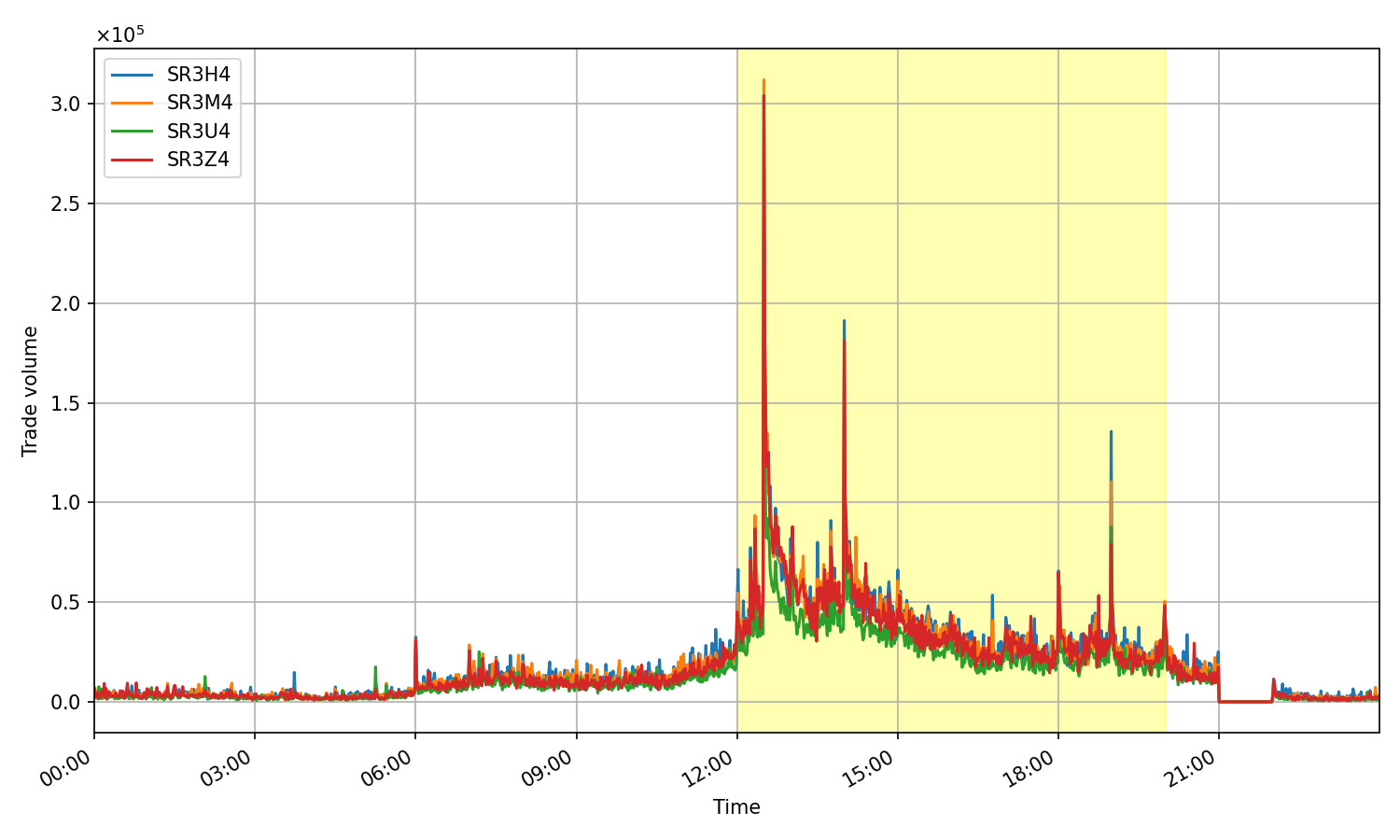}
		\label{fig:dailyTrade}
	\end{subfigure}\\[1ex]
	\begin{subfigure}{\linewidth}
		\centering
		\includegraphics[width=.45\linewidth]{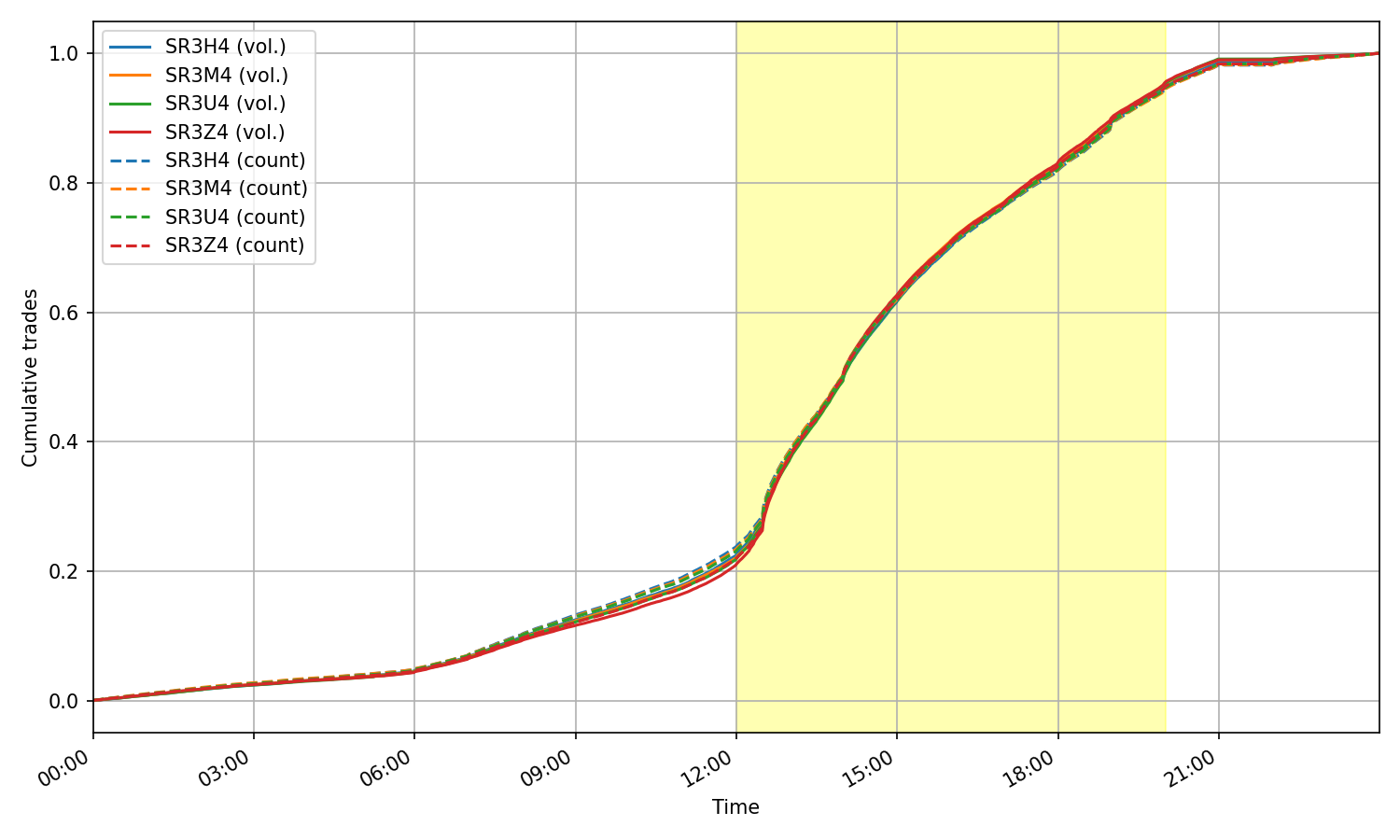}
		\label{fig:dailyTradeCum}
	\end{subfigure}
	\caption{Monthly trade count for different active contracts in data collection (left); Minute-by-minute trading volume on 02/11/2023 (right); Normalized cumulative trading volume and count on 02/11/2023 (bottom)}
	\label{fig:dataSummaries}
\end{figure}

We use trading data over a period of two years for 3-month Secured Overnight Financing Rate (SOFR) futures traded at the CME from January 2022 to January 2024 (see first plot in Figure \ref{fig:dataSummaries}). SOFR replaced the USD LIBOR as the main overnight interest rate in the US and is based on actual repo transactions on US treasuries \citep{heitfield2019inferring}. The 3-month SOFR futures traded at the CME replaced Eurodollar futures and are a main source for price discovery in short-term interest rates. The contracts are used by traders for hedging short-term interest rates exposures and for yield curve and spread strategies. Furthermore, they are highly liquid with trading of over 5 million contracts each day on average \citep{CME2024SOFR}. For execution, the futures are centrally cleared using a limit order book where limit orders are executed according to CME's Allocation algorithm.

We use daily level 2 LOB data from CME, which contains all order book movements but does not allow us to follow individual orders. For the order book, we are given the first 5 levels with active orders for both bid and ask side with tick size $\delta = 0.005$. Additionally, we observe all executed trades from the same source and match this data with the LOB data over the exchange time stamp. We restrict ourselves to the contracts SR4H4, SR4M4, SR4U4 and SR4Z4 (see first plot in Figure \ref{fig:dataSummaries} for monthly trading activity during data collection) and treat them as a single data set\footnote{This is motivated by the fact that all contracts are futures on the same underlying and by the observation in \cite{sirignano2019universal} that models trained on LOB data from multiple assets tend to perform better in replicating price dynamics compared single asset models.}. Furthermore, we only consider the liquid trading times from 12pm to 8pm New York time. For an exemplary day, the second and third plot in Figure \ref{fig:dataSummaries} support this choice. 

After compiling this raw data, we aggregate it by taking a snapshot of the LOB every 250 order book events. We record all trades, order book sizes at trade execution and limit order sizes of aggressions to new ticks occurring between the snapshots to allow us to evaluate limit order trading strategies according to CME's Allocation algorithm. This leads to a data set with around $1.2\times10^7$ samples where each sample contains an initial LOB snapshot and price and the corresponding LOB snapshot and price after 250 order book events. 
Additionally, a sample contains the information as described above on all trades that occurred during the 250 order book events. To simplify the replay of historical LOB dynamics for testing, we order the transitions chronologically by contracts.

\begin{rem}[Event time vs. real time]
Note that in our experiments the time we consider is event time. This means we take each change in the order book as a discrete time step and the elapsed time corresponds to the number of order book changes since the beginning of the time period. We make this modelling decision as it allows the trader to automatically adapt their trading speed with market activity. Algorithm \ref{alg:matchingSim} can equally be applied to real time scenarios. 
\end{rem}

\section{Simulation results}\label{sec:results}

This section presents and discusses different outputs from applying Algorithm \ref{alg:matchingSim} to real LOB data. In parts, we will use evaluation methods for LOB simulations proposed in \cite{coletta2023conditional,cont2023limit,nagy2023generative, hultin2023generative}. First, we will focus on the simulation capability of Algorithm \ref{alg:matchingSim} without any interactions by a trading agent. If we compare the simulation results to actual market replays, we consider out-of-sample performance. This means we divide the data set in two parts with a 80\%-to-20\% split\footnote{We split our data set deterministically in accordance to the ordering introduced in Section \ref{sec:data}. This means that testing data stems mostly from a single contract and chronological consistency of the split is kept within each contract, see e.g.\ \cite{ruf2019neural}.}. We use the smaller set as ground truth and for initial states in resampling. The larger set is used as data source for the resampled transitions. We use this splitting procedure to ensure that the shown simulation results are not biased from reusing the actual market observations.  For investigating market impact, we then consider simulations where a trading agent either places market or limit orders.

If not mentioned otherwise, the results are obtained with the following choices of parameters. As nearest neighbor parameter, we use $K=20$. For the episode length, we choose $T_n=60$  transitions corresponding to $1.5\times 10^4$ order book events. The experiment evaluation is based on $N=10^4$ resamplings. 

\subsection{Resampling statistics}

\begin{figure}[h]
	\begin{subfigure}{.5\textwidth}
		\centering
		\includegraphics[width=.8\linewidth]{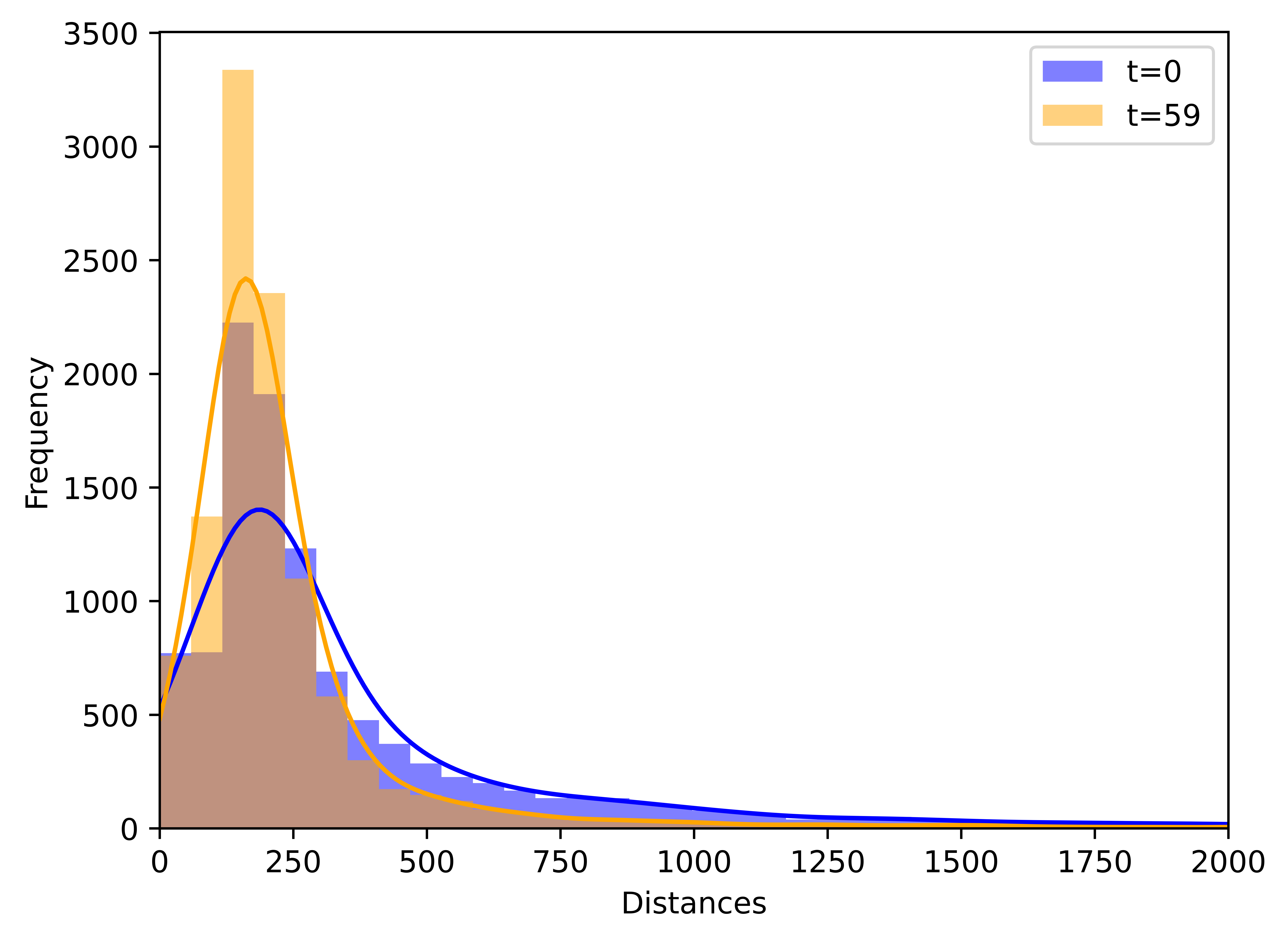}
		\label{fig:NNIndDist}
	\end{subfigure}%
	\begin{subfigure}{.5\textwidth}
		\centering
		\includegraphics[width=.8\linewidth]{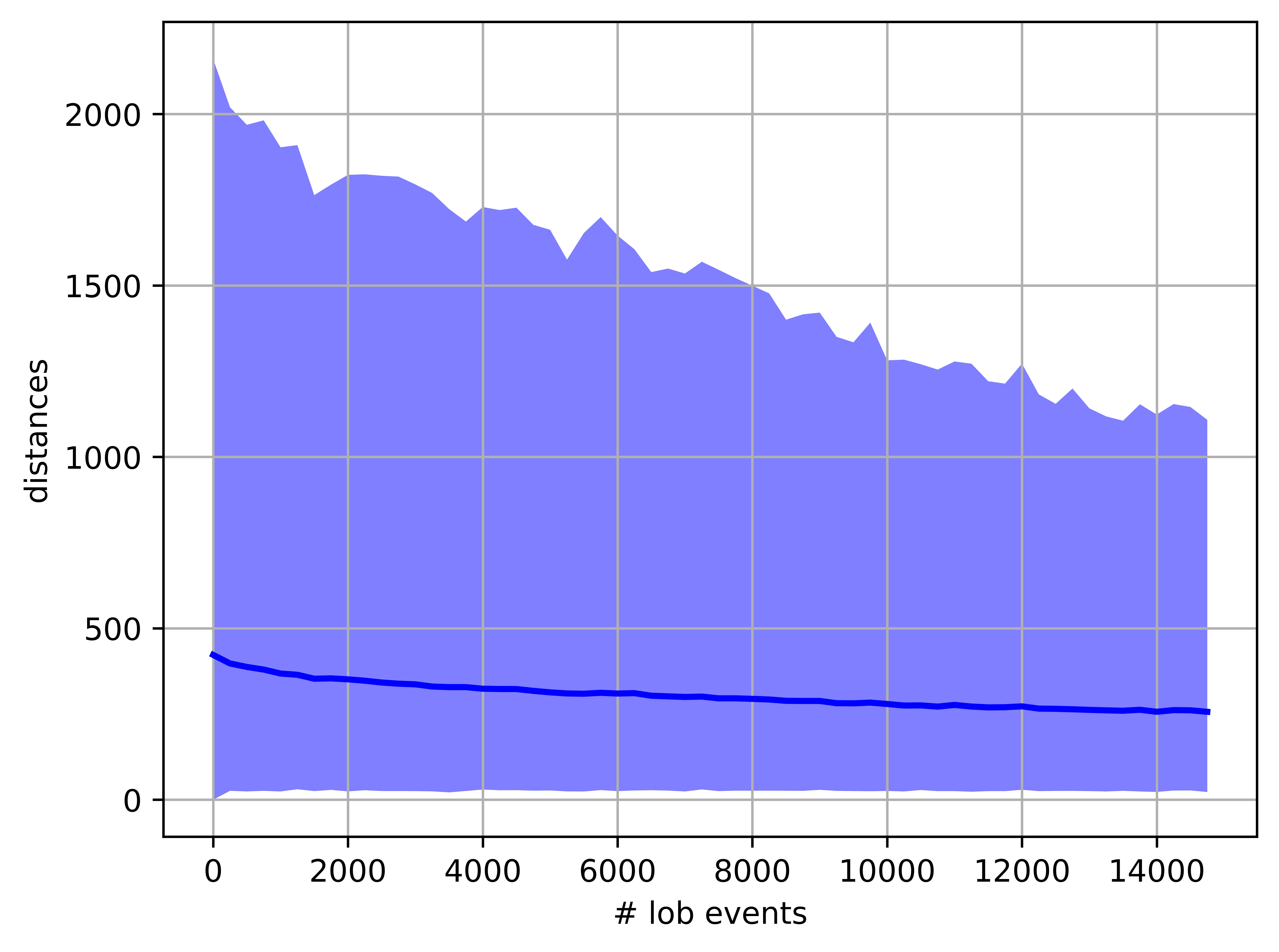}
		\label{fig:NNIndDistDynamic}
	\end{subfigure}
	\caption{Resampling statistics: Distances distribution (left); distances dynamics (right)}
	\label{fig:NNInd}
\end{figure}

We start by reporting several observations on Algorithm \ref{alg:matchingSim}. The first plot in Figure \ref{fig:NNInd} shows the distribution of distances between matched states at the beginning of the resampling path and at the end, i.e.\
\begin{equation*}
	\|\widetilde{S^5_{p_s}}(s)-\overline{S^5_{p_{t_j}^*}}(t_j)\|
\end{equation*}
where $\|\cdot\|$ is the Euclidean norm and $s=0$ or $s=59$. In general, smaller distances in a nearest neighbor matching are associated with a smaller bias in the quantity of interest. We observe that the distributions at both points in time are heavily skewed to the left with outliers to the right. Comparing the distances at the initial and the final matching reveals that the distances at the beginning are larger on average and the right tail is less heavy compared to the last transition. The second plot in Figure \ref{fig:NNInd} confirms this observation. Here, we plot the average distance and its 0.95-quantile against time. We observe that both metrics decrease over time.  This can potentially be explained by the uniformly random initialization of the state as edge cases are more likely to occur in a uniform random framework compared to a nearest neighbor choice where centrality is preferred. This effect could be mitigated by a firmer restriction on less liquid market regimes or an adapted initial sampling scheme. On the other hand, the decreasing distance over time  also implyies a reduced bias contribution of later transitions.

\subsection{Single transitions}
As a next step we compare several properties of a single LOB transition from an actual historic transition to a transition induced by Algorithm \ref{alg:matchingSim}. For all experiments, we use the same initial states for the historical roll-outs and the matched nearest neighbors transitions. 
\begin{figure}[h!]
	\centering
	\begin{tabular}{ccc}
		\begin{subfigure}[t]{0.265\textwidth}
			\centering
			\includegraphics[width=\linewidth]{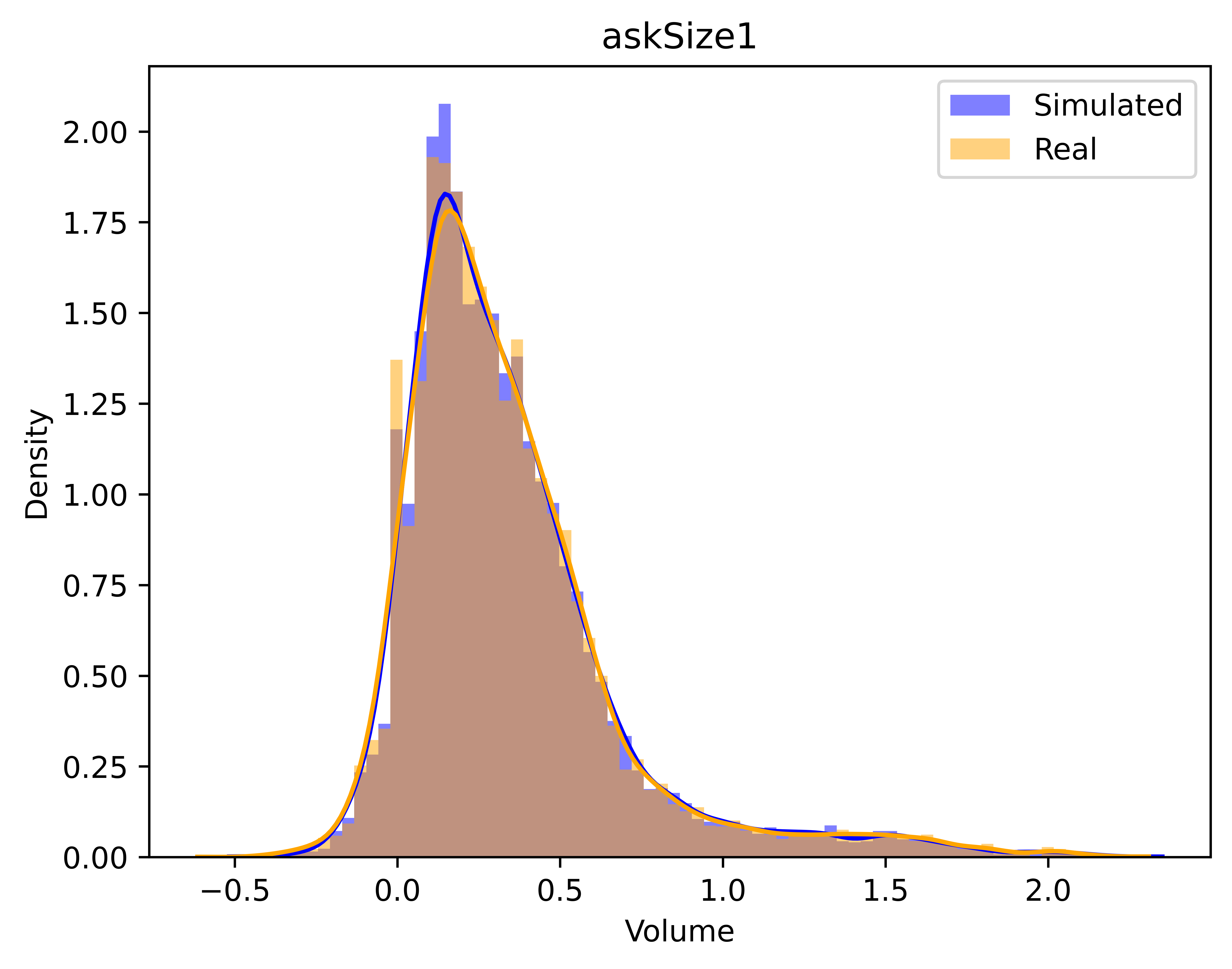}
			\label{fig:hist_ask1}
		\end{subfigure} &
		\begin{subfigure}[t]{0.265\textwidth}
			\centering
			\includegraphics[width=\linewidth]{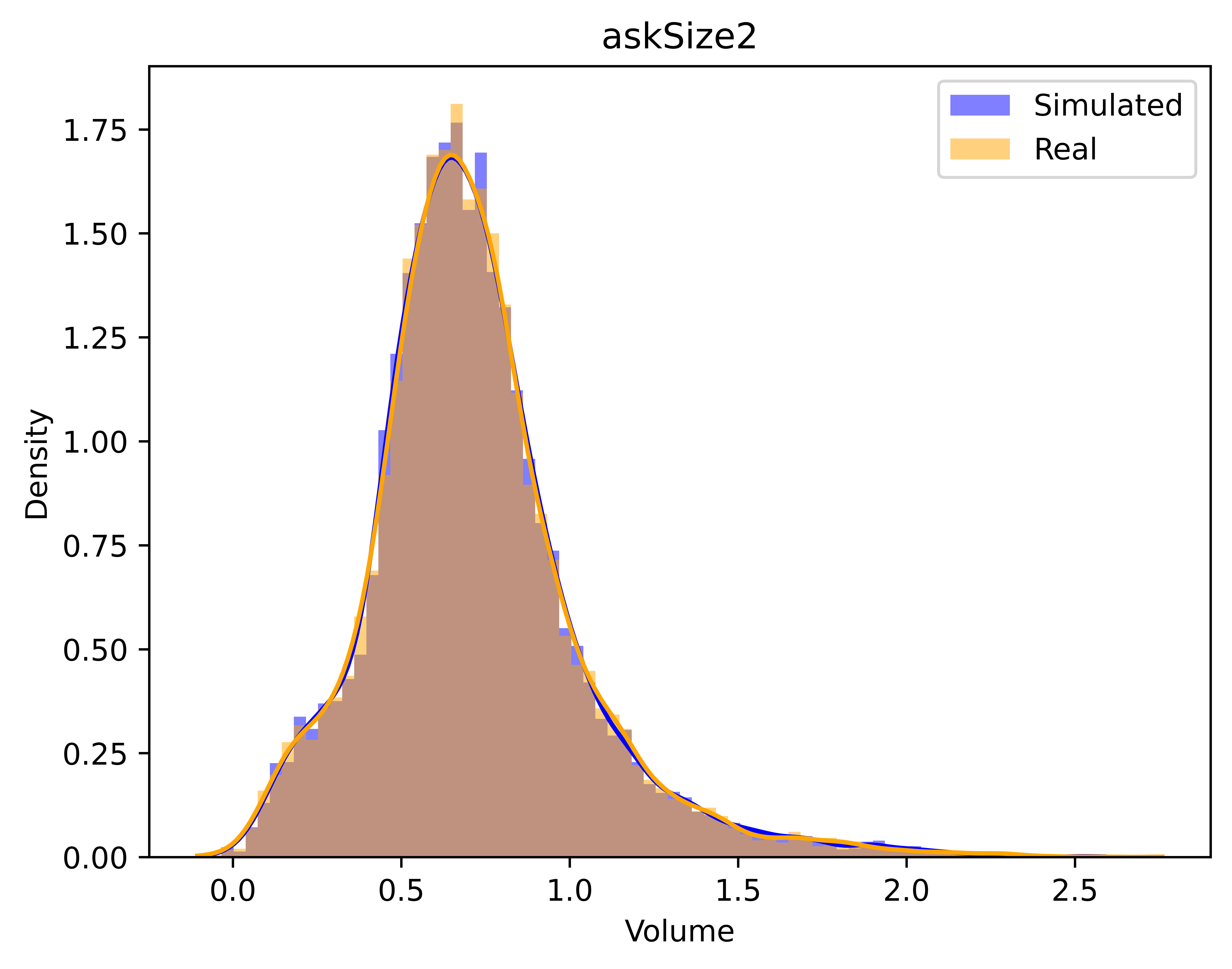}
			\label{fig:hist_ask2}
		\end{subfigure}
		& \multirow{2}{*}[2cm]{
			\begin{subfigure}{0.43\textwidth}
				\centering
				\includegraphics[width=.95\linewidth]{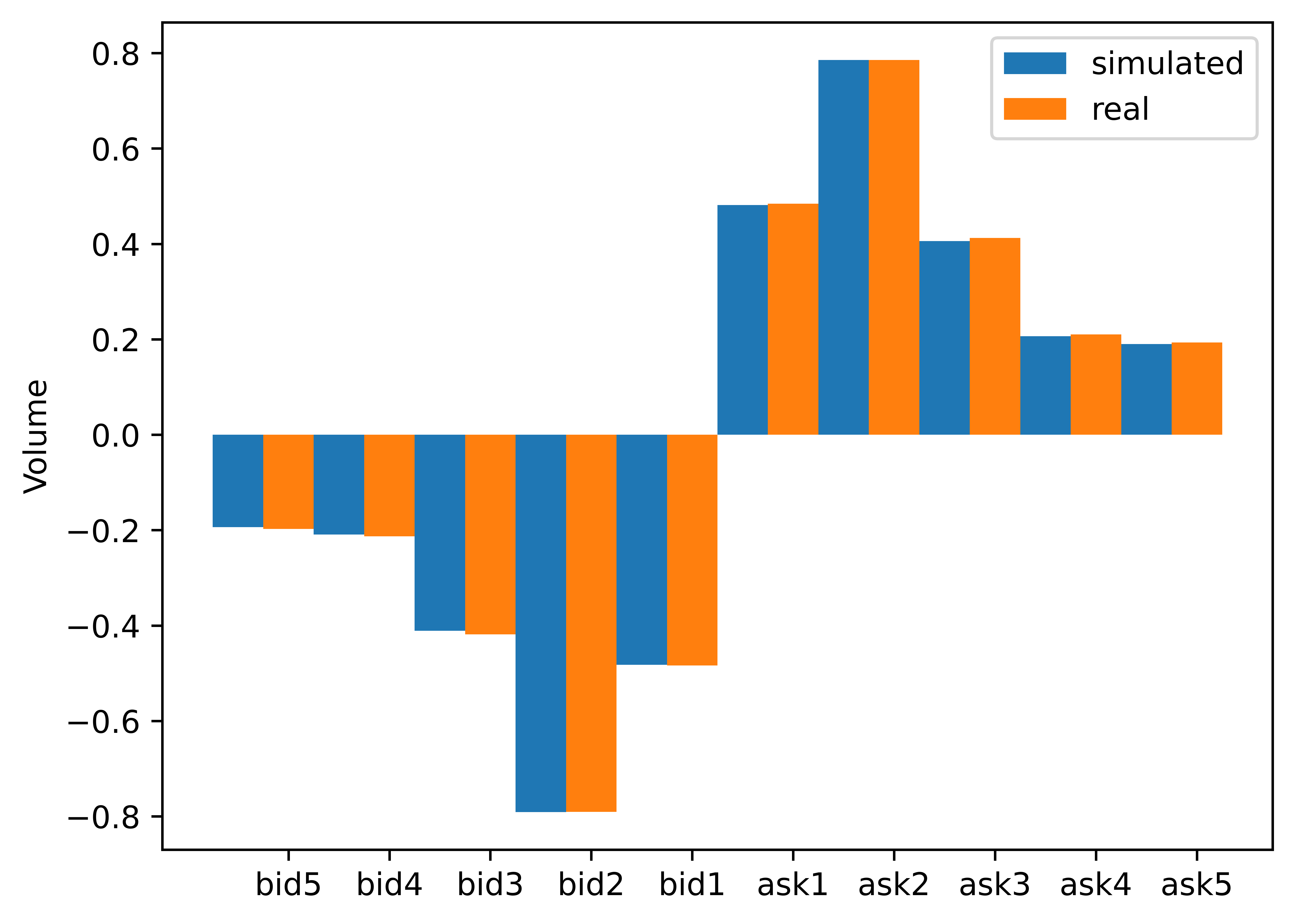}
				\label{fig:averageSizes}
		\end{subfigure}} \\
		\begin{subfigure}[t]{0.265\textwidth}
			\centering
			\includegraphics[width=\linewidth]{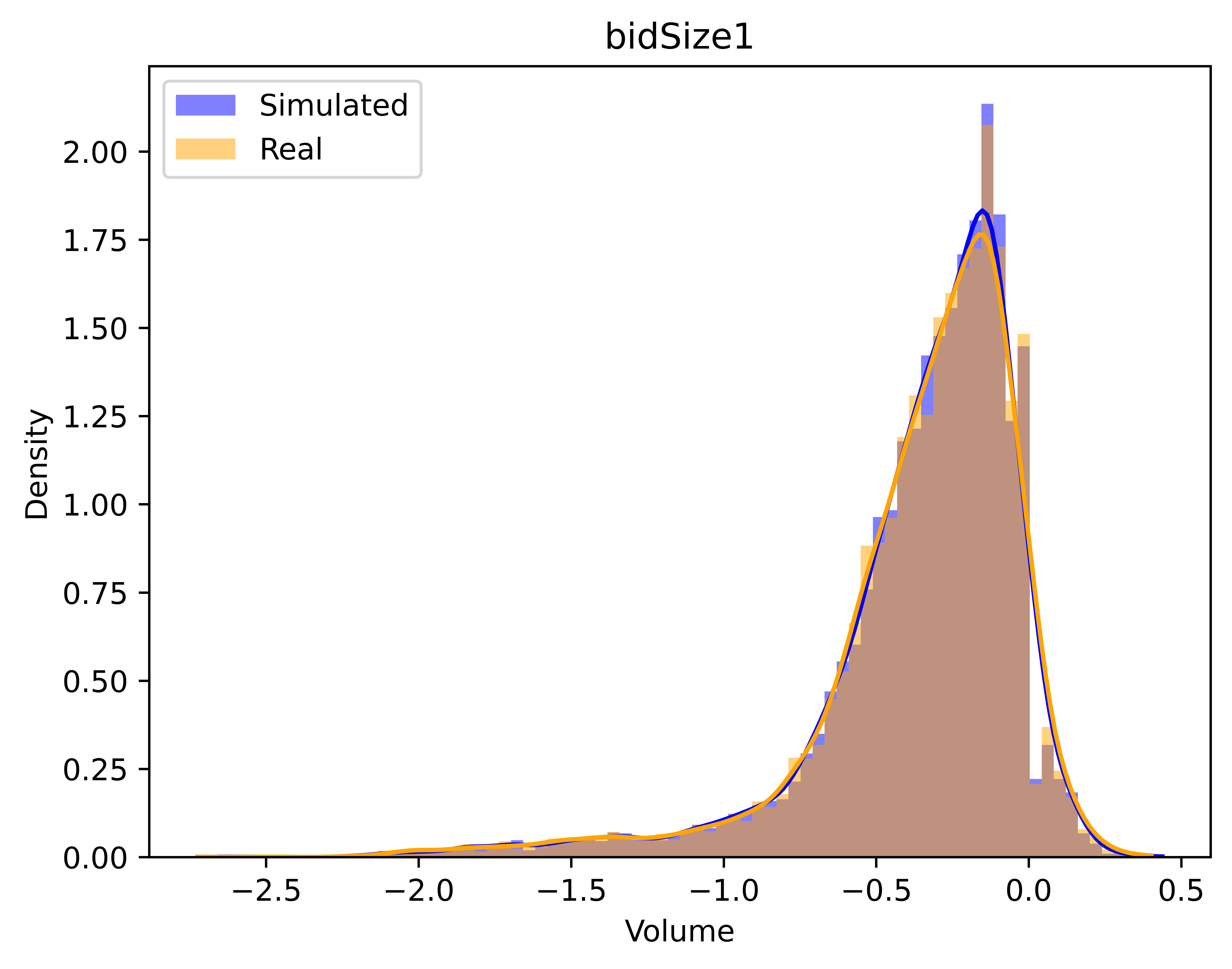}
			\label{fig:hist_bid1}
		\end{subfigure} &
		\begin{subfigure}[t]{0.265\textwidth}
			\centering
			\includegraphics[width=\linewidth]{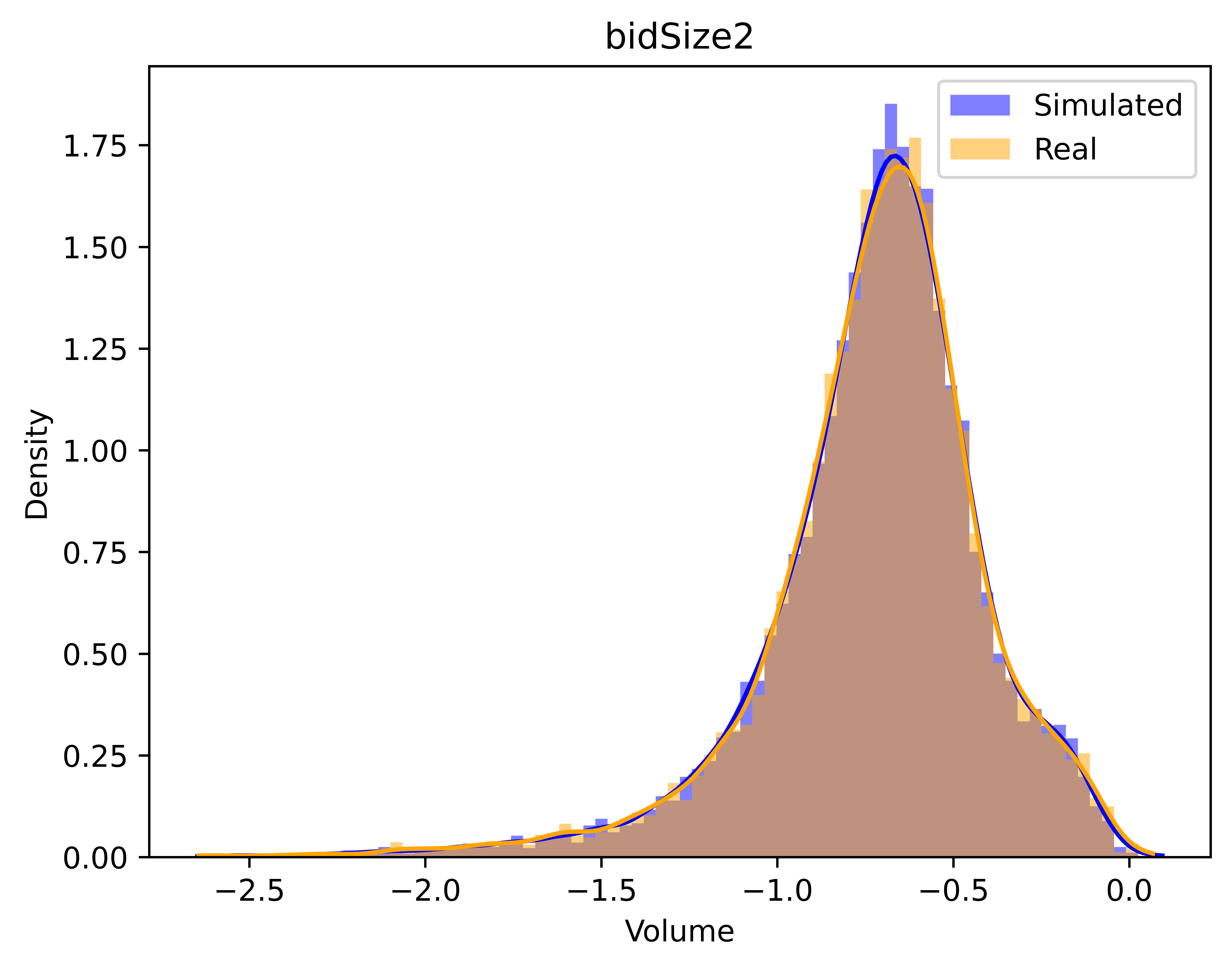}
			\label{fig:hist_bid2}
		\end{subfigure}
		&
	\end{tabular}
	\caption{Volume comparison: Marginal volumes per order book level after one transition (left); Average LOB shape after one transition (right)}
	\label{fig:hist_marginals}
\end{figure}
First, we consider the real and resampled marginal volume distribution at different order book levels. For this, we will use a slightly different formulation for LOB snapshots introduced in \cite{cont2023limit}, i.e., we will fix the dividing price $p^*_0$ from before transitioning and center the LOB after the transition at that price. We use signed volumes that are negative for bid order volumes and positive for ask order volumes: for a  bid price level $\lfloor p_0^*\rfloor-k\delta$ with $k\in\mathbb{N}_{>0}$ and $t=1$, we will consider the volume  $V_{\lfloor p_0^*\rfloor-k\delta}(t)$ with a  negative sign if $\lfloor p_0^*\rfloor-k\delta$ is on the bid side at time $t$, and with a positive sign else. We denote the signed volume by $\tilde{V}_{\lfloor p_0^*\rfloor-k\delta}(t)$. The definition is analogous for ask prices. This formulation allows us to compare the volume at a given price level and include information on order book movements, i.e., if we consider a bid price $\lfloor p_0^*\rfloor-k\delta$  for positive $\tilde{V}_{\lfloor p_0^*\rfloor-k\delta}(t)$ we can infer that the market moved downwards and $\lfloor p_0^*\rfloor-k\delta$ is an ask price at $t=1$.  Furthermore, we normalize the volume $V$ with the square root transformation $\sign(V)\sqrt{|V|}/100$ as in \cite{cont2023limit} to control the skewness of the volume distribution. 

On the left in Figure \ref{fig:hist_marginals}, we give the distribution of signed volumes after one transition for the real market and the simulated market. We report the results for the two best bid and ask price levels from before transitioning, i.e.\ $k=0,1$. Overall the simulated market provides a very good fit to the real market. Furthermore, we observe that for the best price levels (askSize1 and bidSize1) the volumes tend to be smaller than for the second best price levels (askSize2 and bidSize2). If the volume's sign at the best price level changes, the newly established volume tends to be small on the opposing order book side and the probability of a sign change is relatively small. For second best price levels there is almost no mass on the opposite sign implying that market moves of two ticks are highly unlikely. 

On the right in Figure \ref{fig:hist_marginals}, we give the average (normalized) volume after a transition for the real and the simulated LOB on different LOB levels. We report bid volumes as being negative signed and ask volumes as positive. The real and simulated average volumes are almost indistinguishable. Very small differences can be observed on the LOB levels furthest away from the spread on both sides where the simulated market has slightly less volume compared to the real market. Comparing this result to a benchmark which unconditionally and uniform randomly selects a transition from the training data (cf. Figure \ref{fig:hist_marginalsNaive}) reveals that Algorithm \ref{alg:matchingSim} is better in fitting the average volumes for the samples from the test data. The discrepancy between the benchmark and the averages from the test data points to a slight distributional shift between the training and test split of the data. This distributional shift seems to be captured by Algorithm \ref{alg:matchingSim} and shows that our method goes beyond an unconditional law of large numbers. We further discuss this comparison in Appendix \ref{sec:bench}.

\begin{figure}[H]
	\begin{subfigure}{.8\textwidth}
			\centering
		\includegraphics[width=1.0\linewidth]{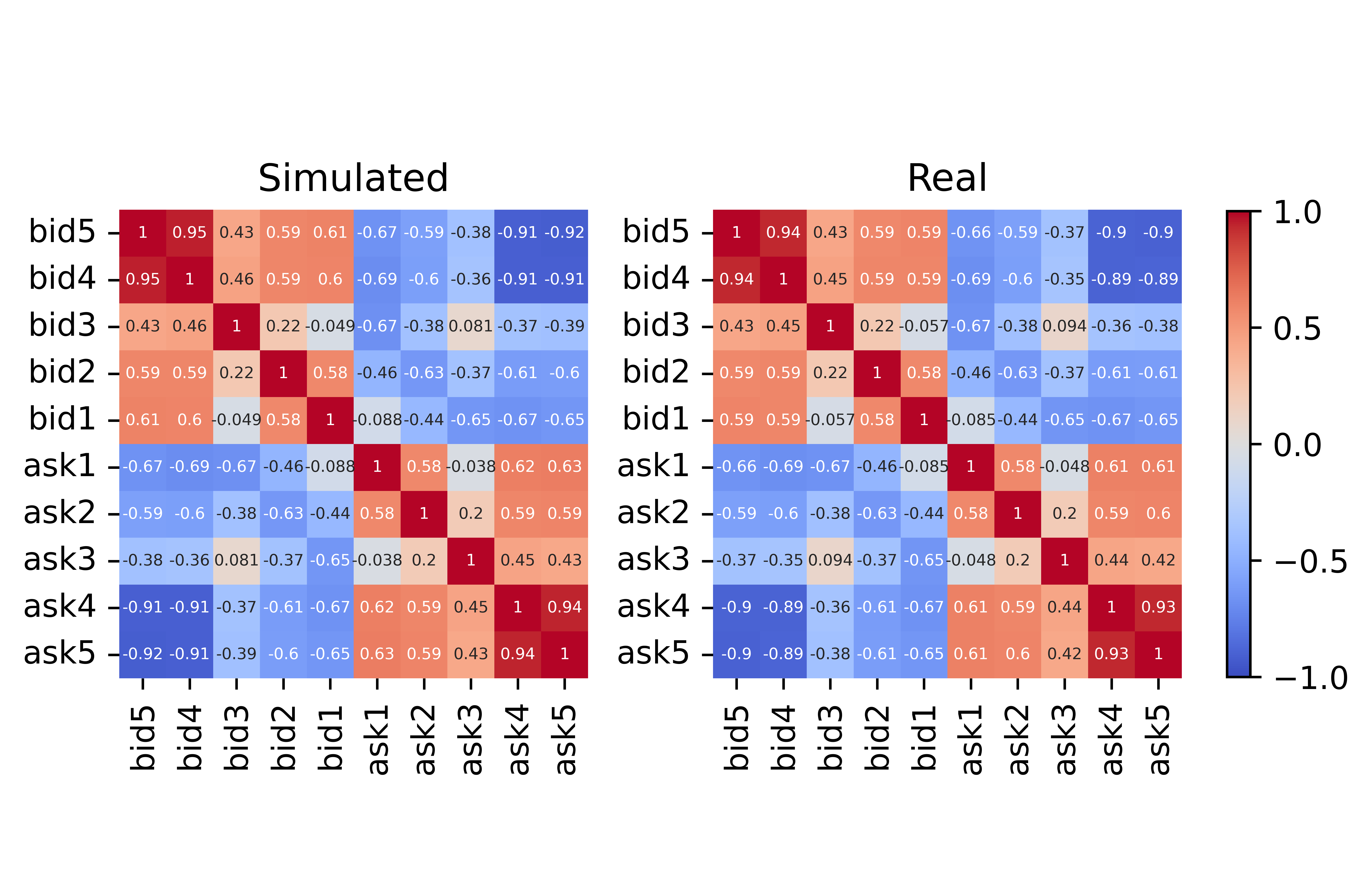}
		\label{fig:corrStatic}
	\end{subfigure}\\
\begin{subfigure}{.8\textwidth}
	\centering
\includegraphics[width=1.0\linewidth]{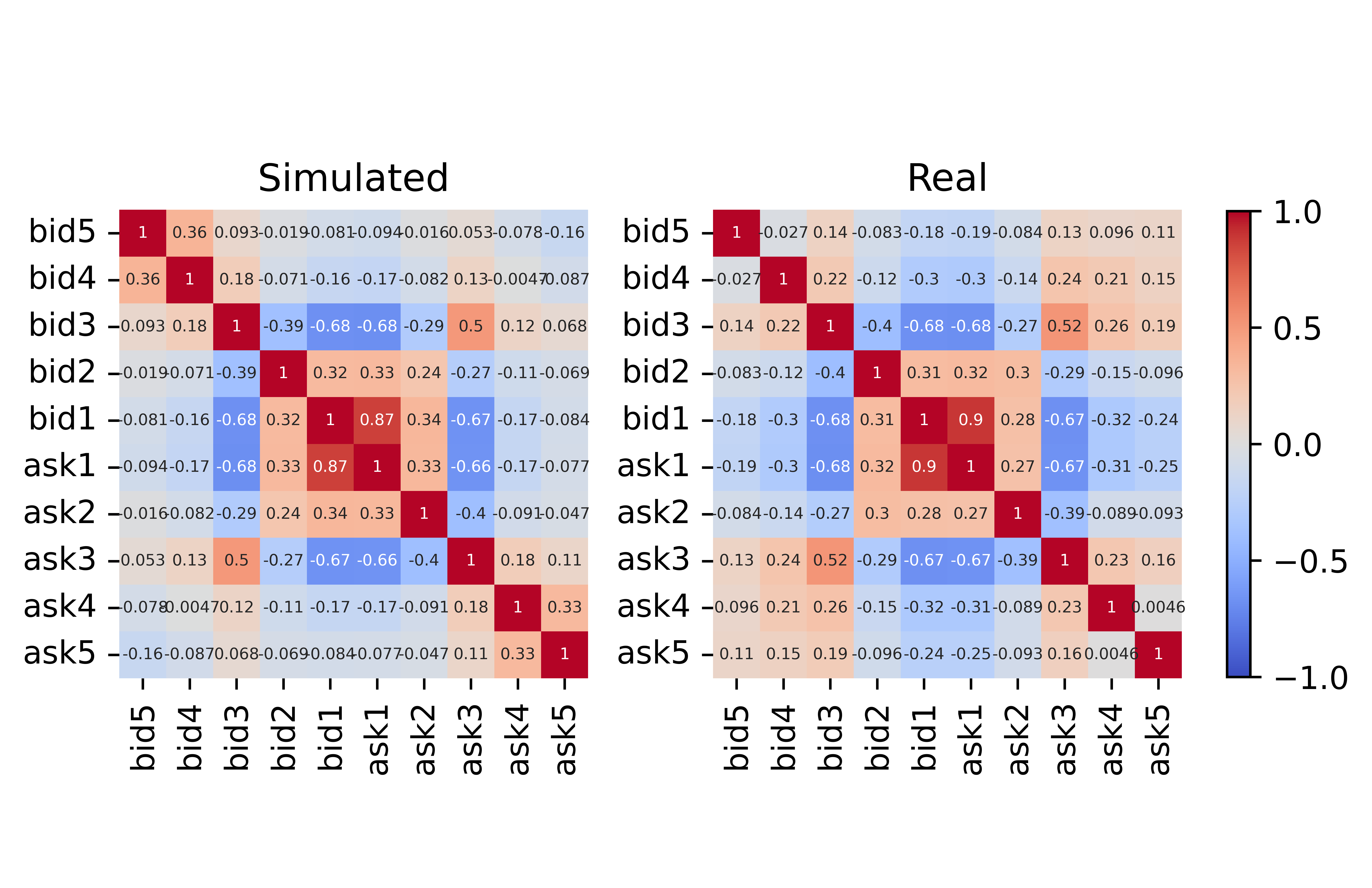}
	\label{fig:corrDiffs}
\end{subfigure}
	\caption{Correlation: Absolute volumes (first row); Volume differences (second row)}
	\label{fig:corrVol}
\end{figure}

Figure \ref{fig:corrVol} compares the correlation structure of LOB volumes. The first row gives the correlation between different order book level volumes after the resampled transitions and after the real transitions. The simulated correlation structure and the real structure correspond very well. Together with the results on the marginals (Figure \ref{fig:hist_marginals}), this points to an overall good fit for the joint distribution of volumes generated by a resampling transition. In the second row of \ref{fig:corrVol}, we consider the correlation of changes in volume for a given order book level over one transition. For central order book levels (bid and ask 1-3), the correlation of the difference coincides well for the real and the simulated transition. Thus, also the interdependence in the change of volume is appropriately captured for the central order book levels. For the outer order book levels, we observe mismatches. This may be related to the observation that volume on these levels are often relatively time persistent and do not tend to change much. As these levels are often less important for price formation, one could remove them from the order book matching depending on the simulation objective.

\subsection{Dynamic comparison}
We turn to evaluating the simulation performance of Algorithm \ref{alg:matchingSim} over multiple time steps.  
\begin{figure}[h!]
	\begin{subfigure}{.5\textwidth}
		\centering
		\includegraphics[width=.8\linewidth]{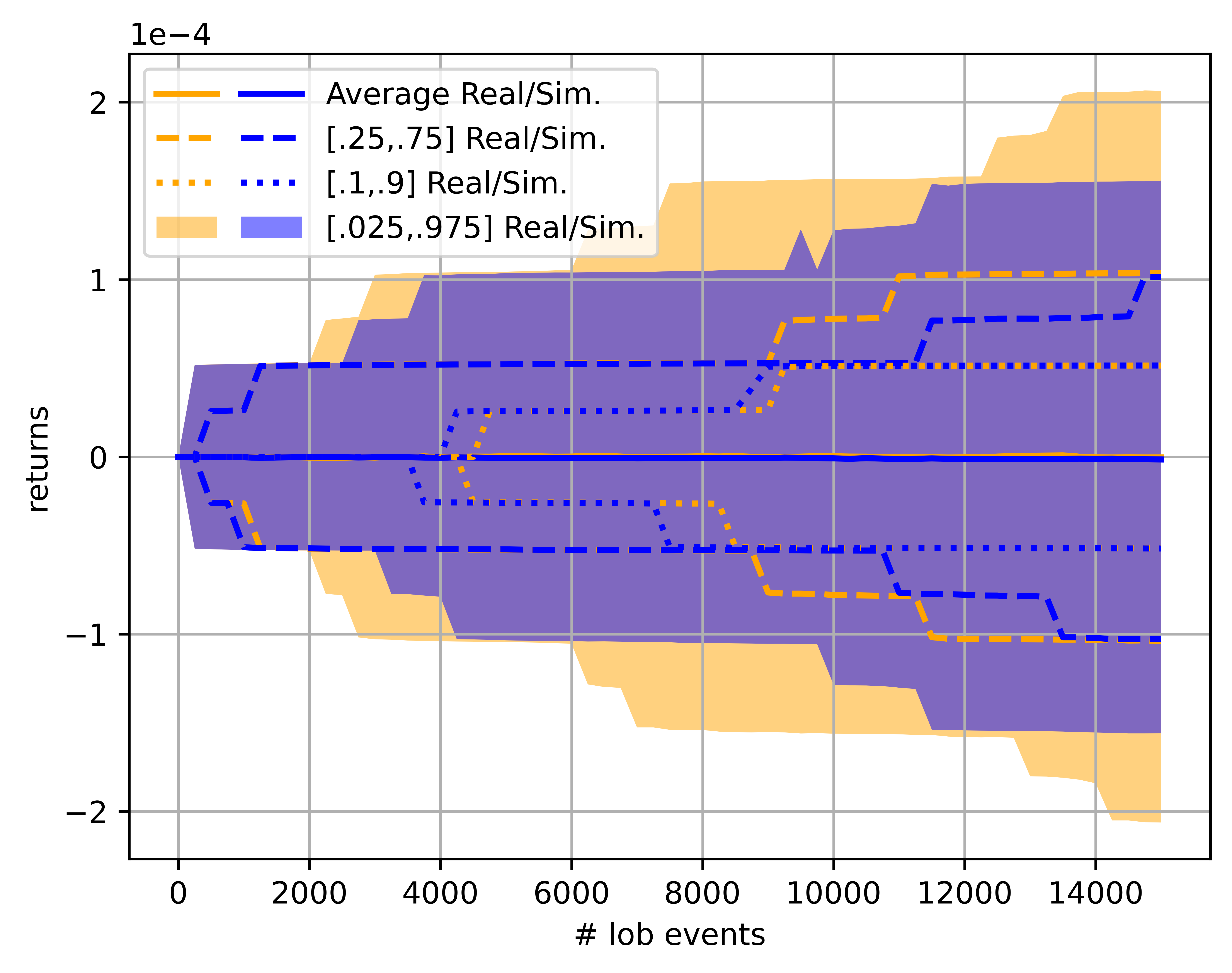}
		\label{fig:dynamicReturnMid}
	\end{subfigure}%
	\begin{subfigure}{.5\textwidth}
		\centering
		\includegraphics[width=.8\linewidth]{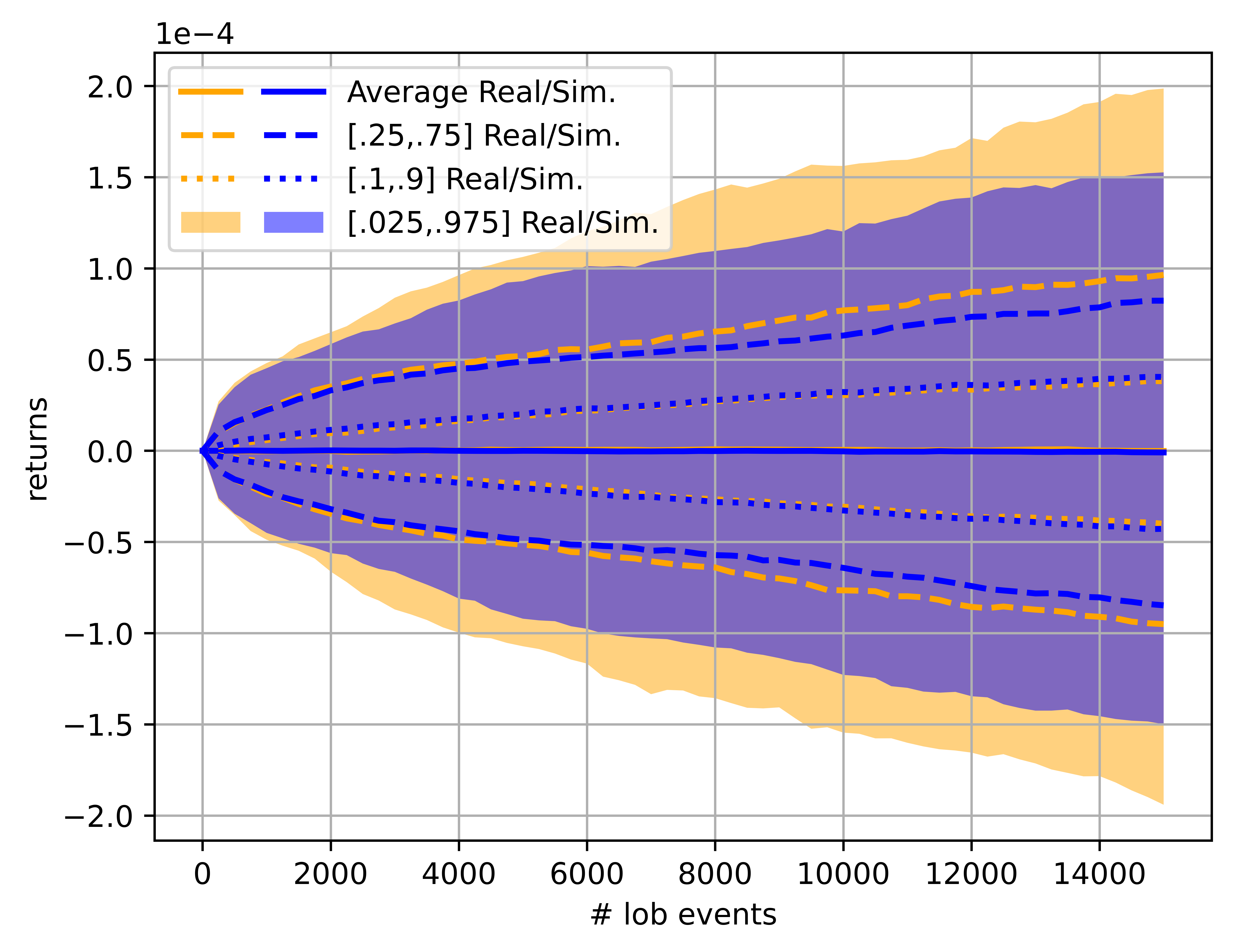}
		\label{fig:dynamicReturnWeighted}
	\end{subfigure}
	\caption{Comparison of real return time series with simulations by Algorithm \ref{alg:matchingSim}: Mid-price (left); Weighted mid-price (right)}
	\label{fig:dynamicReturn}
\end{figure}
First, we compare the unconditional return distribution of the resampled LOB dynamics and the real market returns. For this purpose, we choose $10^4$ real LOB paths with 60 transitions each and calculate the corresponding return paths. Reusing the initial states from the real paths as starting points, we simulate $10^4$ LOB paths using Algorithm \ref{alg:matchingSim} with the same length and calculate the returns along them. Figure \ref{fig:dynamicReturn} plots the average return (line) and several of its quantiles for the real and simulated market. On the left, the mid-price returns are compared and one can see that the averages almost coincide. The quantiles mostly match as well, however one can see a slight tendency that the real market reaches more extreme levels of return slightly faster compared to the simulation. The weighted returns, shown on the right in Figure \ref{fig:dynamicReturn}, do not only reflect tick-sized changes in the price but also represent shifts in the positioning on the highest levels of the LOB. This is especially relevant in pro-rata order book markets as the execution mechanism requires us to have a realistic view on the volumes as well. The averages of the real and simulated return almost coincide again. The return quantiles have a similar coverage. However, the quantiles of the real return tend to be slightly larger compared to the simulated return quantiles.

\begin{figure}[h!]
	\begin{subfigure}{.5\textwidth}
		\centering
		\includegraphics[width=.8\linewidth]{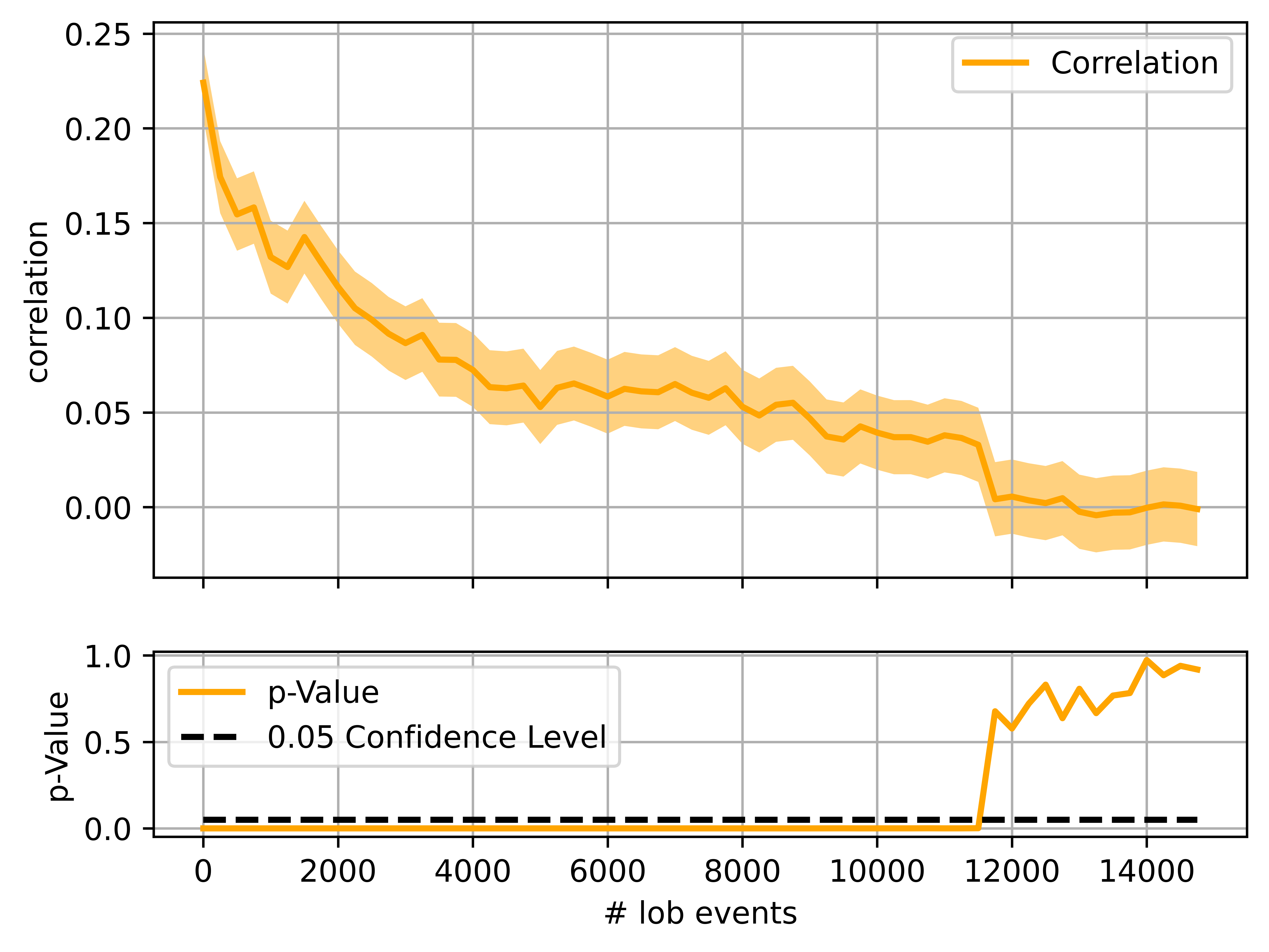}
		\label{fig:dynamicCorrReturnMid}
	\end{subfigure}%
	\begin{subfigure}{.5\textwidth}
		\centering
		\includegraphics[width=.8\linewidth]{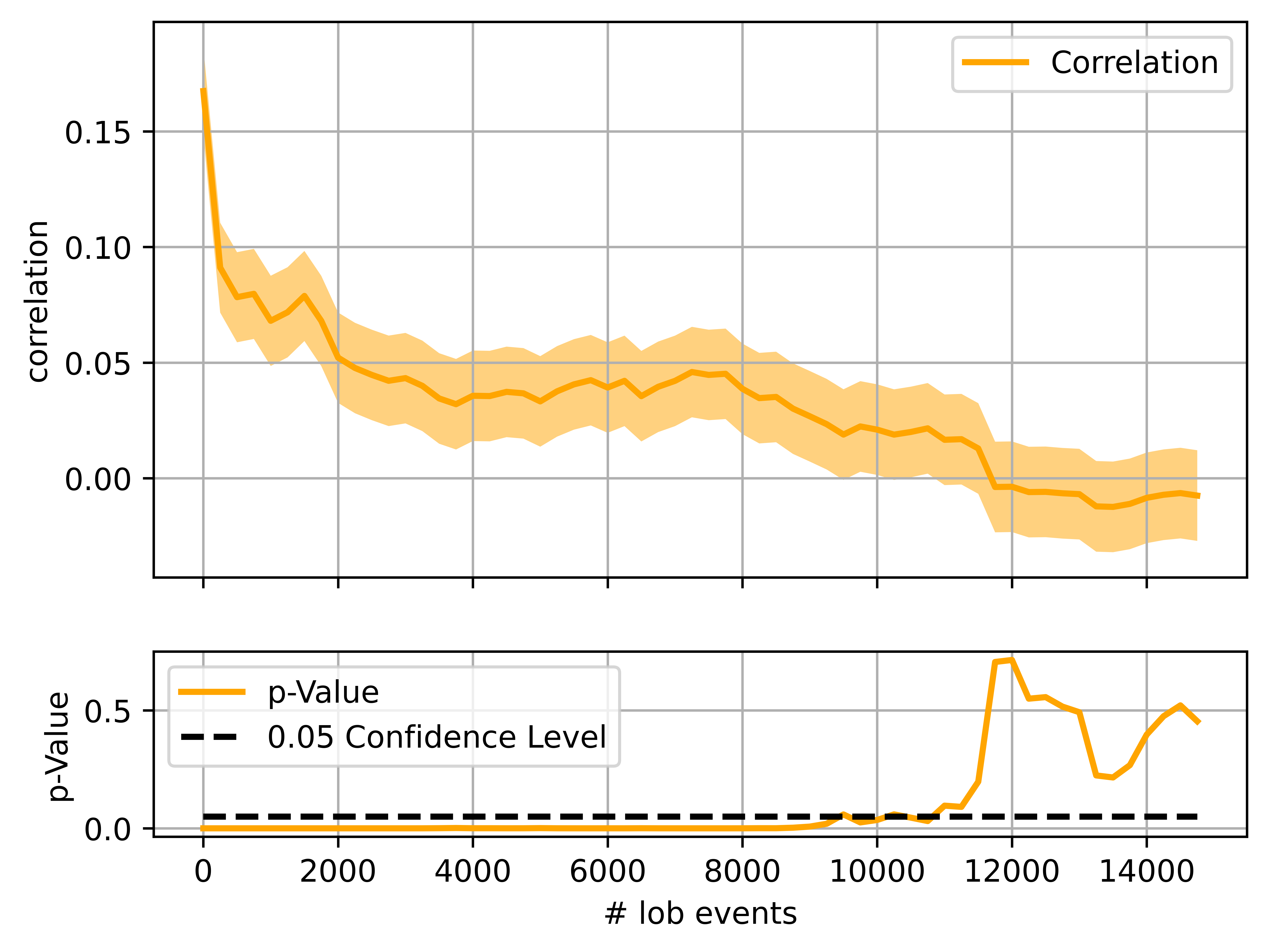}
		\label{fig:dynamicCorrReturnWeighted}
	\end{subfigure}
	\caption{Return correlation: Mid-price (left); Weighted mid-price (right)}
	\label{fig:dynamicCorrReturn}
\end{figure}

To evaluate the effectiveness of simulating the conditionality inherent in order book transitions, we evaluate the predictive power of the initial state on the simulated return. For that purpose, consider an initial LOB snapshot $S^5_{p_0^*}(0)\in\mathcal{D}$, and let $R^r_{k\Delta t}(S^5_{p_0^*}(0))$ denote the real return $k\Delta t$-steps in the future starting at $S^5_{p_0^*}(0)$ and let $R^s_{k\Delta t}(S^5_{p_0^*}(0))$ denote the simulated return $k\Delta t$-steps in the future starting at $S^5_{p_0^*}(0)$. If the choice of the initial state $S^5_{p_0^*}(0)$ has a similar impact on the simulated return $R^s_{k\Delta t}(S^5_{p_0^*}(0))$ as the real return $R^r_{k\Delta t}(S^5_{p_0^*}(0))$ one expects them to be positively correlated, i.e.\ the simulated return is a predictor for the real return. However, this correlation should not account for newly incorporated noise within the dynamical system to allow the simulator to be effective. Further, the impact of the initial state should decrease over time assuming ergodicity and thus also the correlation between the returns. 

In Figure \ref{fig:dynamicCorrReturn}, we estimate the correlation of real and simulated returns starting from the same initial states over time. The left plot compares the correlation of the mid-price returns and the right plot the correlation of the weighted mid-price returns. The shaded area in the plots corresponds to the 95\%-confidence intervals  and the smaller plots in the second row are the corresponding $p$-values for testing that the correlation is not equal to zero. Both plots show the expected behaviour for the correlation. In particular, the correlation between real and simulated returns is significantly larger than zero over a large number of LOB events and is decreasing over time. The weighted mid-price return correlation is not as strongly correlated compared to unweighted return correlation. However, both correlation estimates lose significance at a similar time after around $1.1\times10^4$ LOB events.

\begin{figure}[h!]
	\begin{subfigure}{.5\textwidth}
		\centering
		\includegraphics[width=.8\linewidth]{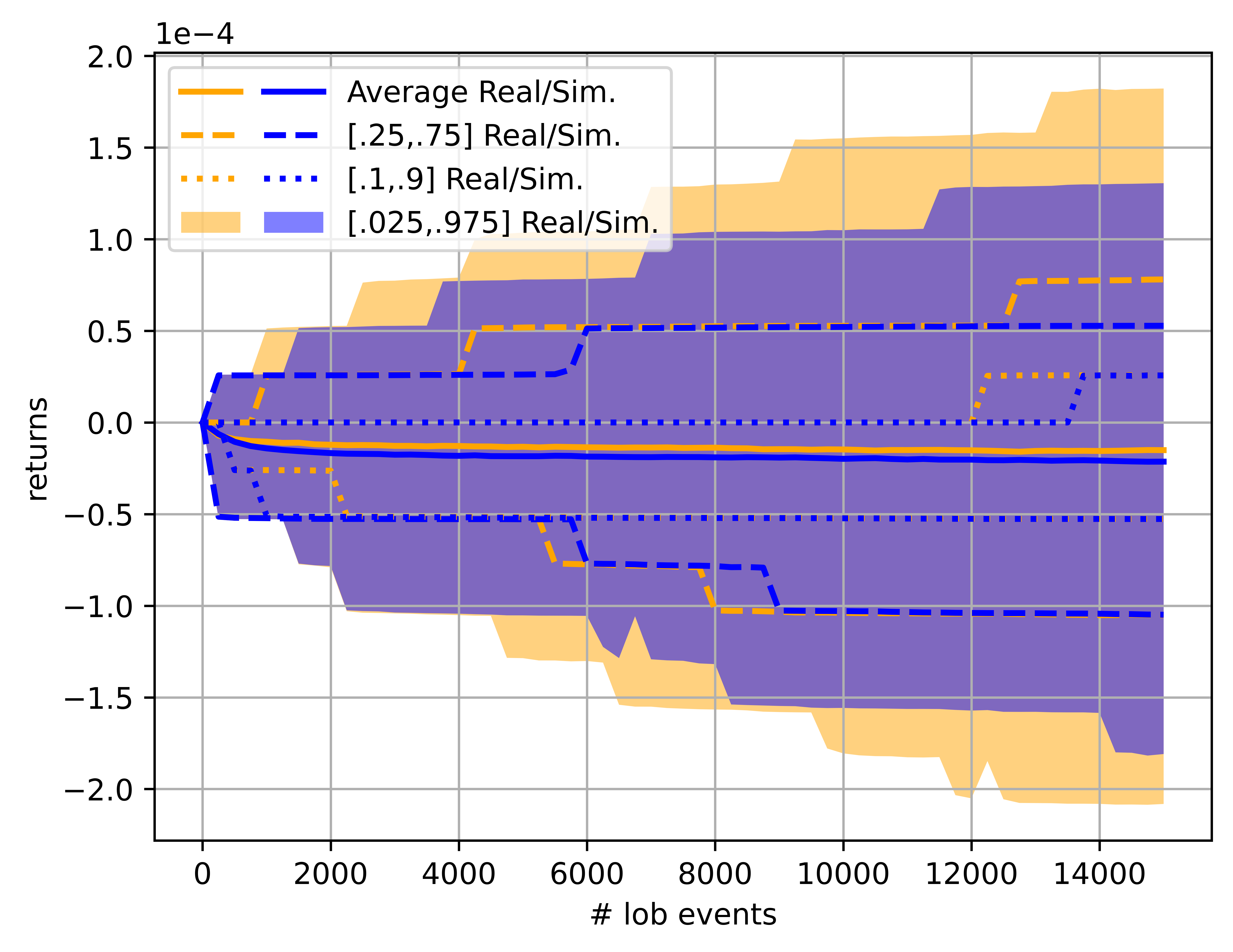}
		\label{fig:dynamicCorrReturnMidOBIlow}
	\end{subfigure}%
	\begin{subfigure}{.5\textwidth}
		\centering
		\includegraphics[width=.8\linewidth]{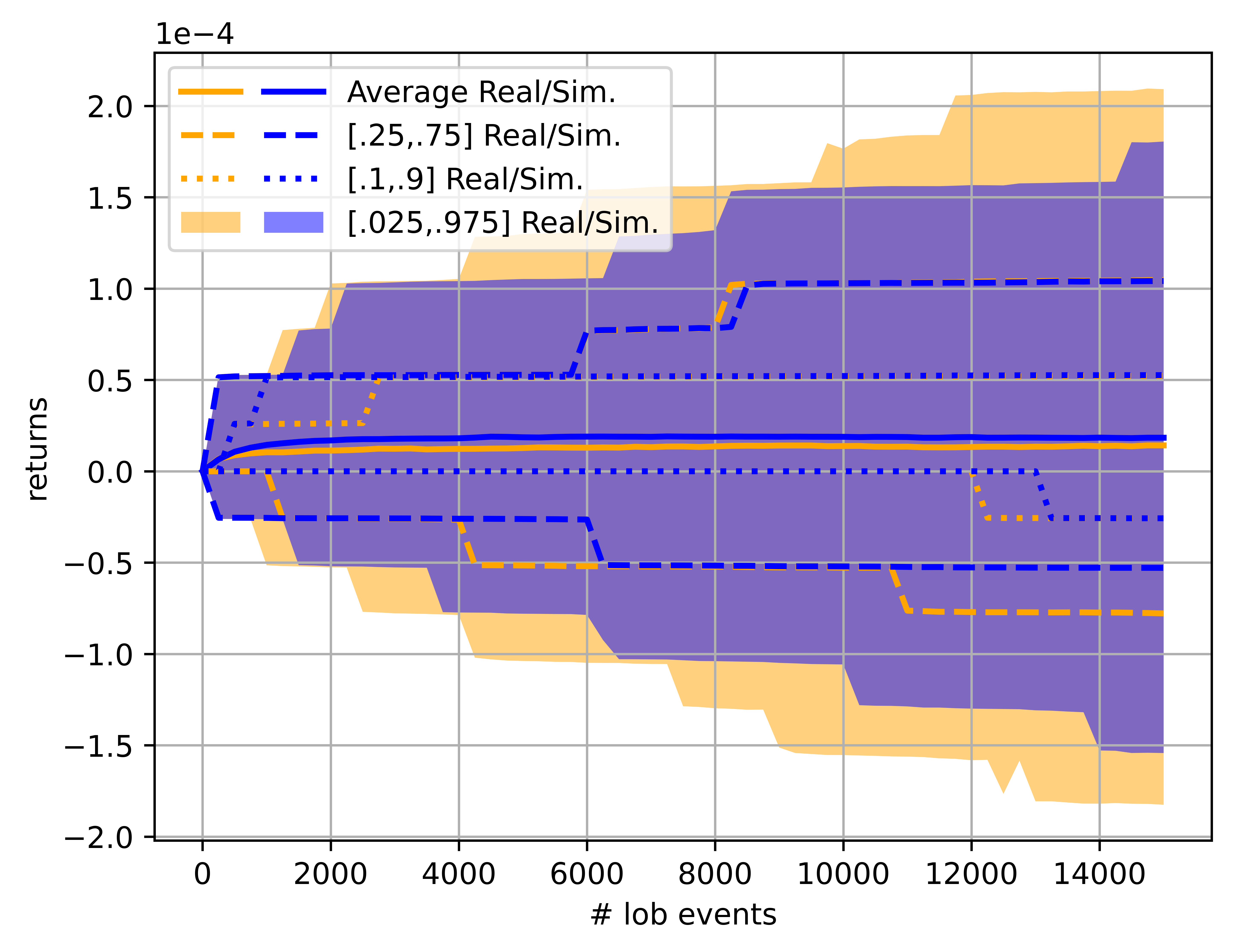}
		\label{fig:dynamicCorrReturnMidOBIhigh}
	\end{subfigure}
	\caption{Comparison of real return time series with simulations by Algorithm \ref{alg:matchingSim} with large order book imbalances in initial state: Mid-price (left); Weighted mid-price (right)}
	\label{fig:dynamicCorrReturnOBI}
\end{figure} 

To enhance the point that Algorithm \ref{alg:matchingSim} captures conditionality, we consider an additional experiment where the initial states show strong order book imbalances. These are known to be predictors for price movements, e.g.\ \cite{cartea2018enhancing}. A large negative value signifies a much lower volume on the bid compared to the ask. Hence, the best bid level is more likely to be depleted, giving a larger probability for downward move in the price. For larger positive values of the order book imbalance the opposite is implied. In Figure \ref{fig:dynamicCorrReturnOBI}, we randomly choose initial states for which the order book imbalance is within the 0-0.05 quantile (left) or the  0.95-1 quantile (right). With these initial states, we compare the real mid-price return time-series with the simulated ones (cf. Figure \ref{fig:dynamicReturn}). As expected, the distribution of the real returns exhibits the impact of the large scale order book imbalances. For large negative values of the order book imbalance at the initial state the mean and the quantiles shift downward (left) at the beginning of the time horizon and then remain at this new level. For large positive values (right), we see an analogous upward shift. The simulated mid-price returns exhibit for both cases the same behavior where the distributional attributes closely match the real returns. This allows the conclusion that Algorithm \ref{alg:matchingSim} is capable of capturing the conditional dependence in the order book transition on the current order book imbalance. 

\subsection{Benchmark comparison}\label{sec:quantComp}

\begin{table}[h]
	\begin{center}
		\begin{tabular}{ |p{4.2cm}||p{2.75cm}|p{2.75cm}|p{2.75cm}|  }
			\hline
			Feature Name & \ref{alg:matchingSim} & CGAN & naive\\
			\hline\hline
			bidSize2  & \textbf{.024 $\pm$ .0062}    & .034 $\pm$ .0089 & .049  $\pm$ .0160\\
			bidSize1  & \textbf{.024 $\pm$ .0053}    &.040 $\pm$ .0078&   .046  $\pm$	.0110\\
			askSize1  & \textbf{.029 $\pm$ .0044}    &.043 $\pm$ .0098 &   .058  $\pm$	.0141\\
			askSize2  & \textbf{.027 $\pm$ .0070}    &.029 $\pm$ .0053&   .054  $\pm$	.0158\\	
			\hline\hline
			OBI ($s=1$)   & \textbf{.033 $\pm$	.0062}     &.035 $\pm$	.0095&  .038 $\pm$ .0126\\
			OBI (s=10)   & .040 $\pm$	.0100     &.062 $\pm$	.0140&  \textbf{.034 $\pm$ .0069}\\
			OBI ($s=30$)   & .045 $\pm$ .0132     &.058 $\pm$	.0153&  \textbf{.042 $\pm$ .0155}\\
			OBI ($s=60$)   & \textbf{.038 $\pm$	.0082}     &.077 $\pm$	.0162&  .042 $\pm$ .0052\\	
			\hline\hline
			mid-price return ($s=1$)   & \textbf{.020 $\pm$ .0054}  &.023 $\pm$ .0060&  .048 $\pm$	.0111\\
			mid-price return ($s=10$)   & \textbf{.040	$\pm$ .0091}   &.047 $\pm$ .0126&  .154 $\pm$ .0165\\
			mid-price return ($s=30$)   & \textbf{.041 $\pm$ .0063}  &.051 $\pm$ .0071&   .171 $\pm$ .0173\\
			mid-price return ($s=60$)   & \textbf{.053 $\pm$ .0121}    &.065 $\pm$ .0138&   .184 $\pm$	.0094\\	
			\hline\hline
			weighted return ($s=1$)   & \textbf{.075 $\pm$ .0140}   &.080 $\pm$ .0123&   .258 $\pm$	0.0183\\
			weighted return ($s=10$)   & \textbf{.066 $\pm$ .0115}     &.091 $\pm$ .0183&   .203 $\pm$	0.0140\\
			weighted return ($s=30$)   & \textbf{.056 $\pm$	.0137}  &.010 $\pm$ .0218&   .196 $\pm$	0.0166\\
			weighted return ($s=60$)   & \textbf{.059 $\pm$	.0164}    &.106  $\pm$ .0201&   .193 $\pm$	0.0067\\			
			\hline
		\end{tabular}
		\caption{Mean and standard deviation for the KS test statistic for Algorithm \ref{alg:matchingSim} and benchmarks.\label{tab:ksBench}}
	\end{center}
\end{table}

The previous analysis of the simulation results for Algorithm \ref{alg:matchingSim} focused on a comparison with real markets. To benchmark our method against other LOB simulation techniques, we follow \cite{hultin2023generative} and compute the two-sample Kolmogorov Smirnov (KS) test statistic for the empirical distribution of the simulation with the real empirical distribution for certain features of interest. The two-sample Kolmogorov Smirnov (KS) test statistic is given by
\begin{equation*}
	\max_x|F_1(x)-F_2(x)|,
\end{equation*}
where $F_1$ and $F_2$ are one-dimensional empirical cumulative distribution functions of the two samples. Thus, a smaller value suggests a better fit of the distributions, with all values in the interval $[0,1]$. 

We evaluate the KS statistic for several different features, namely, we consider the marginal distributions for order book sizes for one transition (as in Figure \ref{fig:hist_marginals}), the weighted mid-price and the mid-price return distribution (cf. Figure \ref{fig:dynamicReturn}) and the distribution of order book imbalances (OBI)\footnote{For the definition of the order book imbalance, see Section \ref{sec:lob}.} at different times $s$. To assess the realized values of the test statistic, we compare Algorithm \ref{alg:matchingSim} to benchmark models, specifically an out-of-the-box implementation of the LOB generator (CGAN) suggested in \cite{cont2023limit} and a simple randomized replay of LOB transitions (naive). Implementation details and the corresponding comparison of the benchmarks with the real market can be found in Appendix \ref{sec:bench}. Similar to \cite{hultin2023generative}, for reporting, we calculate the KS statistic for 1000 samples of the simulation and compare them to 1000 samples from the real distribution. We repeat this procedure 10 times and report the mean and the standard deviation for the KS statistic. 

Table \ref{tab:ksBench} shows that Algorithm \ref{alg:matchingSim} consistently outperforms our benchmarks for most variables. In particular, we highlight that all features related to returns outperform both the CGAN and the naive benchmark. We note that features related to returns are of particular interest as they capture the conditional dynamics of the order book by the evolution of prices in contrast to the volume-based measures. For these volume-based features, an unconditional central limiting behaviour is potentially sufficient as evidenced by the strong performance of the naive benchmark for order book imbalances after multiple transitions. Furthermore, we observe that the CGAN benchmark outperforms the naive benchmark for the majority of features and, in particular, on all return features. 

\subsection{Market order impact}
Next, we investigate the impact of market orders by a trading agent on the market simulation. In general, markets are expected to respond to market orders of sufficient size in an adverse way. For example, if a trader places a large sell order, the market price will tend to decrease. The impact of market orders is a widely researched topic, e.g.\ \cite{webster2023handbook}, as it can have a significant negative effect on profits from trading. It is suggested e.g.\ in \cite{grinold2000active,gatheral2010no, toth2011anomalous} that the market impact is proportional to the square root of the order size $P$ divided by the daily trading volume $\overline{V}$, 
\begin{equation*}
	\text{price impact}\propto\sqrt{\frac{P}{\overline{V}}}. 
\end{equation*}
Other empirical findings, e.g.\ \cite{moro2009market, bacry2015market}, similarly suggest that the impact of order size follows a concave power law, 
\begin{equation*}
	\text{price impact}\propto\left(\frac{P}{\overline{V}}\right)^\gamma. 
\end{equation*}
with $\gamma\in[0,1]$, where the exact choice of $\gamma$ depends on the market.

We will study the scenario where a trader seeks to liquidate a long positions of varying sizes using market orders. To do this, the trader will split a large parent order in smaller child orders and disperse them over time. If $P$ is the size of the parent order and $\widetilde{t}_n$ the maximum trading time, then a single child order will be of size $P/\widetilde{t}_n$ for LOB snapshot times $s\leq \widetilde{t}_n$. After time $\widetilde{t}_n$, no orders will be executed. In our experiments, we trade over half of the entire trading time horizon, i.e.\ $\widetilde{t}_n=30$. In total, we will sample $5000$ paths per order size\footnote{For all succeeding experiments, we will use $N=5000$ unless mentioned otherwise.}.

\begin{figure}[h!]
	\begin{subfigure}{.33\textwidth}
		\centering
		\includegraphics[width=.8\linewidth]{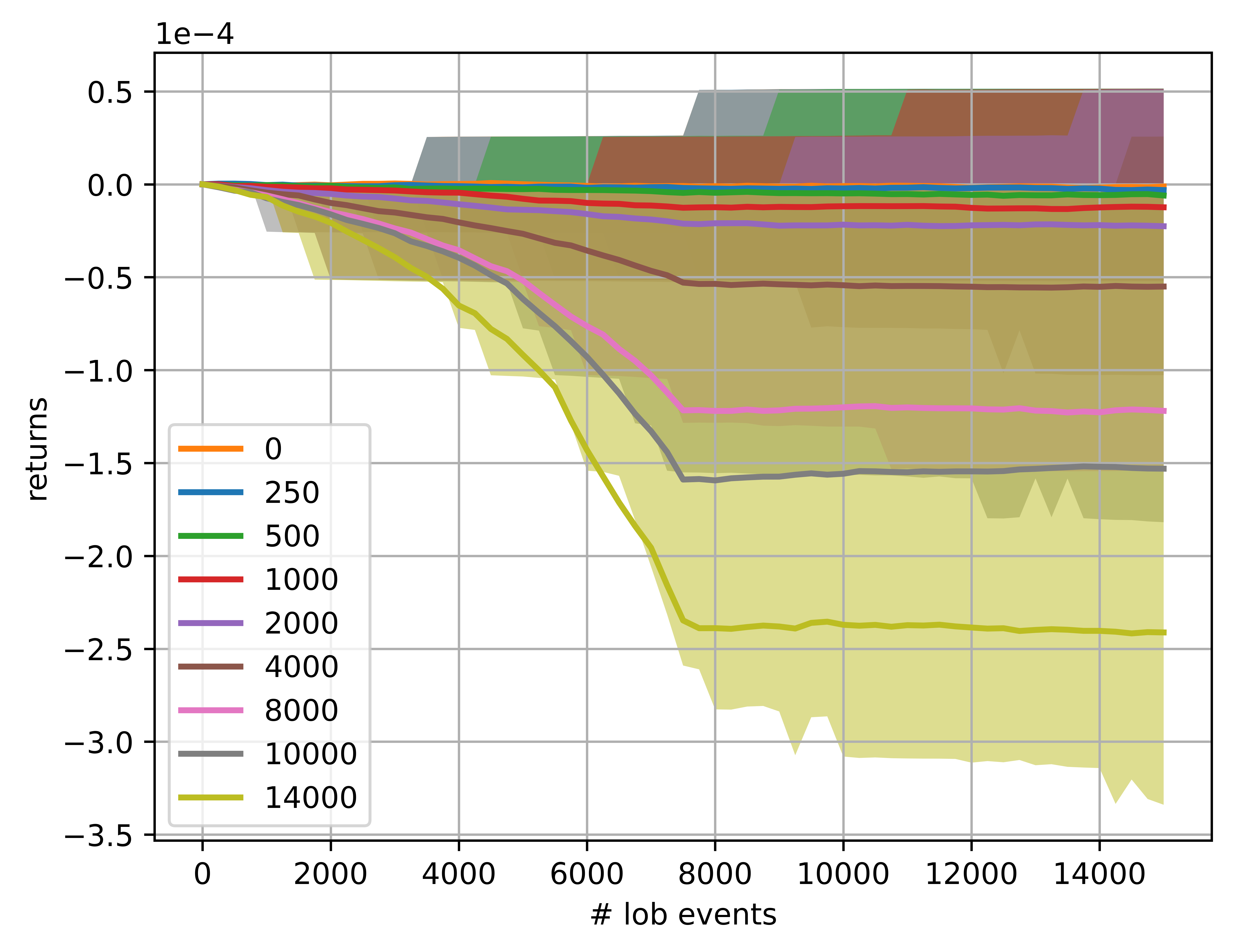}
		\label{fig:MOImpactReturn}
	\end{subfigure}%
	\begin{subfigure}{.33\textwidth}
		\centering
		\includegraphics[width=.8\linewidth]{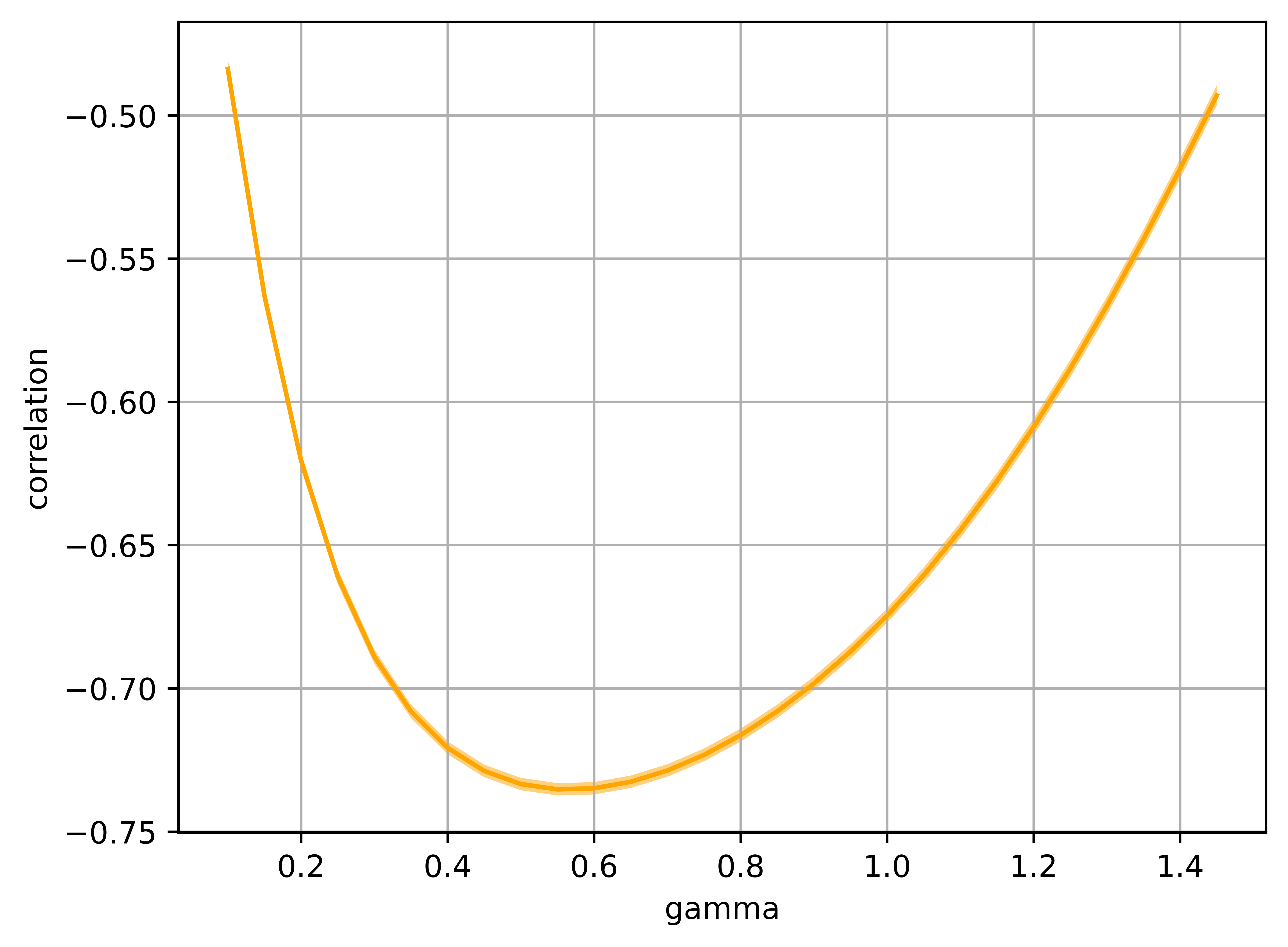}
		\label{fig:MOImpactRVal}
	\end{subfigure}
	\begin{subfigure}{.33\textwidth}
		\centering
		\includegraphics[width=.8\linewidth]{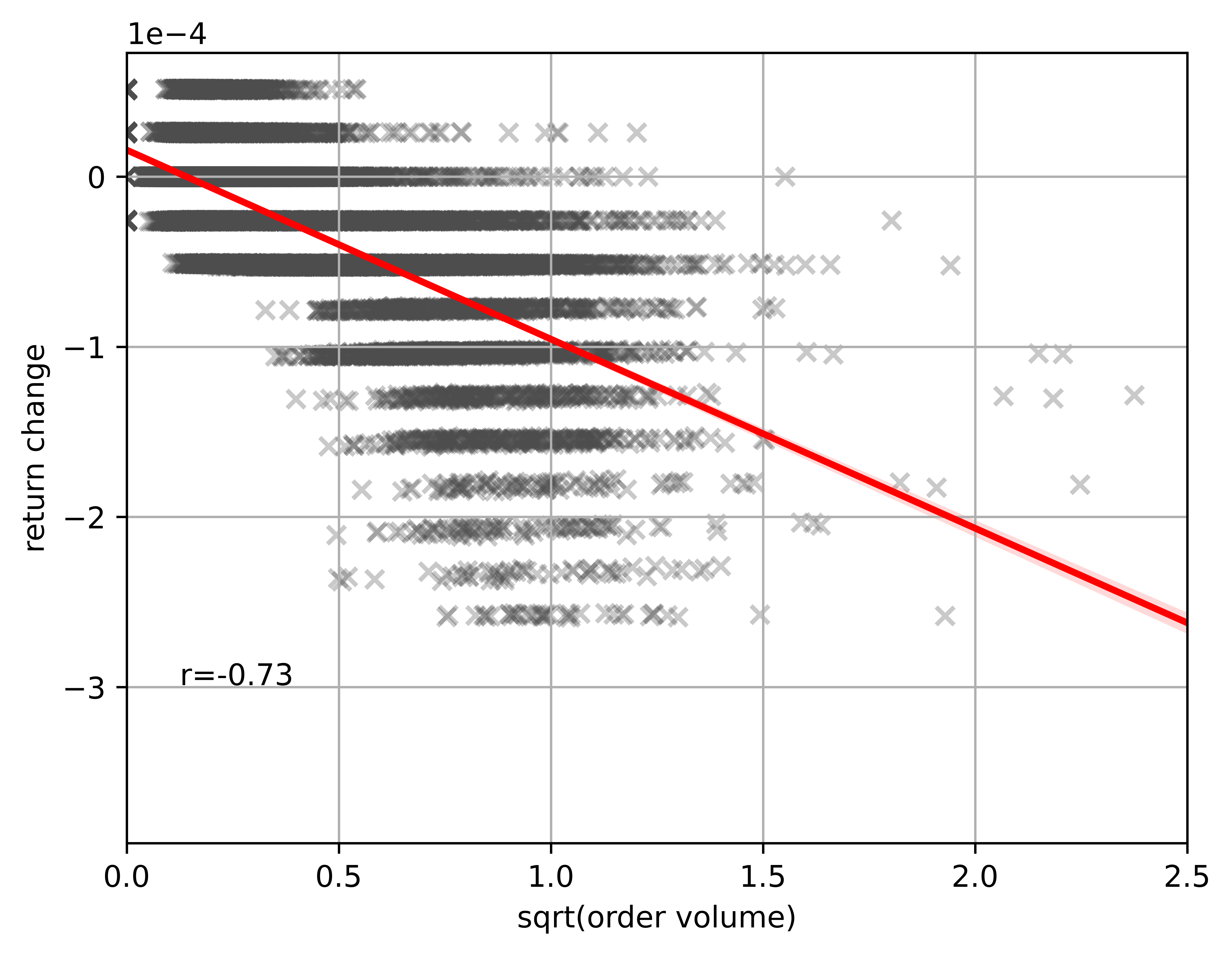}
		\label{fig:MOImpactRegression}
	\end{subfigure}
	\caption{Impact of market orders: Mid-price returns for different parent order sizes (left); Explanatory power for returns depending on functional shape (middle); $\sqrt{\cdot}$-order volume return fit (right)}
	\label{fig:MOImpact}
\end{figure}
The first plot in Figure \ref{fig:MOImpact} shows the evolution of the average mid-price market returns for different parent order sizes over time. The shaded areas reflect the central quartiles (0.25-0.75). One can immediately see the adverse effect of market sell orders on the price as the average return decreases over time up until trading stops, after which the prices seem to remain constant. This corresponds to the simulation results for market orders in \cite{cont2023limit, coletta2023conditional}. The average price impacts are ordered by the size of market orders and the return ranges increase for increasing market orders. Especially for very large orders, the average return is at places smaller than the 0.25-quantile, suggesting that outliers impact the mean paths. 

We note that regimes with extremely large market orders of the size tested in our experiments are very rare in our data set and, thus, coverage under resampling is not as good as in ordinary market regimes. This leads to poor tail behaviour for the resamped scenarios. Thus, we will focus on the central quartiles of market responses to exclude outliers in the following analysis. To ensure comparability, we do this across all position sizes. A possible alternative could be the implementation of a mixture model that combines our non-parametric approach with a parametric model for regions with insufficient coverage similar to \cite{gottesman2019combining}. 

Next, we investigate the shape of the market impact. For this purpose, we calculate the correlation between the mid-price return at time $\widetilde{t}_n$ and the power of the weighted parent order size $\left(P/\overline{V}\right)^\gamma$ for different values of $\gamma$. As a proxy for the daily volume $\overline{V}$, we use the volume of the entire order book at the initial starting point. This is necessary since our data set contains trading regimes with varying levels of liquidity due to the seasonality of trading in particular futures contracts. The second graph in Figure \ref{fig:MOImpact} plots this correlation against different values of $\gamma$. The resulting image is roughly parabola-shaped with a minimum value of $0.74$ at around $\gamma=0.55$. This suggests that the weighted parent order size has the most explanatory power for the mid-price return if $\gamma=0.55$.The varying  thickness of the line represents the 99\%-confidence interval for this estimate. Hence, the correlation value at $\gamma=0.55$ is significantly smaller than at $\gamma=1$, giving strong evidence for a concave function shape. To show the fit of the classical $\sqrt{\cdot}$-law, we regress $\left(P/\overline{V}\right)^{0.5}$ against the mid-price returns and plot the fit in the right-most graph in Figure \ref{fig:MOImpact}. The varying thickness of the line indicates the 95\%-confidence interval for the estimate. The plot reveals a good fit to a square-root model with nearly optimal explanatory power out of the power-law model class. One should note, however, that the confidence bands reveal a slight heteroscedastic error with increased errors for large values of weighted order volumes.

\subsection{Limit order impact}
Similar to market orders, we expect an adversarial effect of large limit orders on the price, i.e.\ \cite{ hautsch2012market, eisler2012price, said2017market}.


\begin{figure}[h!]
	\begin{subfigure}{.33\textwidth}
		\centering
		\includegraphics[width=.8\linewidth]{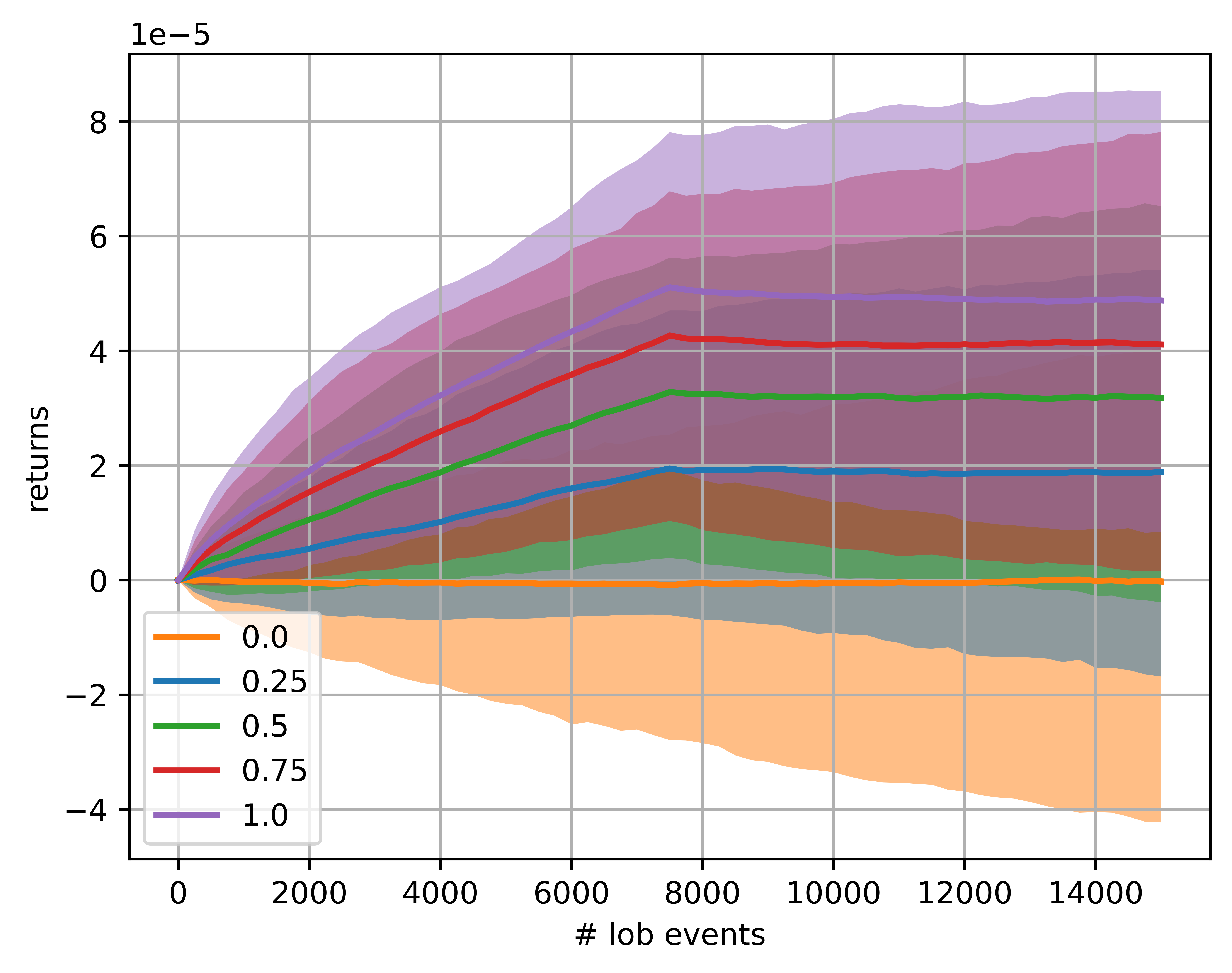}
		\label{fig:LOImpactAsk1}
	\end{subfigure}%
	\begin{subfigure}{.33\textwidth}
		\centering
		\includegraphics[width=.8\linewidth]{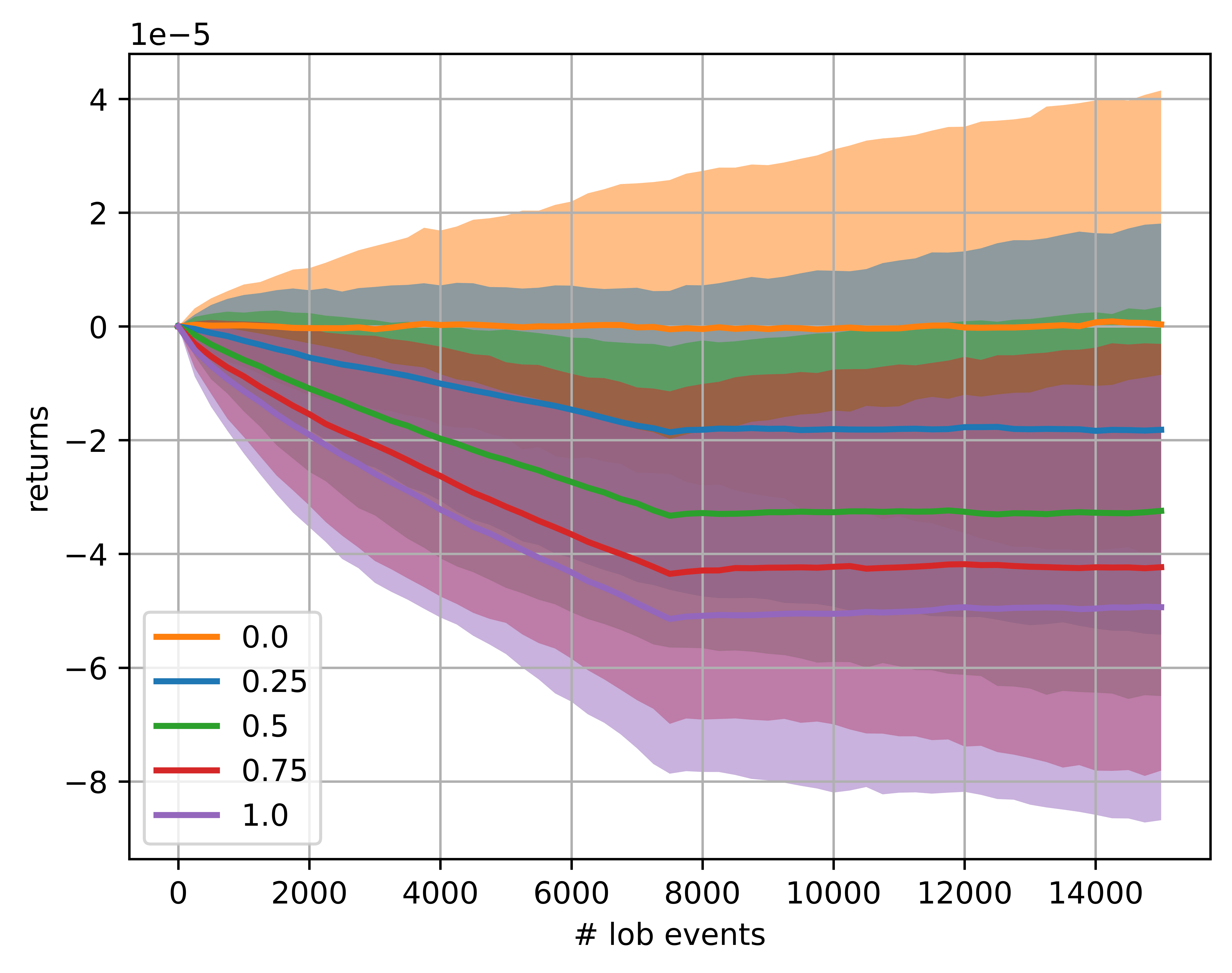}
		\label{fig:LOImpactBid1}
	\end{subfigure}
	\begin{subfigure}{.33\textwidth}
		\centering
		\includegraphics[width=.8\linewidth]{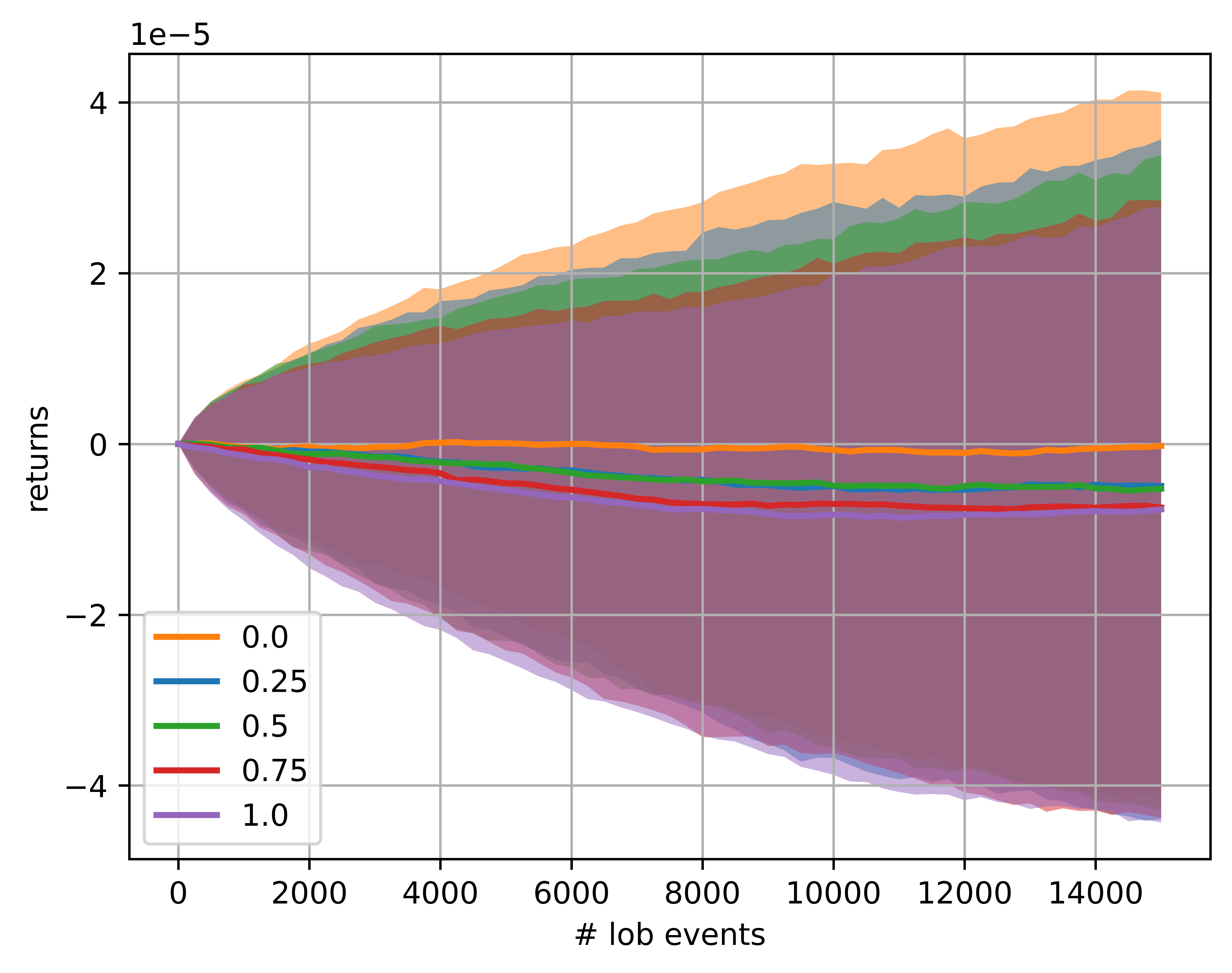}
		\label{fig:LOImpactBid2}
	\end{subfigure}
	\caption{Impact of limit orders with different parent order sizes: Best bid (left); Best ask (middle); Second best ask (right)}
	\label{fig:LOImpact}
\end{figure}

For the experiments, we assume that a trader continuously places limit orders over time at particular levels of the centered LOB.  Similarly, as for market orders, we fix the size of the parent order $P$ and a maximum trading time $\widetilde{t_n}$, such as a single child order corresponds to order size $P/\widetilde{t_n}$ for LOB snapshot times $s\leq \widetilde{t_n}$. After time $\widetilde{t_n}$, no limit orders will be placed, and we choose  $\widetilde{t_n}=30$. Instead of using mid-price returns, we will consider weighted mid-prices to reflect repositioning on the highest order book level. Furthermore, we will use sizes relative to the overall order book volume at initialization for parent orders to have a comparable distortion for the weighted returns. 

We exhibit our results in Figure \ref{fig:LOImpact}. The plots show for different order book levels the weighted return paths for different parent order sizes and the shaded areas correspond to the central quartiles (0.25-0.75). For the case where the orders are placed on the best bid and ask level, one can directly see the adverse effect of the limit orders on the weighted return. Namely, for large ask orders, the weighted return decreases while for large bid orders it increases. The average return impacts are ordered by the size of the orders. Once trading stops, the return levels stabilize at the new price level (cf.\ \cite{cont2023limit}). For limit orders placed on the second best ask level, the impact is not perfectly ordered by order size any more and is smaller in scale compared to order placement at the best level. This is consistent with the observation that the SOFR futures market only tends to move by single tick increments (cf.\ Figure \ref{fig:hist_marginals}) as the relatively small variability allows traders to reposition themselves once a new price level is reached.

\section{Policy evaluation} \label{sec:polEval}

In this section, we show how Algorithm \ref{alg:matchingSim} can be used to evaluate and choose trading strategies. A central theme for placing limit orders in markets with a pro-rata type execution mechanism is choosing the right quoting amount. In particular, it is well known that market participants tend to place limit orders with a volume larger than they would like to execute (overquoting), e.g.\ \cite{field2008pro, guilbaud2015optimal}.  Market participants overquote because it increases their relative volume share on a given order book level and with it the probability of being allocated an incoming market order. The trade-off present with overquoting is the possibility that too many trades are executed against the volume posted by the trader. Balancing this trade-off via the quoting volume is key for execution and market making strategies in markets with narrow spreads and high trading volumes such as the SOFR futures market. 

First, we want to show that the market simulations generated by Algorithm \ref{alg:matchingSim} capture the necessity of overquoting for achieving per-transition trade targets and that the resulting market shows favourable characteristics for overquoting. For this purpose, we introduce a trading agent to Algorithm \ref{alg:matchingSim} who keeps a constant position on the best bid level in the order book over time. If the market price moves, the agent cancels their remaining volume at the old level and places a limit order at the new best bid level. If the limit order is (partially) executed, the trading agent replenishes their order to the prescribed order size. 

\begin{figure}[h!]
	\begin{subfigure}{.5\textwidth}
		\centering
		\includegraphics[width=.8\linewidth]{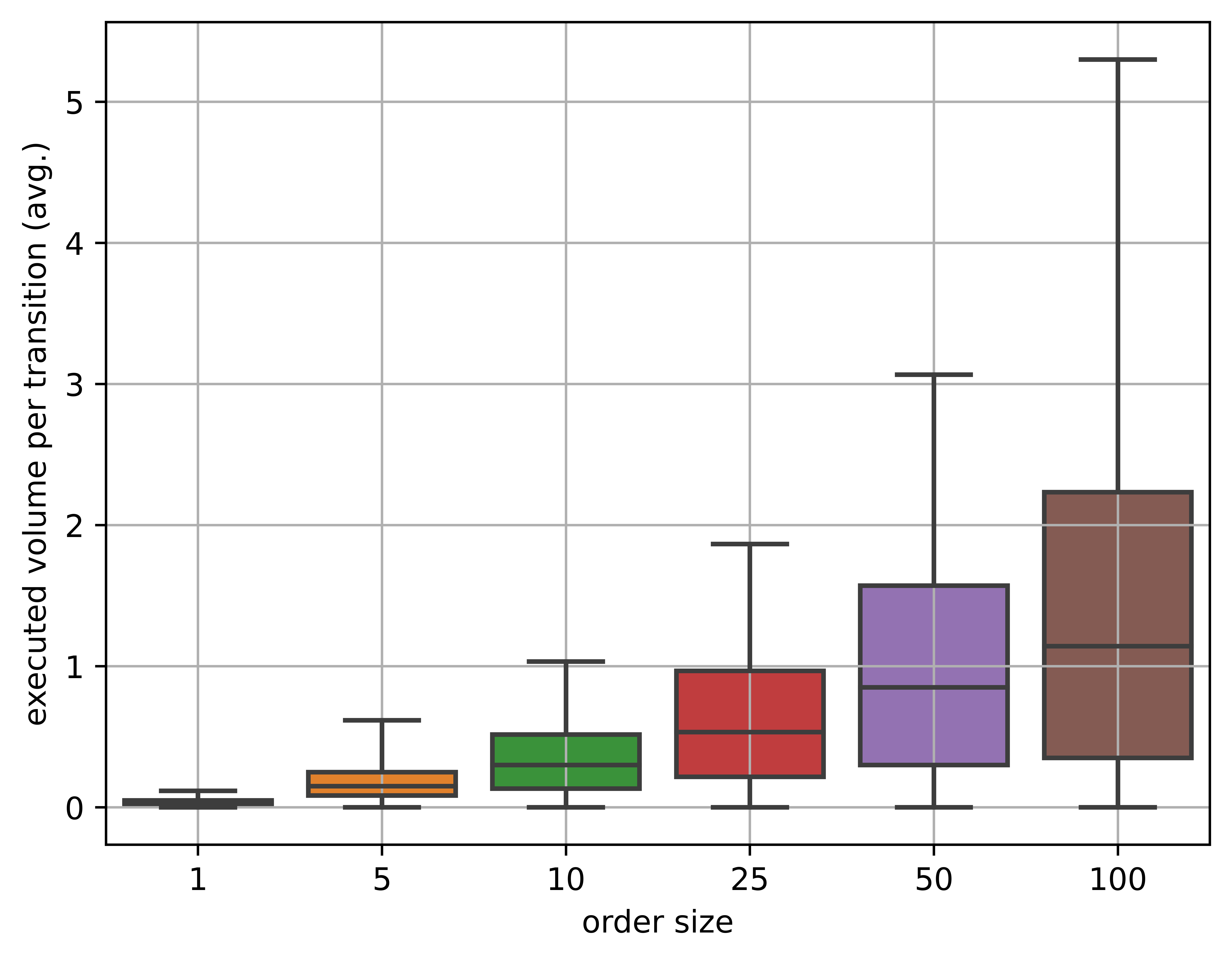}
		\label{fig:peAnFillsExec}
	\end{subfigure}%
	\begin{subfigure}{.5\textwidth}
		\centering
		\includegraphics[width=.8\linewidth]{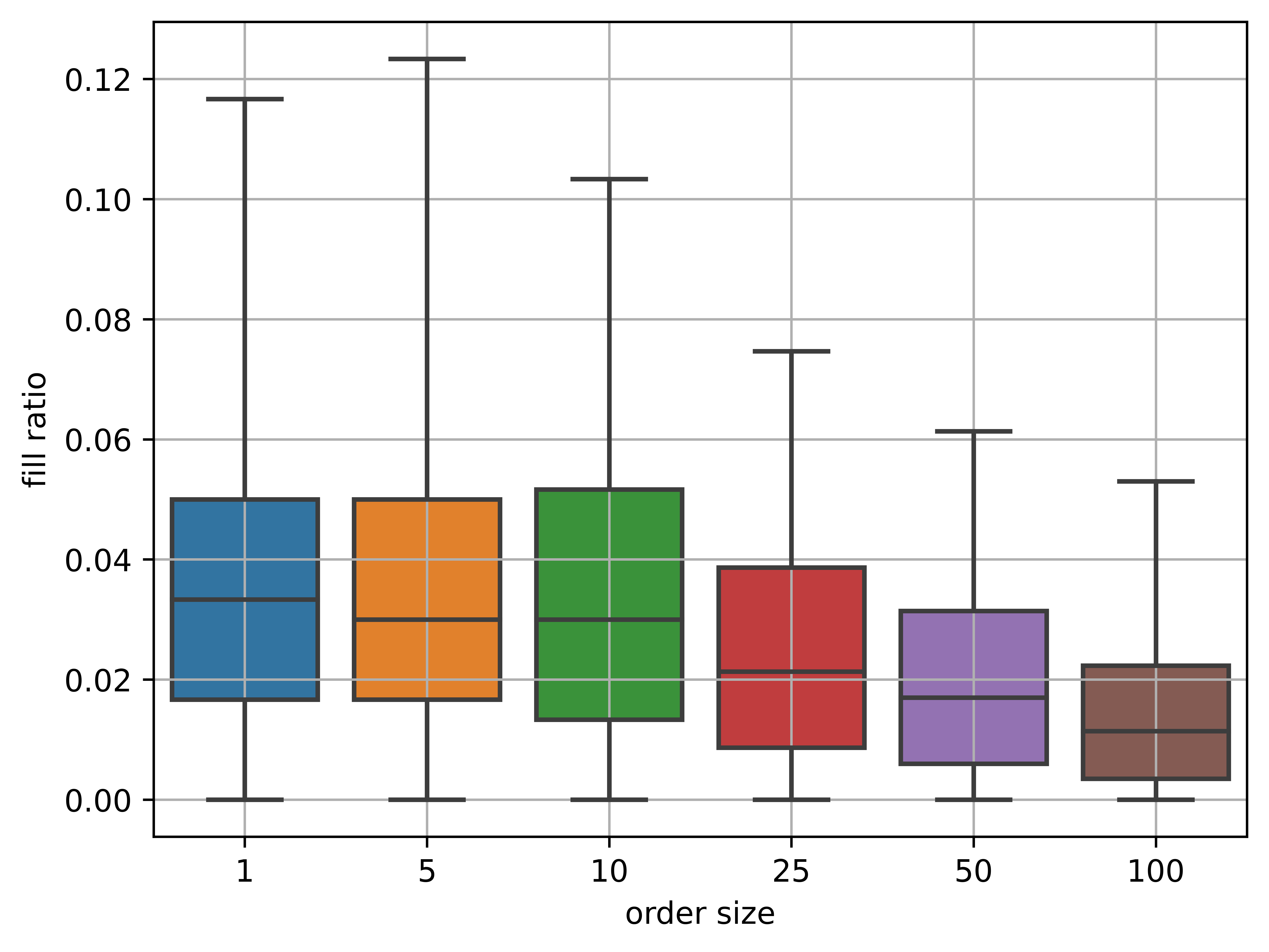}
		\label{fig:peAnFills}
	\end{subfigure}
	\caption{Constant order sizes: Time-averaged executed volumes per transition (left); Fill ratios (right)}
	\label{fig:peAnCons}
\end{figure} 

In Figure \ref{fig:peAnCons}, we report the results for this experiment. In the left-hand plot, we use box plots to report the distributions of executed volumes normalized by the number of transitions for different order sizes. The whiskers highlight the 0.025 and the 0.975 quantile, the box represent the first and third quartile and the median is given by the line contained in the box\footnote{We use this convention for all box plots in this paper.}. First, one can see that all reported quantiles, except the 0.025 one, increase for an increasing order size. If the trading agent only places an order with volume 1 each period the executed volume per transition is close to zero. On the other hand, if order sizes are between 50 and 100 contracts, the median executed volume per transition is around one. However, the upper quartile and the 0.975-quantile can be significantly higher. To achieve a time-averaged executed volume per transition of one contract, a trader would  need to increase their order size substantially above one, which brings the risk of overshooting the target and having to trade more volume than desired. Note that this behavior is consistent with the empirical findings and the theoretical model described in \cite{field2008pro} and that the exact choice of order size would depend on the trader's risk profile. The box plot on the left in Figure \ref{fig:peAnCons} shows the ratio between filled orders and total posted volume over time (fill ratio). Overall, the fill ratios and their variation tend to decrease for increasing order sizes. This behavior of the market further encourages overquoting as it counteracts the risk of overfilling.  
\begin{figure}[h!]
	\begin{subfigure}{.33\textwidth}
		\centering
		\includegraphics[width=.8\linewidth]{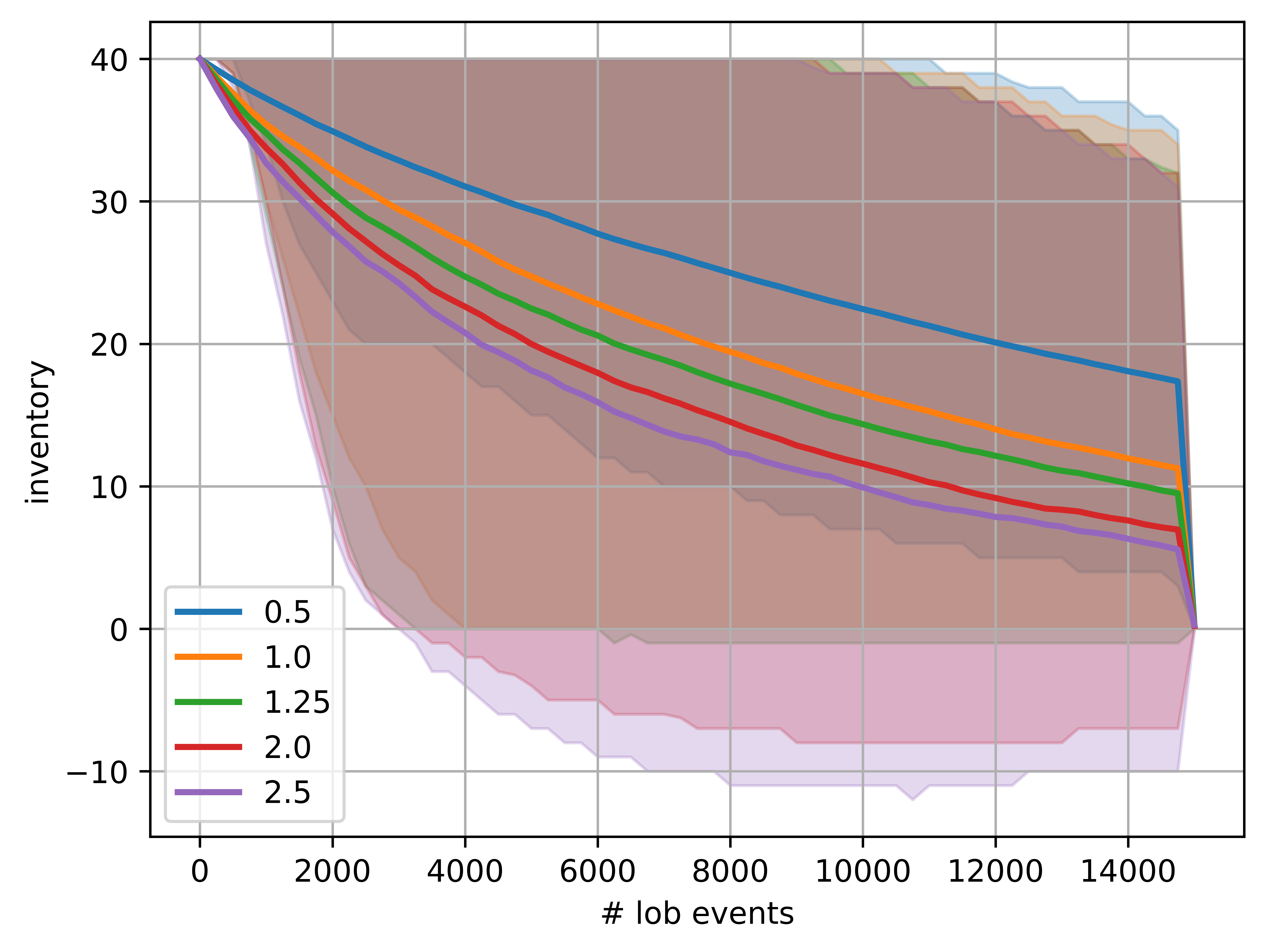}
		\label{fig:peAnInv}
	\end{subfigure}%
	\begin{subfigure}{.33\textwidth}
		\centering
		\includegraphics[width=.8\linewidth]{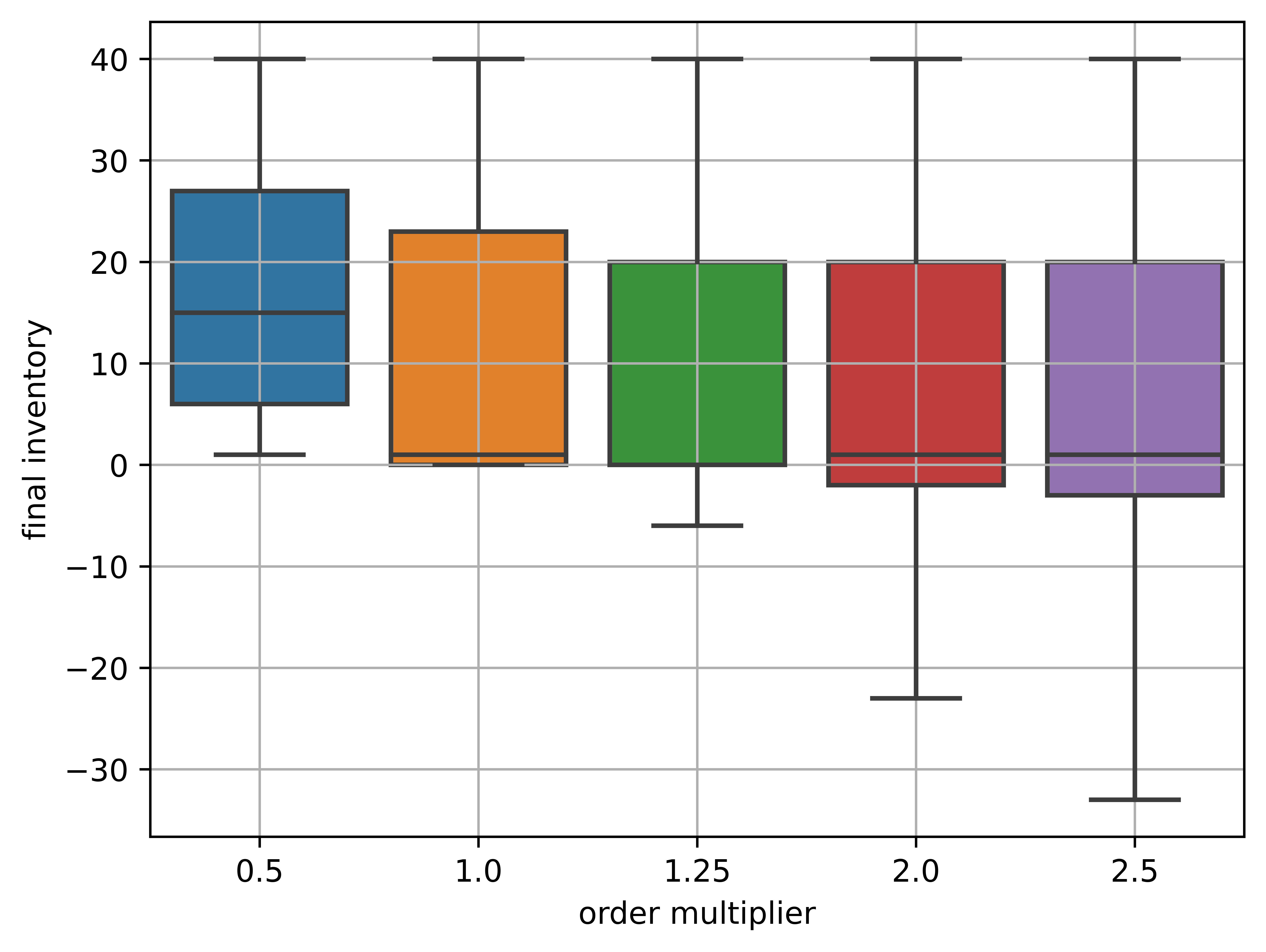}
		\label{fig:peAnFinalInv}
	\end{subfigure}
	\begin{subfigure}{.33\textwidth}
		\centering
		\includegraphics[width=.8\linewidth]{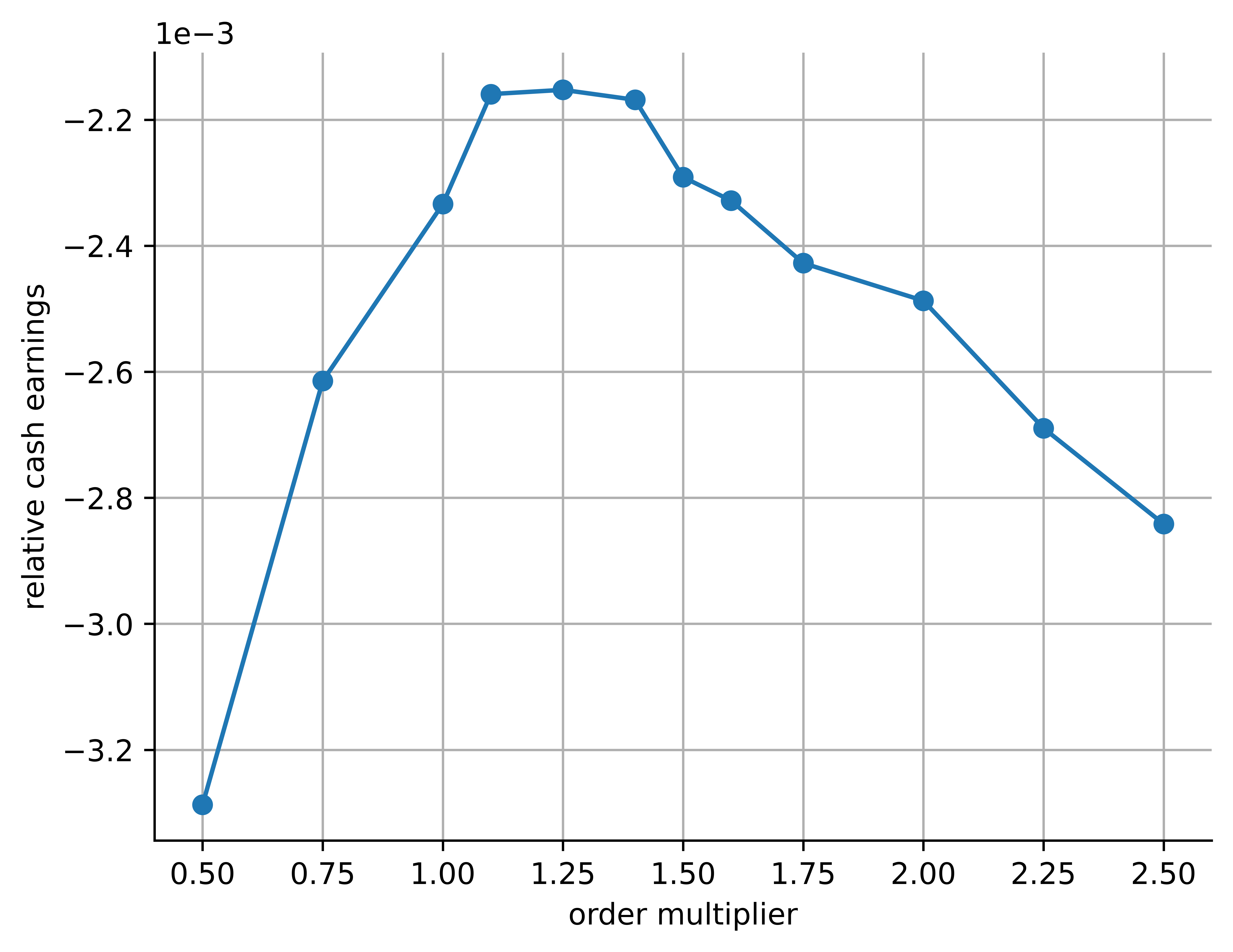}
		\label{fig:peAnCash}
	\end{subfigure}
	\caption{Trading strategy calibration: Inventory paths for varying order multipliers $k$ (left); Final inventory distribution for varying order multipliers $k$ (middle); Relative cash earnings for varying order multipliers $k$ (right)}
	\label{fig:peAn}
\end{figure} 

In our second experiment, we demonstrate how Algorithm \ref{alg:matchingSim} can be used to calibrate a trading strategy. For this purpose, we consider a trader who seeks to unwind their long position using limit orders over a certain time horizon\footnote{Note that the first experiment and \cite{field2008pro} differ in context to this second experiment. Instead of focussing on a one shot/per-transition trade target, this experiment takes into account the time dependency of the execution problem limiting its comparability.}. We consider a simple execution strategy whereby the trader quotes a multiple of the remaining inventory on the best ask level, i.e.\ the quoted volume on the best ask will be $k I_{s}$, with $I_s$ the inventory after the $s$-th transition and $k$ is the order multiplier. Should the inventory become negative, meaning the trader has a short position, all remaining volume on the best ask will be cancelled. Instead, the trader will place an order of size $k|I_{s}|$ on the best bid level. All remaining inventory at the last transition will be executed via a market order. To calibrate the trading strategy, we run Algorithm \ref{alg:matchingSim} for different values of $k$ ranging from $0.5$ to $2.5$ with the starting inventory $I_0=40$.

The results of this experiment are reported in Figure \ref{fig:peAn}. The left plot shows mean inventory paths (lines) and the 90\%-quantile intervals for inventories (shaded areas) for different values of the order multiplier. The mean inventory paths and quantile intervals are decreasing and are ordered by the value of the order multiplier. The lower quantile values decrease especially fast for the medium and larger order multipliers, hinting that a single trade is potentially enough to execute all the inventory. Further note that for values where the order multiplier is larger than one, the trader may overshoot their trading target and may end up with a short position that then needs to be unwinded during the remaining trading period. The upper quantile limit on the other hand remains close to $40$ over a longer period of time for all order multipliers. This implies that there is a possibility that no trade is executed against our trader's limit orders. The upper quantiles then start decreasing after some time for all $k$. For larger $k$, this decrease starts earlier, signifying that larger orders are more likely to be (partially) executed in this pro-rata type market. 

The middle plot in Figure \ref{fig:peAn} shows the distribution of inventories at the last time step before the market orders are executed for different order multipliers. For $k=0.5$, it is infeasible that the entire inventory is executed via limit orders and the median is relatively large. This means that a large proportion of inventory will be executed via market orders. If the order multiplier is equal to one, the trader cannot overshoot the execution target with limit orders. The median final inventory is close to zero, but,  compared to larger $k$, the upper quartile is larger, implying a higher reliance on market orders. For $k=1.25$, the lower quartile and the median coincide at a zero inventory and the trade target is overshot only in the tails. Furthermore, the upper quartile is lower compared to $k=1.0$, implying less reliance on market sell orders. For large values of $k$, we observe that the trade target is consistently overshot, which will require market buy orders to unwind short positions. On the other hand, the upper quartile remains almost unchanged compared to $k=1.25$. Finally, note that the upper whisker is for all order multipliers equal to the initial inventory. This points to the possibility that no order is filled via limit orders for all multipliers. 

Since executing market orders requires paying the spread, we expect that order multipliers requiring large market orders in the final period will exhibit a higher cost of execution. To investigate this hypothesis, we compare the average earnings per asset for different values of the order multiplier. We quantify the earnings by dividing the total earned cash by the initial inventory and then subtracting the initial price. The corresponding plot can be found in Figure \ref{fig:peAn} on the right. First, note that for all order multipliers the relative average cash earnings are negative, implying a cost of execution on average. The scale of the execution costs varies around half a tick size. Highest earnings (or lowest execution costs) are reached at $k=1.25$, while they decrease consistently for smaller and larger values in $k$. Considering the box plot on final inventories and its discussion, this observation is consistent with the hypothesis that order multipliers requiring larger market orders will perform worse. In particular, for $k=1.25$, market sell orders are less often needed compared to smaller $k$ and this benefit seems to outweigh the occasional costs associated to overshooting. On the other hand, for larger $k$, the additional costs of overshooting outweigh any potential benefit of additional unit execution. From the view point of strategy calibration, this experiment implies that a risk neutral trader should choose $k=1.25$ to minimize expected execution costs.

\section{State space extension}\label{sec:state}

In this section, we provide a strategy on how to extend Algorithm \ref{alg:matchingSim} to a more complex state space. In particular, we expand the state space by multiple features beyond LOB snapshots. Adding these additional features will increase the dimensionality beyond the level usually seen as appropriate for nearest neighbor search due to the curse of dimensionality; see e.g.\ \cite{gyorfi2002distribution}. We show how this issue can be overcome with simple dimension reduction techniques and how this can lead to beneficial results. 

In addition to the current volume levels in the order book, we include features containing information on recent price moves and trading activity in the LOB state. Namely, we add mid-price returns and trade imbalances\footnote{The trade imbalance is given as the volume of all buyer-initiated trades divided by the volume of all trades over a given period of time, see e.g.\ \cite{coletta2023conditional}.} over different time windows for each sample in the data set. This is done by slightly modifying the data set generation. In addition to the procedure described in Section \ref{sec:data}, we look up for each LOB snapshot the mid-price that was observed a certain number of LOB events before the snapshot was taken. With the previous mid-price and the mid-price of the current snapshot we calculate the mid-price return for each sample. We do this for a differing number of preceding order book events. In this experiment, we choose the number of order book events for mid-price returns by 250, 1250, and 5000 order book events. This gives us 3 additional matching dimensions in the nearest neighbor search. Similarly, we add the trade imbalances where the feature is calculated for the previous 250, 2500, and 12500 order book events for each snapshot. In total, we then have a 16-dimensional state space. For normalization, we take the square root of the order book volumes and then apply a $z$-score transformation to all variables. With the normalized features, we calculate the weight vector for a principal component analysis (PCA) and reduce the dimension of the data to the first eight principal components, which explain around 80\% of the data's variability. 

\begin{table}[h]
	\begin{center}
		\begin{tabular}{ |p{3.0cm}|p{2.7cm}||p{4.2cm}|p{2.7cm}|  }
			\hline
			Feature Name & \ref{alg:matchingSim} (ext.) & Feature Name & \ref{alg:matchingSim} (ext.)\\
			\hline\hline
			bidSize2  & .027 $\pm$ .0060    & mid-price return ($s=1$)   & \textbf{.015 $\pm$ .0027} \\
			bidSize1  & .026 $\pm$	.0058    & mid-price return ($s=10$)   & \textbf{.023 $\pm$ .0054} \\
			askSize1  & \textbf{.025 $\pm$	.0044}    &  mid-price return ($s=30$)   & \textbf{.036 $\pm$ .0075} \\
			askSize2  & .028 $\pm$	.0079    & mid-price return ($s=60$)   & .054 $\pm$ .0100\\				
			\hline\hline
			OBI ($s=1$)& \textbf{.030 $\pm$ .0061}& weighted return ($s=1$) & .084  $\pm$ .0128 \\
			OBI ($s=10$)& 	\textbf{.034 $\pm$ .0079}& 	weighted return ($s=10$) &   \textbf{.057  $\pm$	.0137} \\
			OBI ($s=30$)& \textbf{.040 $\pm$ .0100}& weighted return ($s=30$) &   .059  $\pm$	.0167\\
			OBI ($s=60$)& .043 $\pm$ .0108 &  weighted return ($s=60$) &   \textbf{.055  $\pm$	.0071} \\	
			\hline			
		\end{tabular}
	\end{center}
	\caption{Mean and standard deviation for the KS test statistic for the extended \ref{alg:matchingSim} Algorithm. \label{tab:stateExt}}
\end{table}

For comparability, we conduct the same benchmark analysis based on KS statistics as in Section \ref{sec:quantComp} and report the results in Table \ref{tab:stateExt}. We highlight the variables that improved in comparison to Table \ref{tab:ksBench}, and we see that Algorithm \ref{alg:matchingSim} with an extended state space improves on the majority of tested features. Note that we investigated only one particular combination of features and dimension reduction. Other combinations may be more effective and their choice may depend on the task at hand. We leave the exploration of more principled approaches for LOB representations in the context of order book simulations for future work.

\section{Further research directions}\label{sec:conc}


The results presented in this paper point to multiple directions for possible future research. First, the effectiveness of $K$-NN resampling for LOB simulations suggests that the algorithm should be considered for simulating and evaluating strategies in other stochastic control environments where direct testing is costly and historical data available. Other applications in finance such as hedging or portfolio management with market impact could be considered. On the algorithm, we believe that further investigations in how to include dimension reduction and the combination with other generative models can be fruitful. In particular, we would be interested in how to choose the state space for the algorithm optimally such that most of the relevant information is contained while keeping its dimension controlled. A systematic investigation on the choice of metric for the nearest neighbor search for LOBs would equally be of interest. Finally, in the context of trade evaluation in LOBs, we believe that several extensions could be of interest. Combining the algorithm with a simulator for FIFO executions would allow trade evaluation in major asset classes such as equities. For pro-rata type markets, more complex trading strategies could be considered, i.e.\ liquidation with limit and market orders or the inclusion of alpha signals. 

\section*{Funding, disclosures and acknowledgements}
Michael Giegrich is supported by the EPSRC Centre for Doctoral Training in Mathematics of Random Systems: Analysis, Modelling and Simulation (EP/S023925/1). Roel Oomen is employed as a Managing Director at Deutsche Bank A.G. This paper was not produced, reviewed or edited by the DB Research Department. The views and opinions rendered in this paper reflect the author's personal views about the subject. No part of the author's compensation was, is, or will be directly related to the views expressed in this paper. We thank the participants of the computational finance session at the 12th Bachelier World Congress and of the Deutsche Bank Quant Talk seminar for their useful comments.

\begin{small}
\renewcommand{\baselinestretch}{1}
\bibliography{refs}
\end{small}
\appendix
\section{Benchmarks}\label{sec:bench}
This section contains implementation details and results for the comparison of the benchmarks with the real market. For the results, we use the same data and follow the same data split. The resulting plots are generated in the same way as in the main part.
\subsection{Naive sampling}
The first benchmark is a naive sampling of LOB transitions from the data. The sampling is uniform over the data set and unconditional. To report the returns, randomly selected transitions are chained together. This benchmark solely relies on the unconditional law of large numbers and does not capture any conditional behavior of the LOB dynamics. 

\begin{figure}[h!]
	\centering
	\begin{tabular}{ccc}
		\begin{subfigure}[t]{0.25\textwidth}
			\centering
			\includegraphics[width=.8\linewidth]{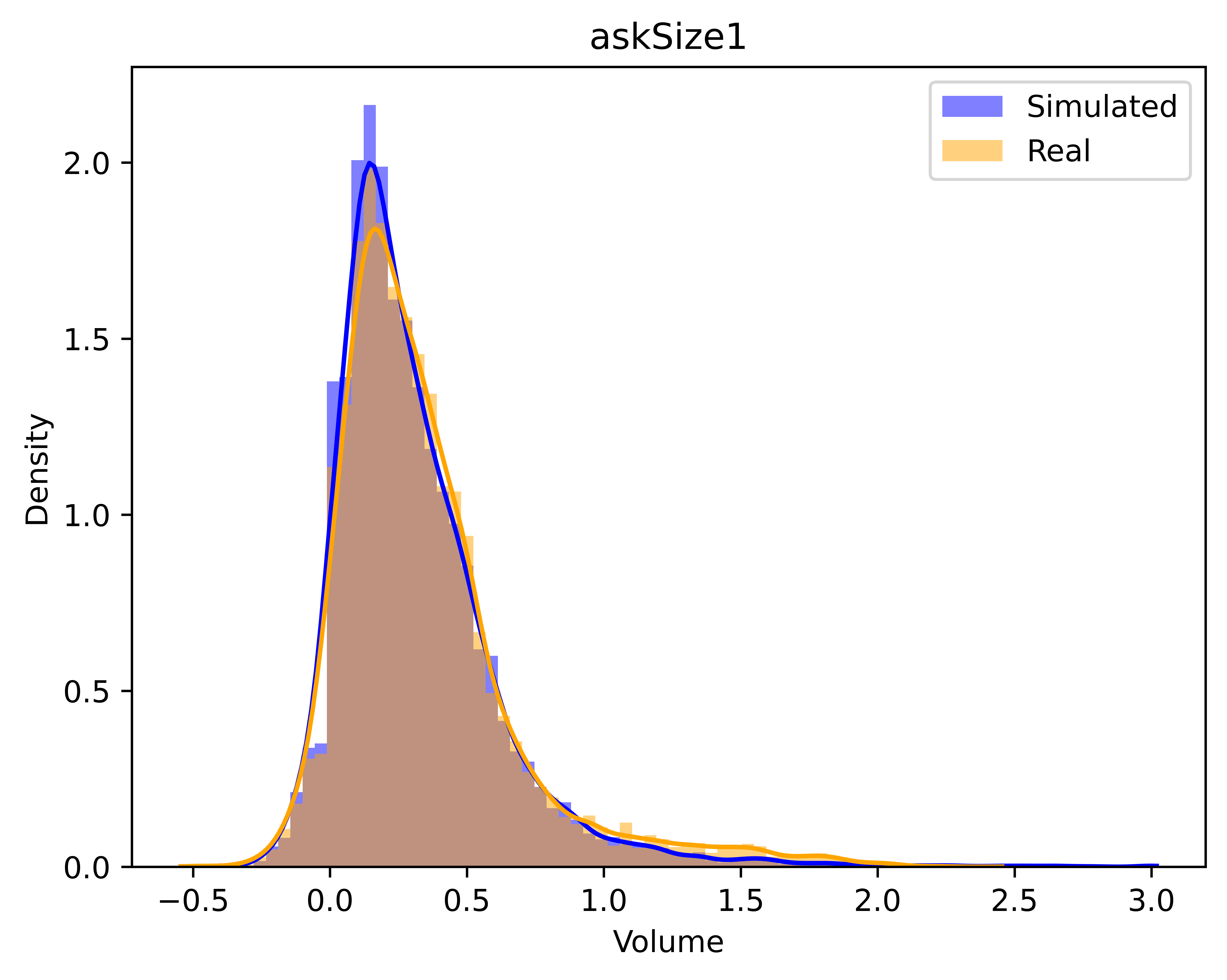}
			\label{fig:histNaive_ask1}
		\end{subfigure} &
		\begin{subfigure}[t]{0.25\textwidth}
			\centering
			\includegraphics[width=.8\linewidth]{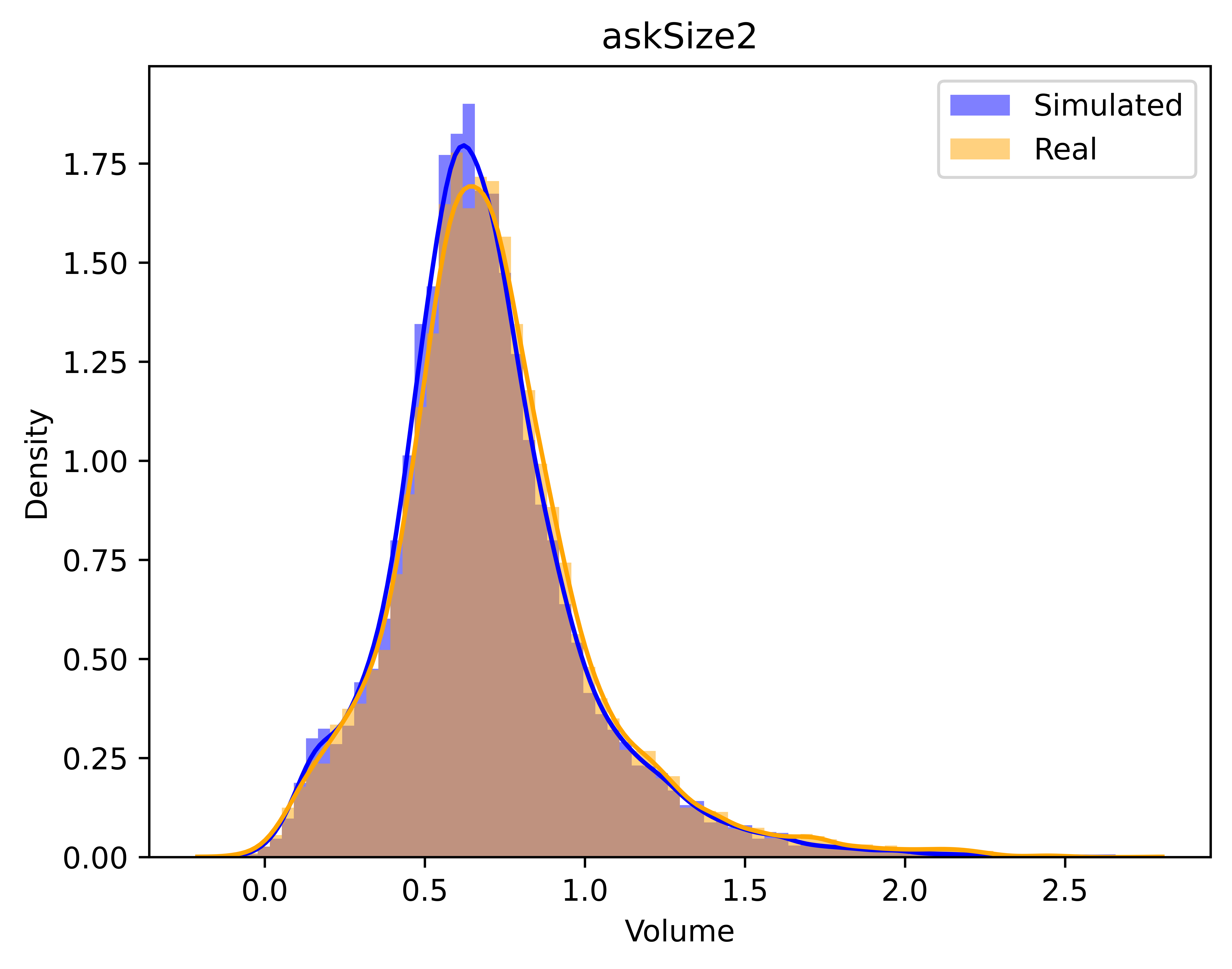}
			\label{fig:histNaive_ask2}
		\end{subfigure}
		& \multirow{2}{*}[2cm]{
			\begin{subfigure}{0.5\textwidth}
				\centering
				\includegraphics[width=.8\linewidth]{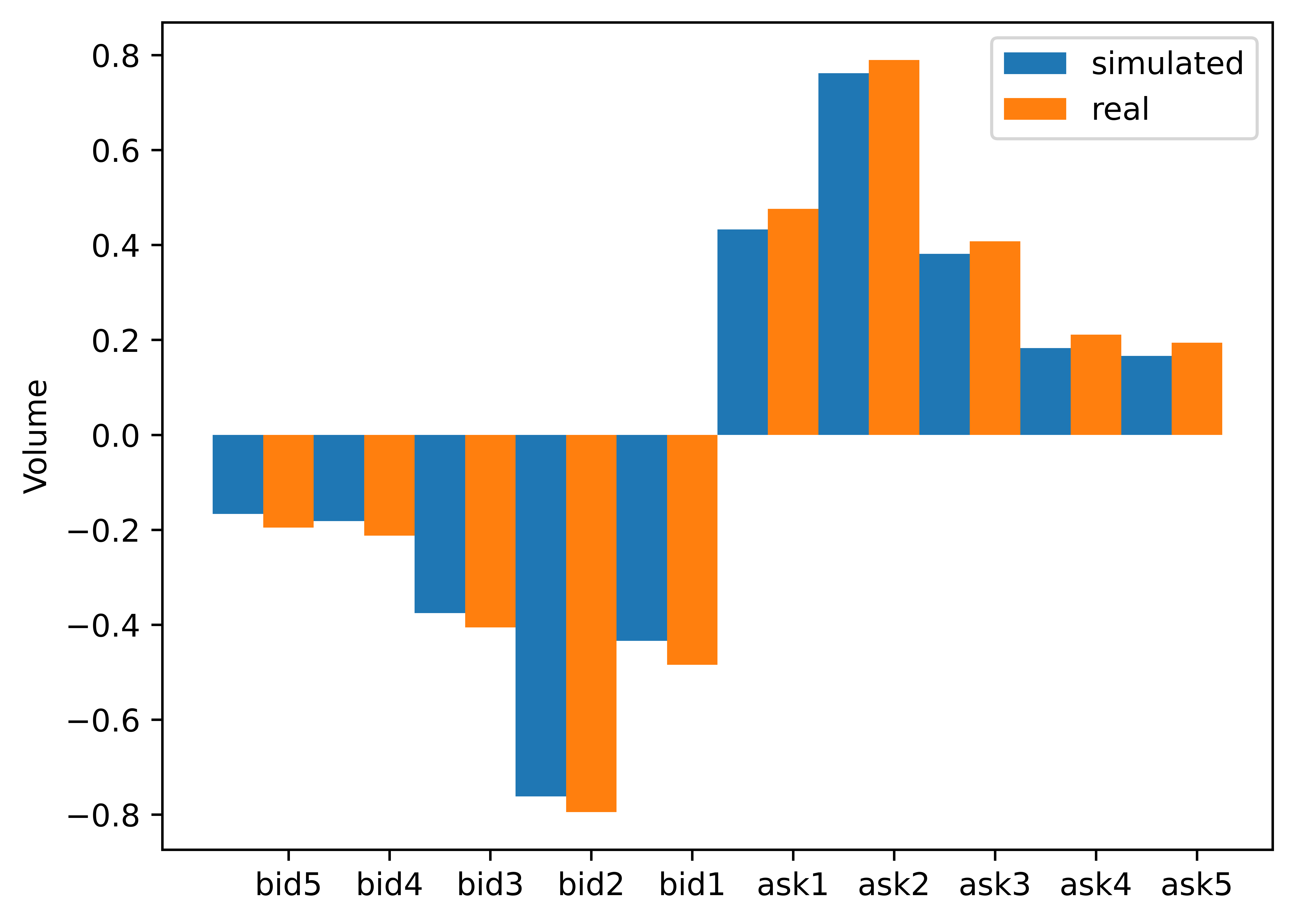}
				\label{fig:averageSizesNaive}
		\end{subfigure}} \\
		\begin{subfigure}[t]{0.25\textwidth}
			\centering
			\includegraphics[width=.8\linewidth]{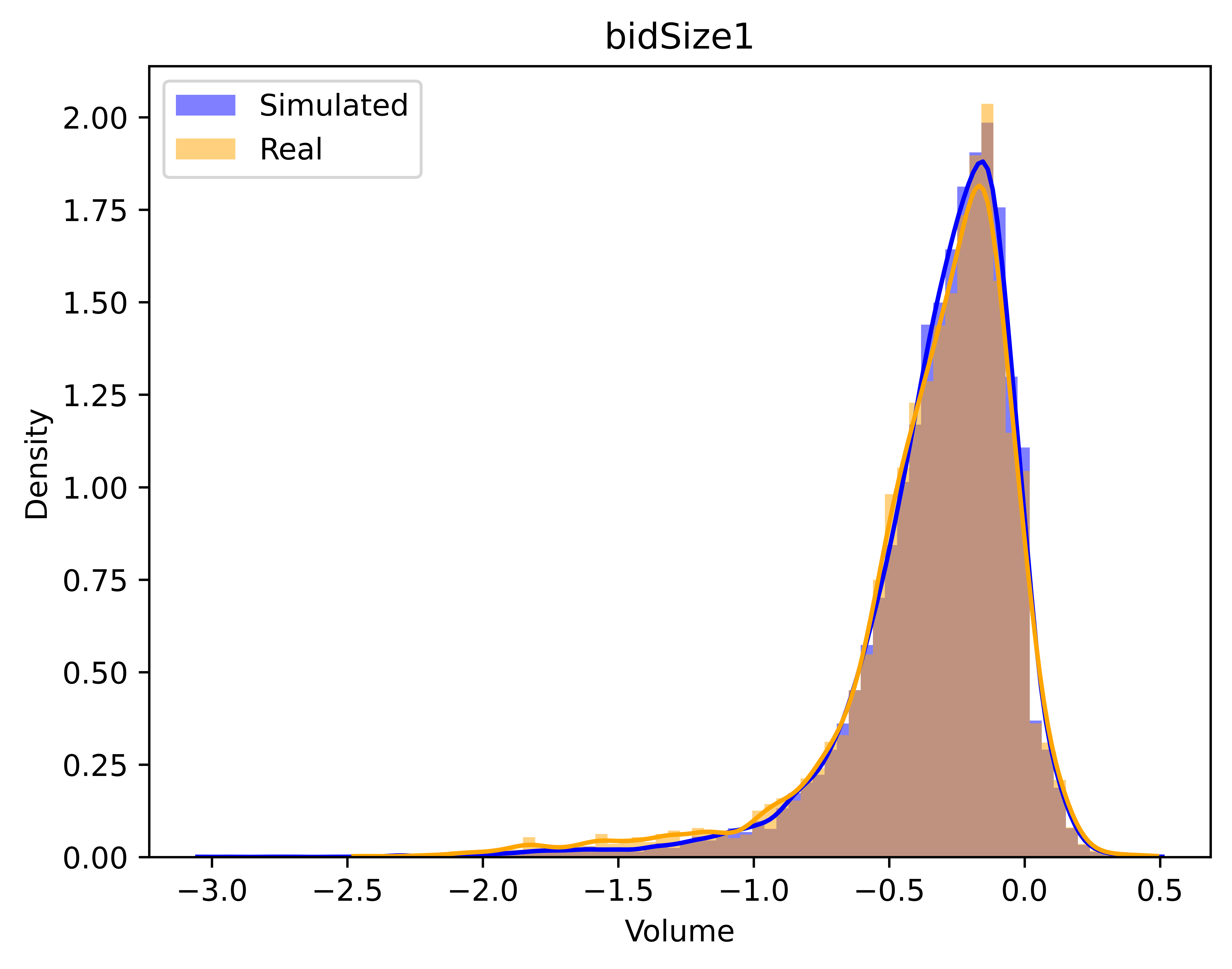}
			\label{fig:histNaive_bid1}
		\end{subfigure} &
		\begin{subfigure}[t]{0.25\textwidth}
			\centering
			\includegraphics[width=.8\linewidth]{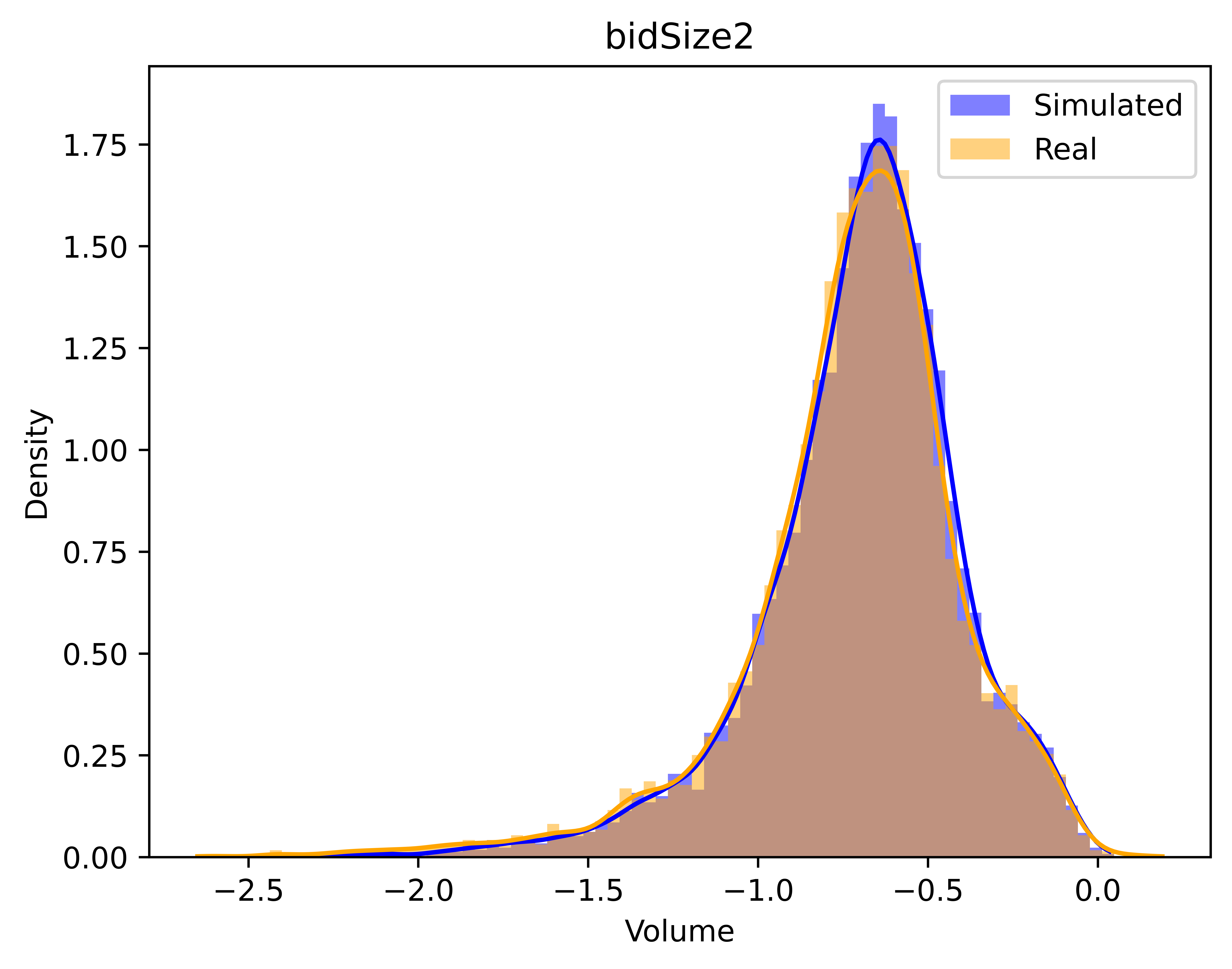}
			\label{fig:histNaive_bid2}
		\end{subfigure}
		&
	\end{tabular}
	\caption{Volume comparison for naive sampling: Marginal volumes per order book level after one transition (left); Average LOB shape after one transition (right)}
	\label{fig:hist_marginalsNaive}
\end{figure}

Figure \ref{fig:hist_marginalsNaive} corresponds to Figure \ref{fig:hist_marginals} in the main part. In Figure \ref{fig:hist_marginalsNaive}, we observe that naive sampling reproduces the marginal volume distributions relatively well although with more discrepancies compared to Algorithm \ref{alg:matchingSim} (cf. Table \ref{tab:ksBench}). The average LOB volumes have a similar shape but are slightly lower  compared to the transitions from the test data. The discrepancies reveal a slight distributional shift between the training data that was used for naive sampling and the test data used for the real transitions. The LLN should give the same results if the sampling distributions are the same (up to the sampling error). This sampling error becomes negligible especially for the average LOB shapes with $10^4$ samples. While naive sampling is not capable to account for this shift, Algorithm \ref{alg:matchingSim} with its conditional resampling is able to adapt.

\begin{figure}[h!]
	\begin{subfigure}{.8\textwidth}
		\centering
		\includegraphics[width=1.0\linewidth]{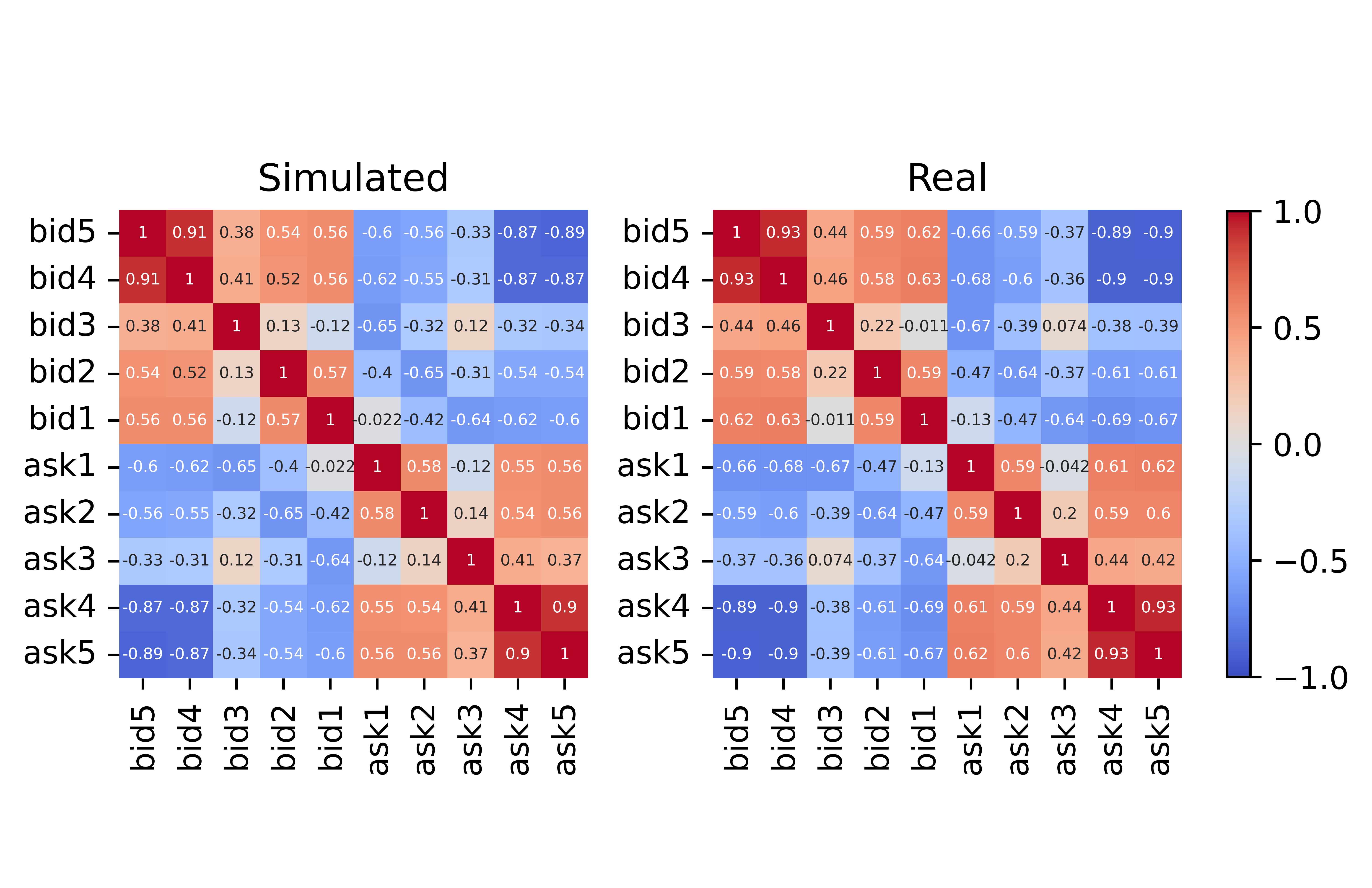}
		\label{fig:corrStaticNaive}
	\end{subfigure}\\
	\begin{subfigure}{.8\textwidth}
		\vspace*{-1.75cm}
		\centering
		\includegraphics[width=1.0\linewidth]{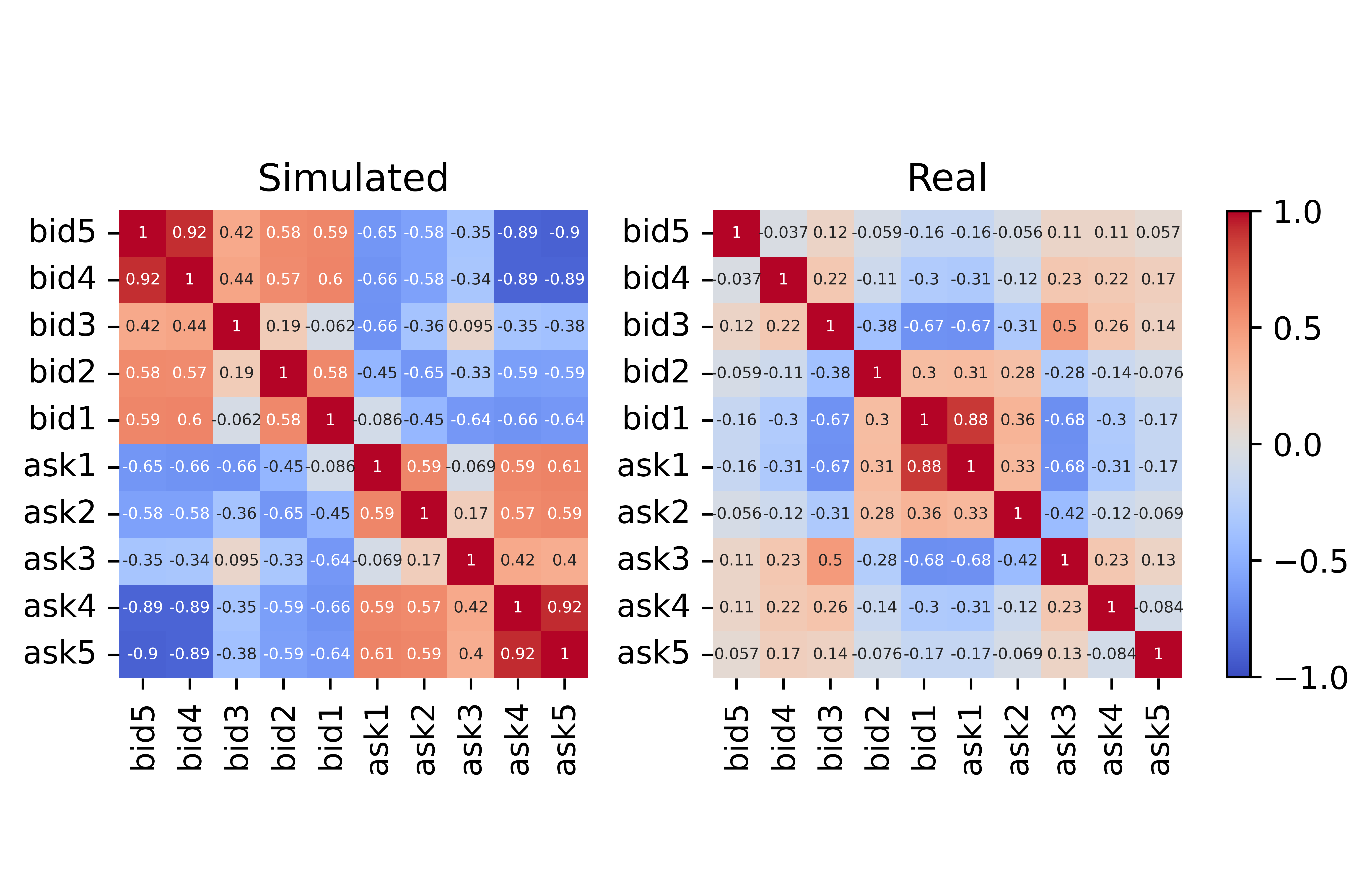}
		\label{fig:corrDiffsNaive}
	\end{subfigure}
	\caption{Correlation for naive sampling: Absolute Volumes (first row) Volume differences (second row)}
	\label{fig:corrVolNaive}
\end{figure}

Figure \ref{fig:corrVolNaive} corresponds to Figure \ref{fig:corrVol} in the main part. The first plot shows that the unconditional LLN is relatively efficient in reproducing the static correlation structure between different order book levels. However, the naive method is unable to reproduce the correlation structure of changes in volumes, as can be seen in the second plot of Figure \ref{fig:corrVolNaive}. As the naive sampling approach can be seen as independent of the initial state, a simple calculation on the correlation coefficient reveals that the correlation of the differences simply corresponds to the correlation of the static volumes for the naive sampling method. This reveals an important distinction between naive sampling and Algorithm \ref{alg:matchingSim}, which is capable of reproducing the correlation structure of changes in volume.

\begin{figure}[h!]
	\begin{subfigure}{.5\textwidth}
		\centering
		\includegraphics[width=.8\linewidth]{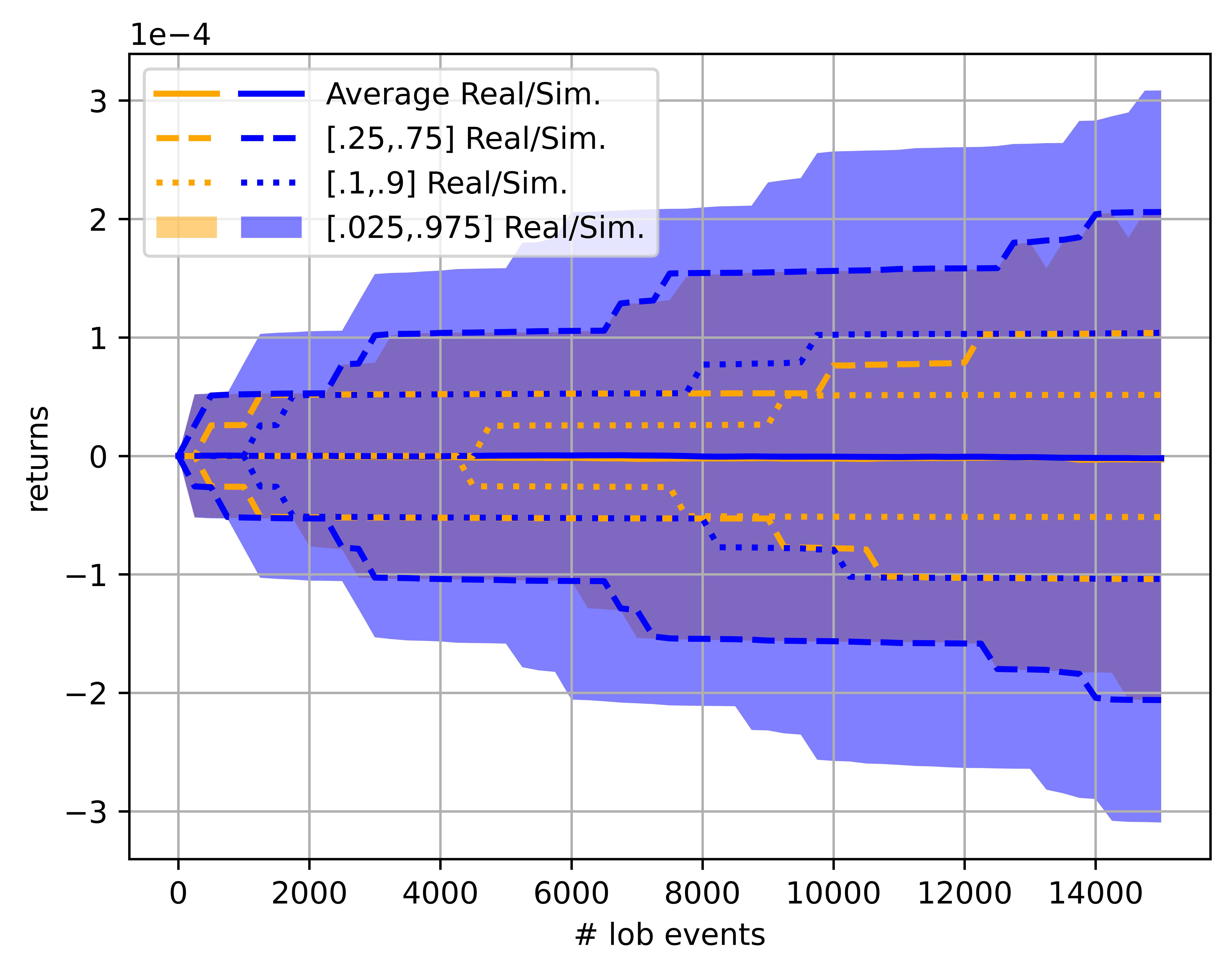}
		\label{fig:dynamicReturnMidNaive}
	\end{subfigure}%
	\begin{subfigure}{.5\textwidth}
		\centering
		\includegraphics[width=.8\linewidth]{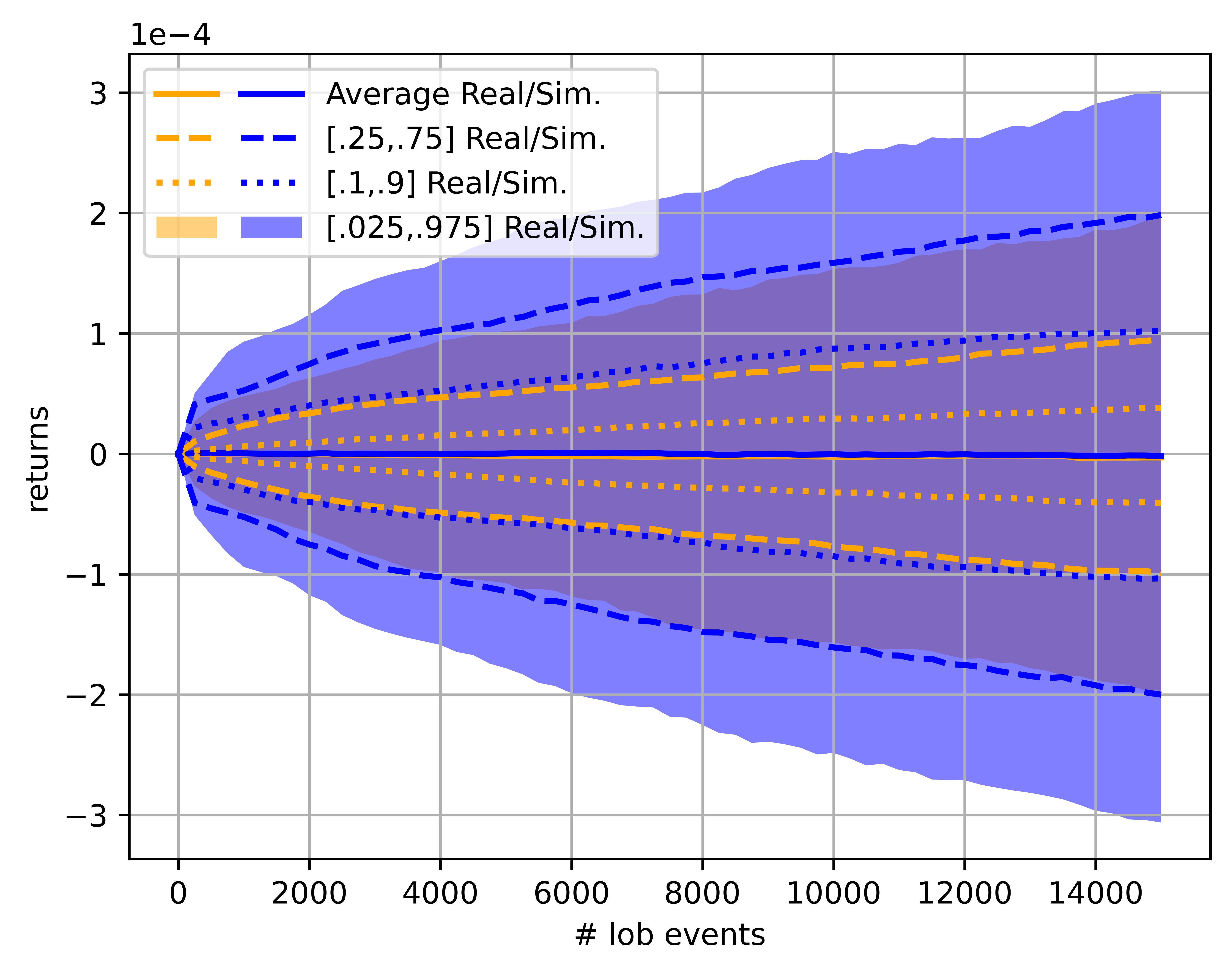}
		\label{fig:dynamicReturnWeightedNaive}
	\end{subfigure}
	\caption{Unconditional Return for naive sampling: Mid-price (left); Weighted mid-price (right)}
	\label{fig:dynamicReturnNaive}
\end{figure}

Figure \ref{fig:dynamicReturnNaive} corresponds to Figure \ref{fig:dynamicReturn} in the main part. In Figure \ref{fig:dynamicReturnNaive}, we compare the mid and the weighted mid-return of the naive sampling method with real return series. We observe that for both plots the average returns coincide at zero. Note that all quantiles of the simulated returns expand faster compared to the quantiles of the real return time series. The slower expansion of the real return distribution indicates that a price movement in the same direction is less likely to occur right after another. In contrast to Algorithm \ref{alg:matchingSim}, the naive sampling method does not capture this conditional price movement behavior. 

\begin{figure}[h!]
	\begin{subfigure}{.5\textwidth}
		\centering
		\includegraphics[width=.8\linewidth]{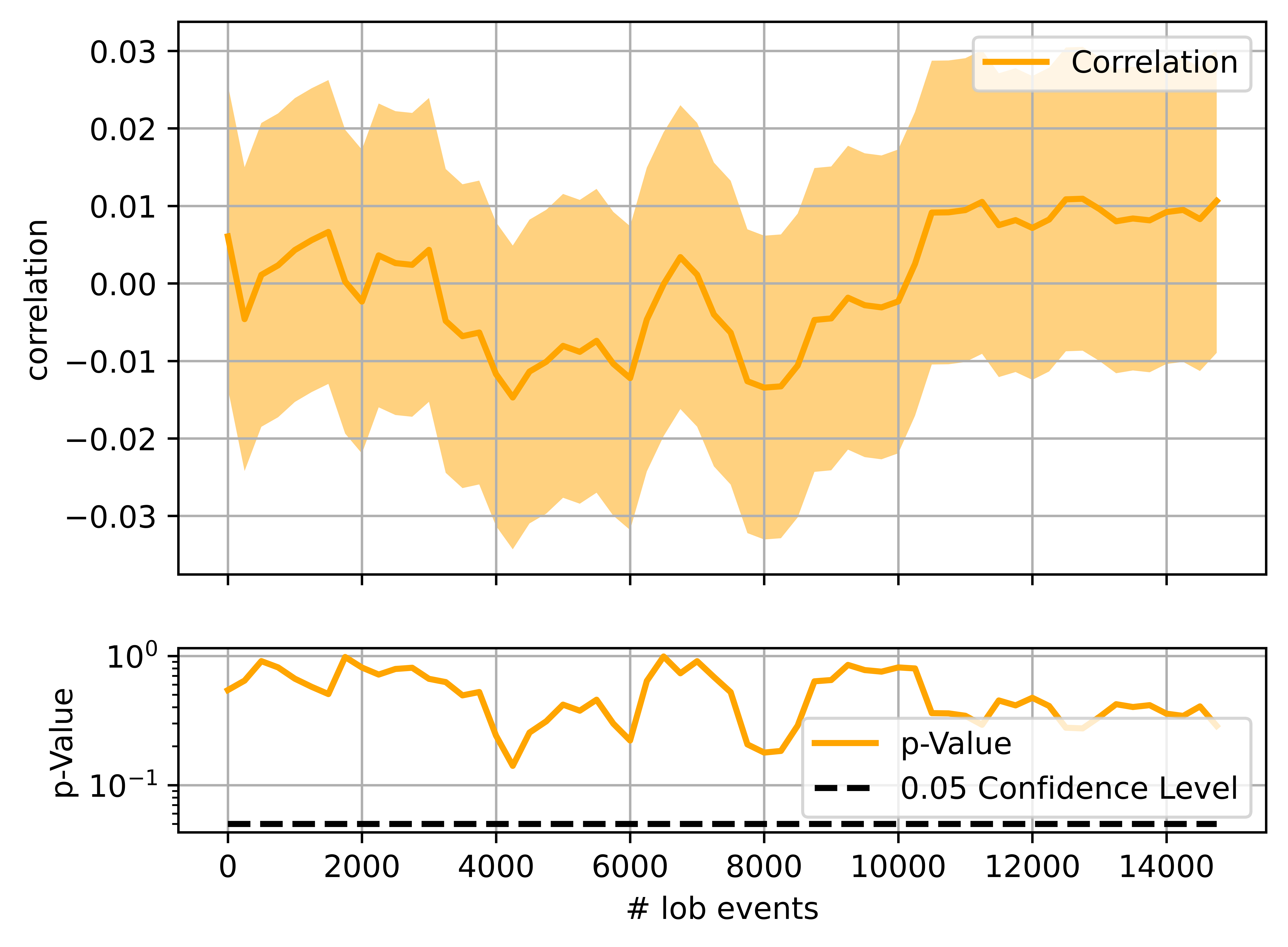}
		\label{fig:dynamicCorrReturnMidNaive}
	\end{subfigure}%
	\begin{subfigure}{.5\textwidth}
		\centering
		\includegraphics[width=.8\linewidth]{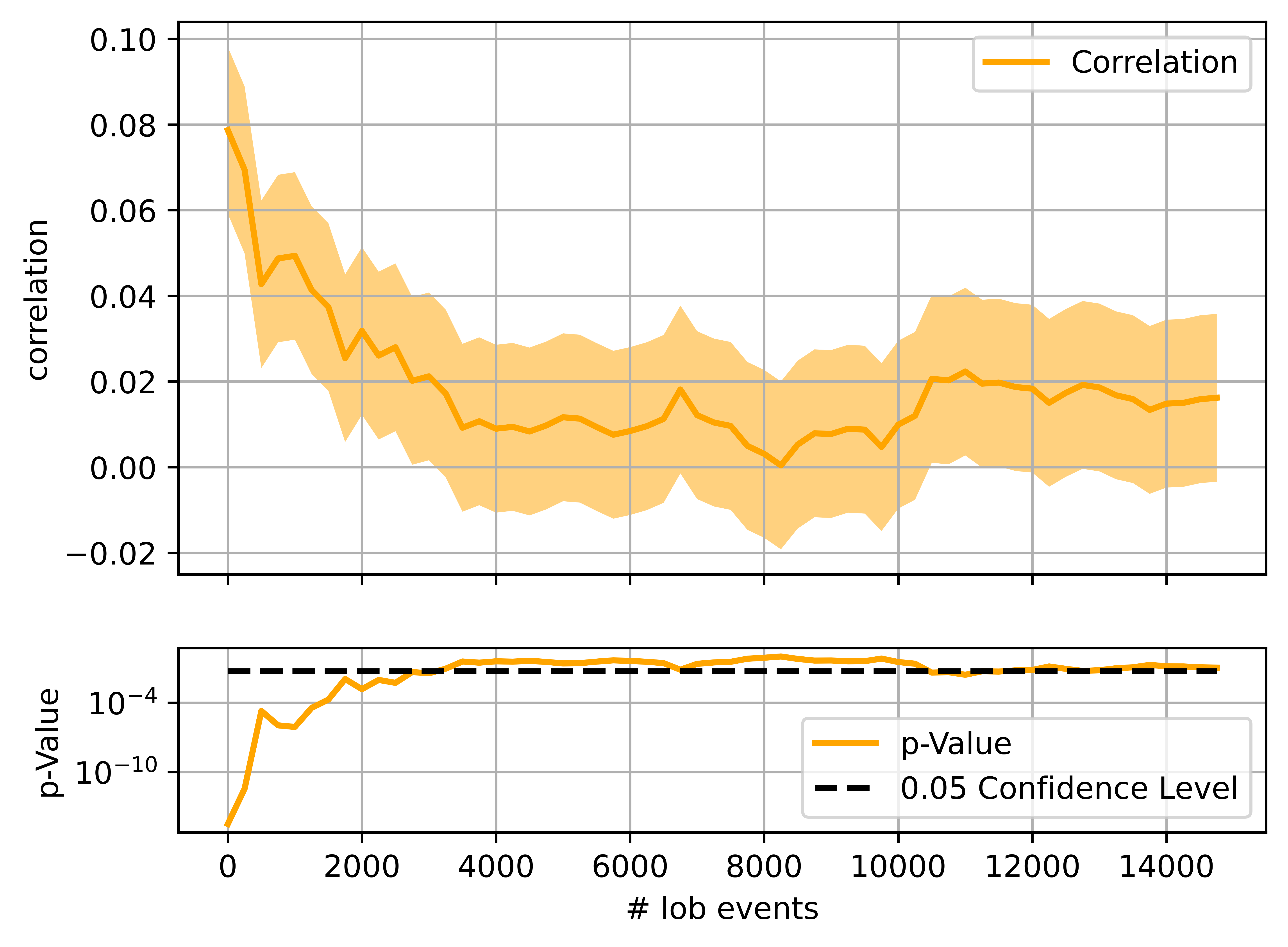}
		\label{fig:dynamicCorrReturnWeightedNaive}
	\end{subfigure}
	\caption{Return Correlation for naive sampling: Mid-Price (left); Weighted mid-price (right)}
	\label{fig:dynamicCorrReturnNaive}
\end{figure}
Figure \ref{fig:dynamicCorrReturnNaive} corresponds to Figure \ref{fig:dynamicCorrReturn} in the main part. The first plot in Figure \ref{fig:dynamicCorrReturnNaive} reveals that there is no correlation of the mid-price return for the naive sampling method. The second plot for weighted mid-price returns shows a small positive correlation that lasts for a short duration. After that, the correlation becomes statistically insignificant. Considering the covariance structure of the returns, this is most likely explained by the correlation of the initial weighted price with the real return time series. We note in particular that the weighted price contains more information on the order book shape compared to the unweighted mid-price. In comparison to Algorithm \ref{alg:matchingSim}, we observe that the correlation for naive sampling vanishes faster and is smaller in scale. This further points to the fact that Algorithm \ref{alg:matchingSim} is capable of extracting the conditional influence of the initial state beyond the information contained in the initial weighted price.

\subsection{CGAN}
The second benchmark we implement is the generative adversarial neural network for conditional LOB simulation proposed in \cite{cont2023limit}. Wherever possible, we follow the implementation as stated in the original paper and refer to that paper for implementation details. In cases where the original paper does not specify concrete parameter values, we conducted a heuristic hyperparameter search\footnote{By hyperparameter search, we choose the noise seed with 10 dimensions and a batch size of 256 samples.}. We present the model giving the best performance for the KS test statistics as reported in Table \ref{tab:ksBench}. Furthermore, the results presented below confirm that the CGAN method is applicable to our data set and that the chosen model is a good LOB simulator. 

For this method, we only simulate the three highest levels of the LOB instead of the five highest. As the CGAN method encodes price movements within a given window of price levels, the two extra order book levels are needed as a buffer in the generation of the training data set. Furthermore, this implementation is consistent with \cite{cont2023limit} who also consider the three highest levels of the order book.
\begin{figure}[h!]
	\centering
	\begin{tabular}{ccc}
		\begin{subfigure}[t]{0.25\textwidth}
			\centering
			\includegraphics[width=.8\linewidth]{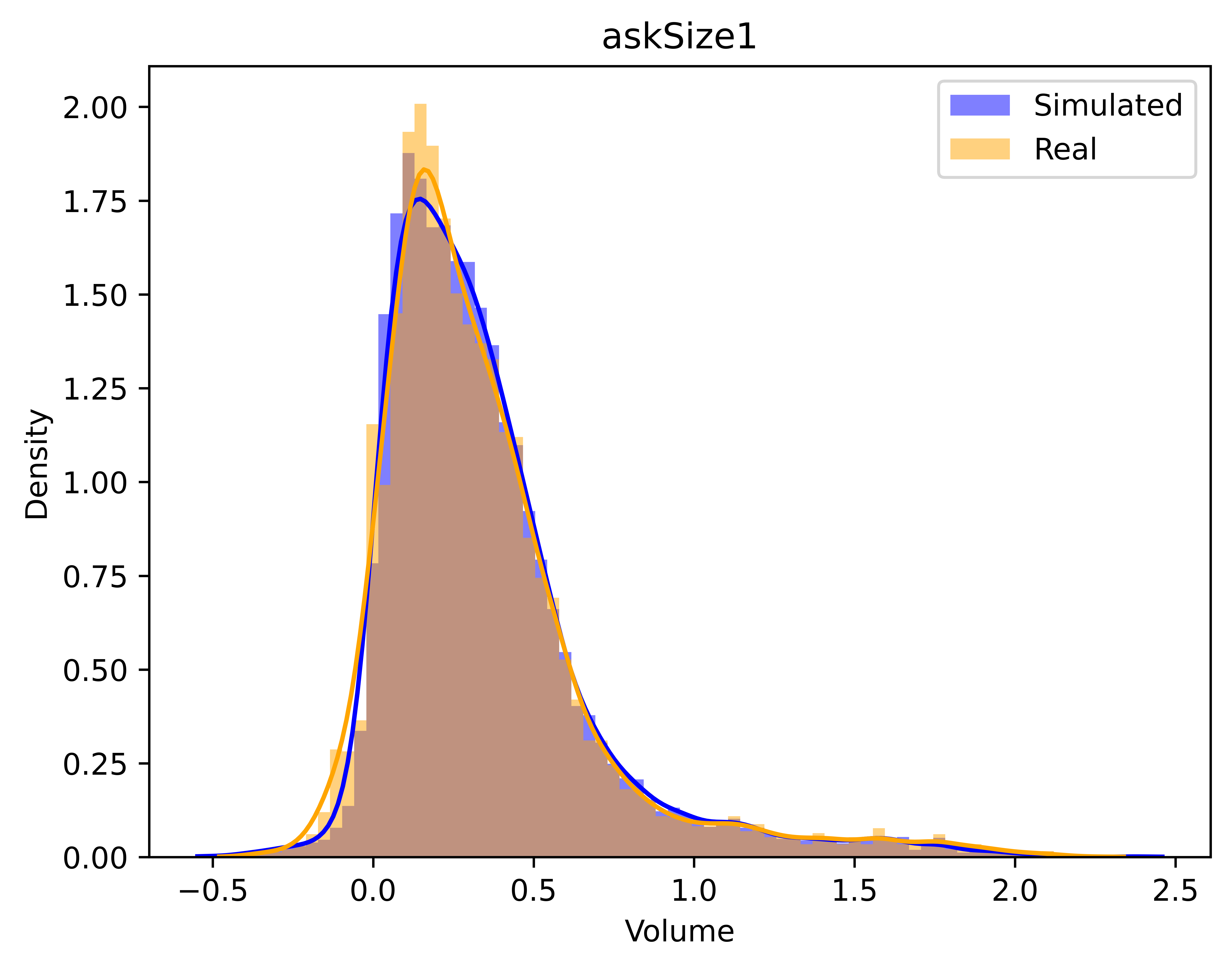}
			\label{fig:histCGAN_ask1}
		\end{subfigure} &
		\begin{subfigure}[t]{0.25\textwidth}
			\centering
			\includegraphics[width=.8\linewidth]{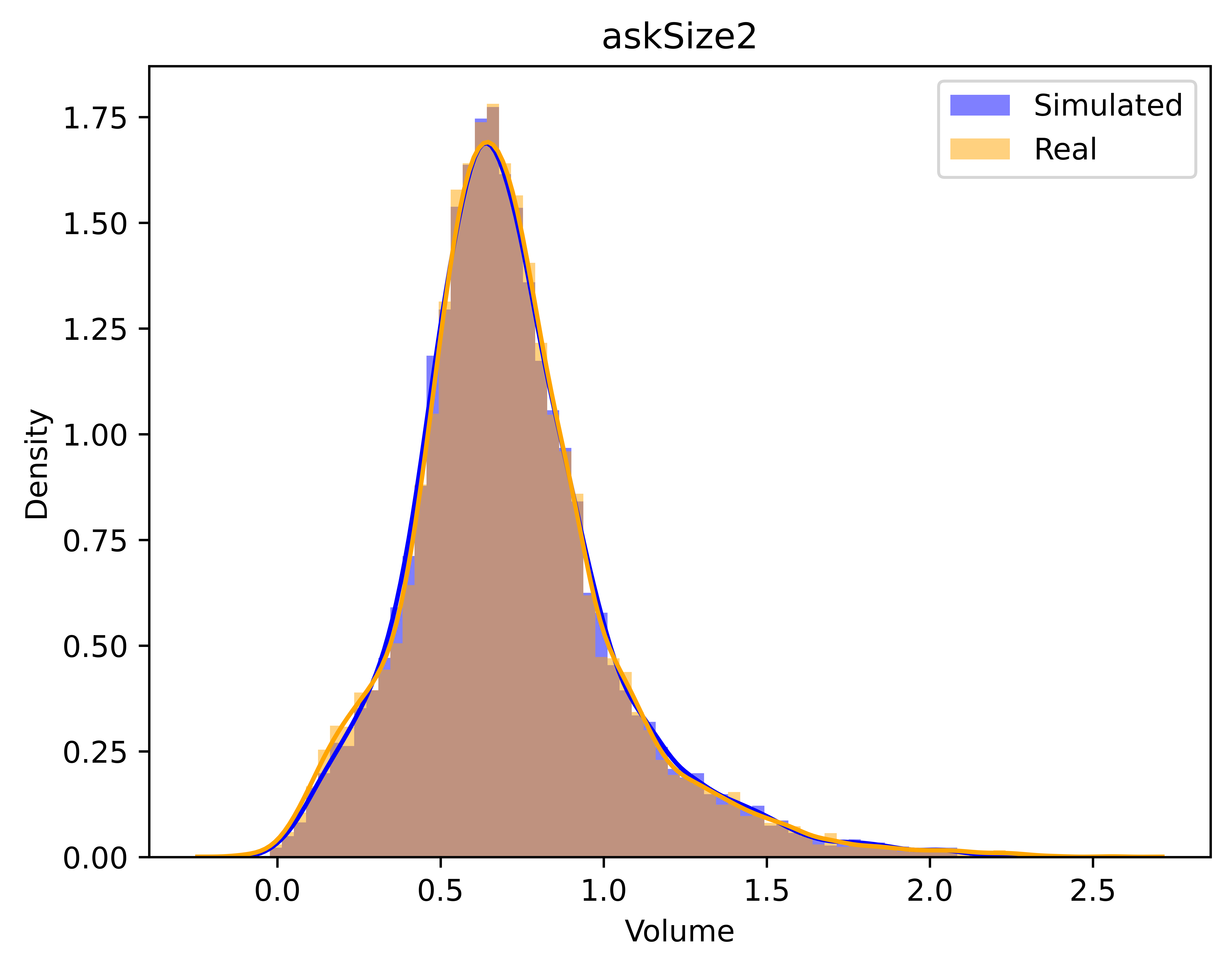}
			\label{fig:histCGAN_ask2}
		\end{subfigure}
		& \multirow{2}{*}[2cm]{
			\begin{subfigure}{0.5\textwidth}
				\centering
				\includegraphics[width=.8\linewidth]{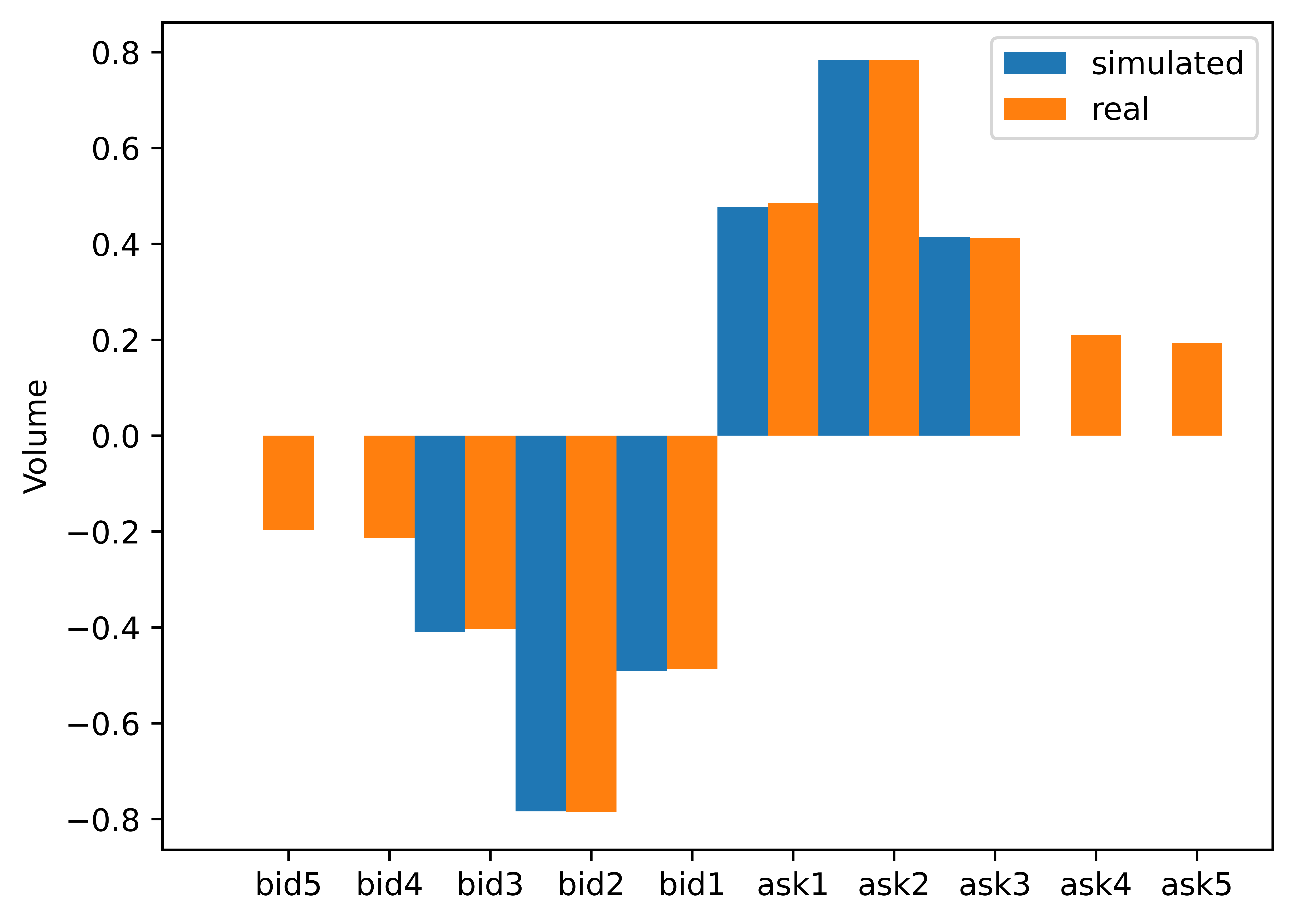}
				\label{fig:averageSizesCGAN}
		\end{subfigure}} \\
		\begin{subfigure}[t]{0.25\textwidth}
			\centering
			\includegraphics[width=.8\linewidth]{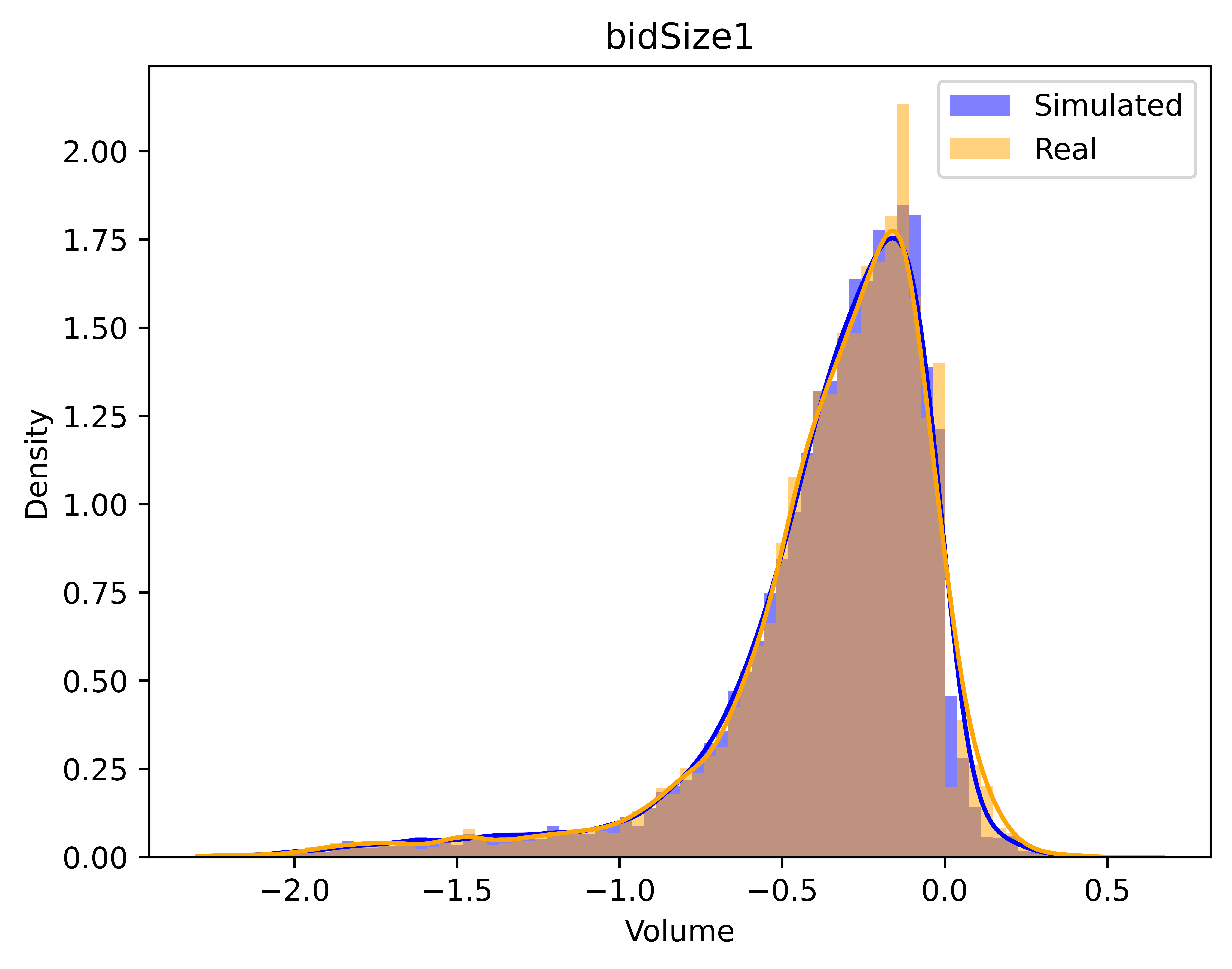}
			\label{fig:histCGAN_bid1}
		\end{subfigure} &
		\begin{subfigure}[t]{0.25\textwidth}
			\centering
			\includegraphics[width=.8\linewidth]{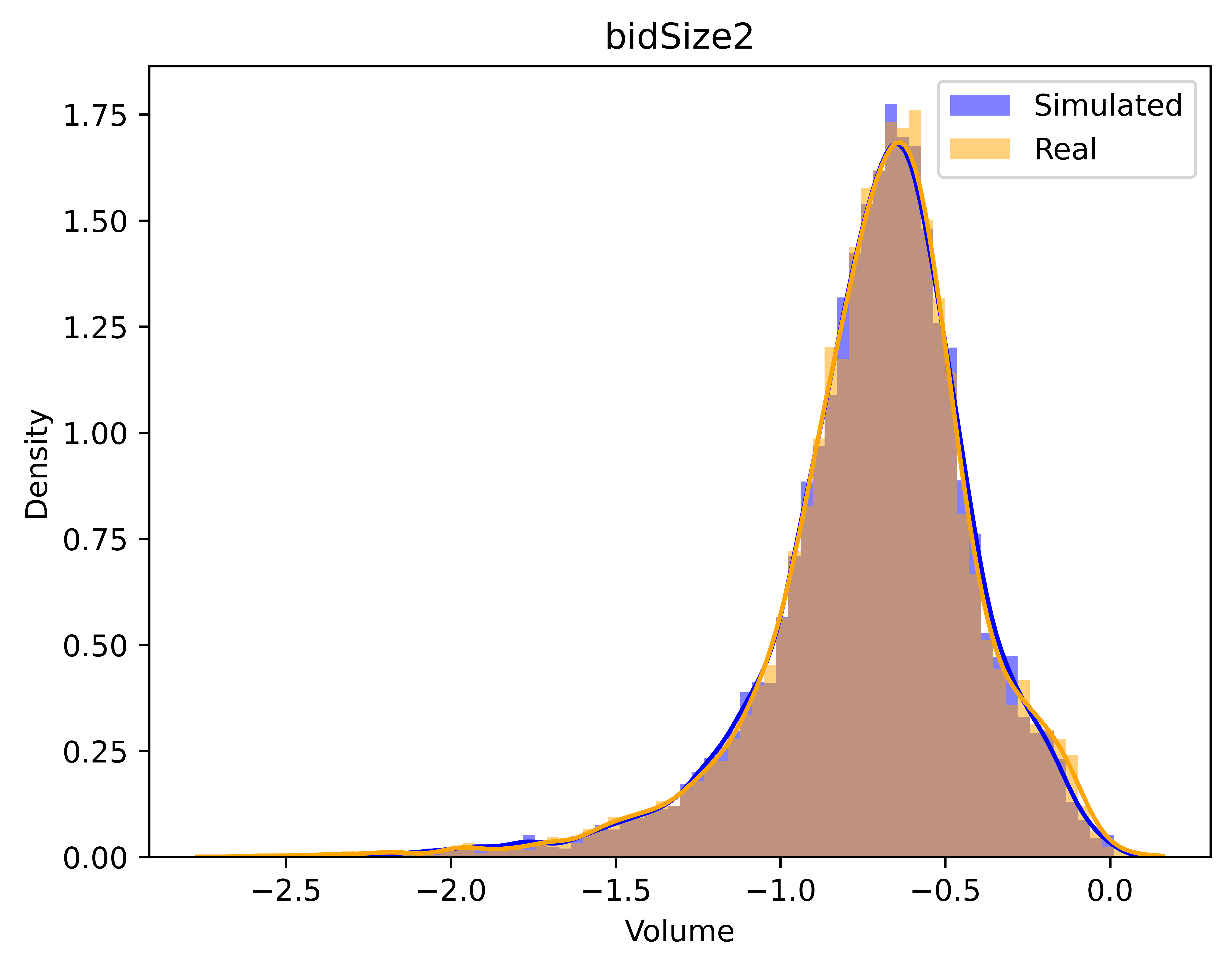}
			\label{fig:histCGAN_bid2}
		\end{subfigure}
		&
	\end{tabular}
	\caption{Volume Comparison for CGAN: Marginal volumes per order book level after one transition (left) Average LOB shape after one transition}
	\label{fig:histCGAN_marginals}
\end{figure}

Figure \ref{fig:histCGAN_marginals} corresponds to Figure \ref{fig:hist_marginals} in the main part. The left-hand plots show that the CGAN method is well capable of reproducing the marginal distributions of volumes for our data set. The right-hand plot demonstrates that also the average order book shape produced by the CGAN method fits the real transitions well. Note that the missing averages for the forth and fifth highest order book level are due to the encoding of price changes as explained above.  

\begin{figure}[!h]
	\begin{subfigure}{.8\textwidth}
		\centering
		\includegraphics[width=1.0\linewidth]{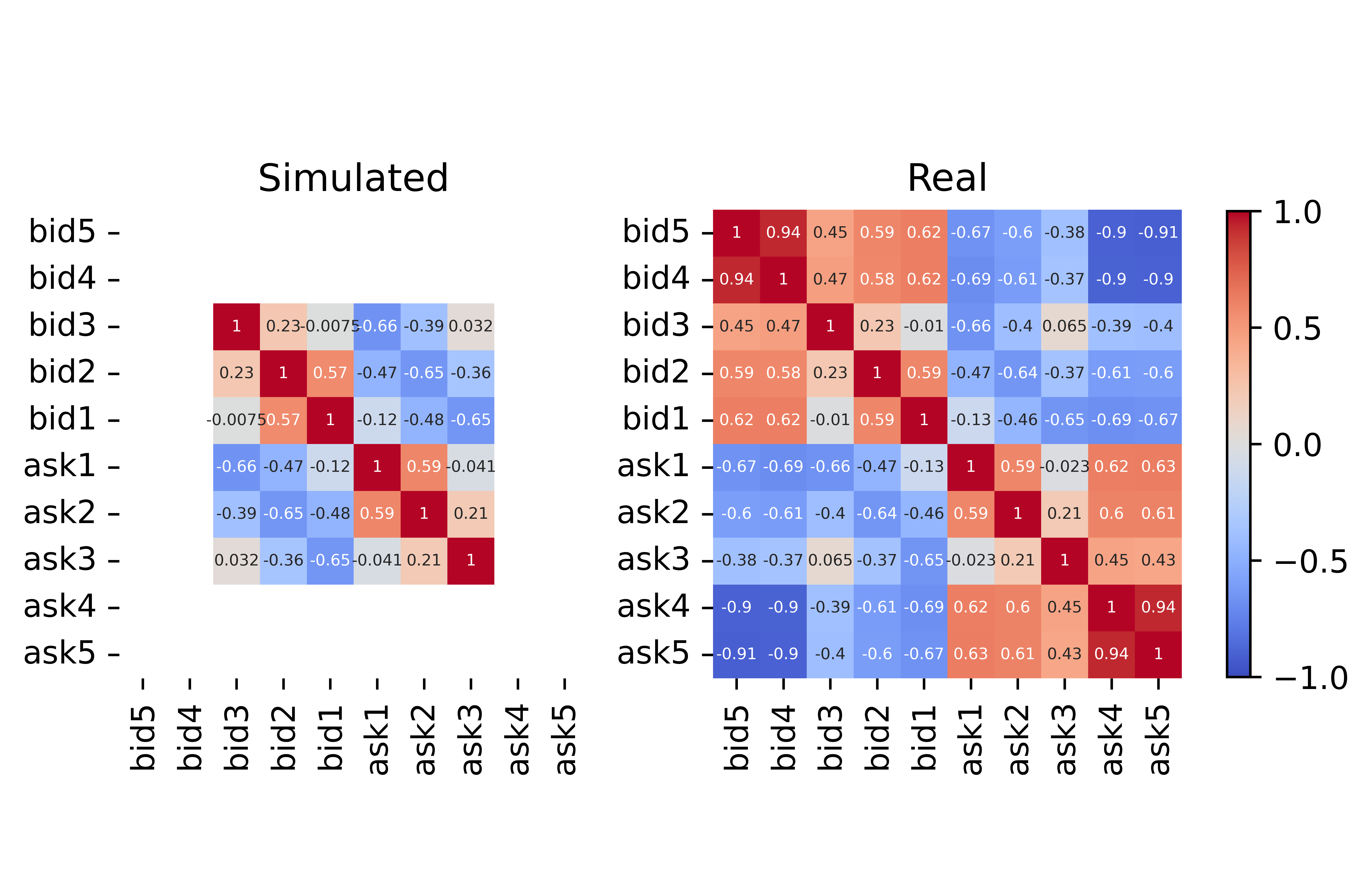}
		\label{fig:corrStaticCGAN}
	\end{subfigure}\\
	\begin{subfigure}{.8\textwidth}
		\centering
		\vspace*{-1.75cm}
		\includegraphics[width=1.0\linewidth]{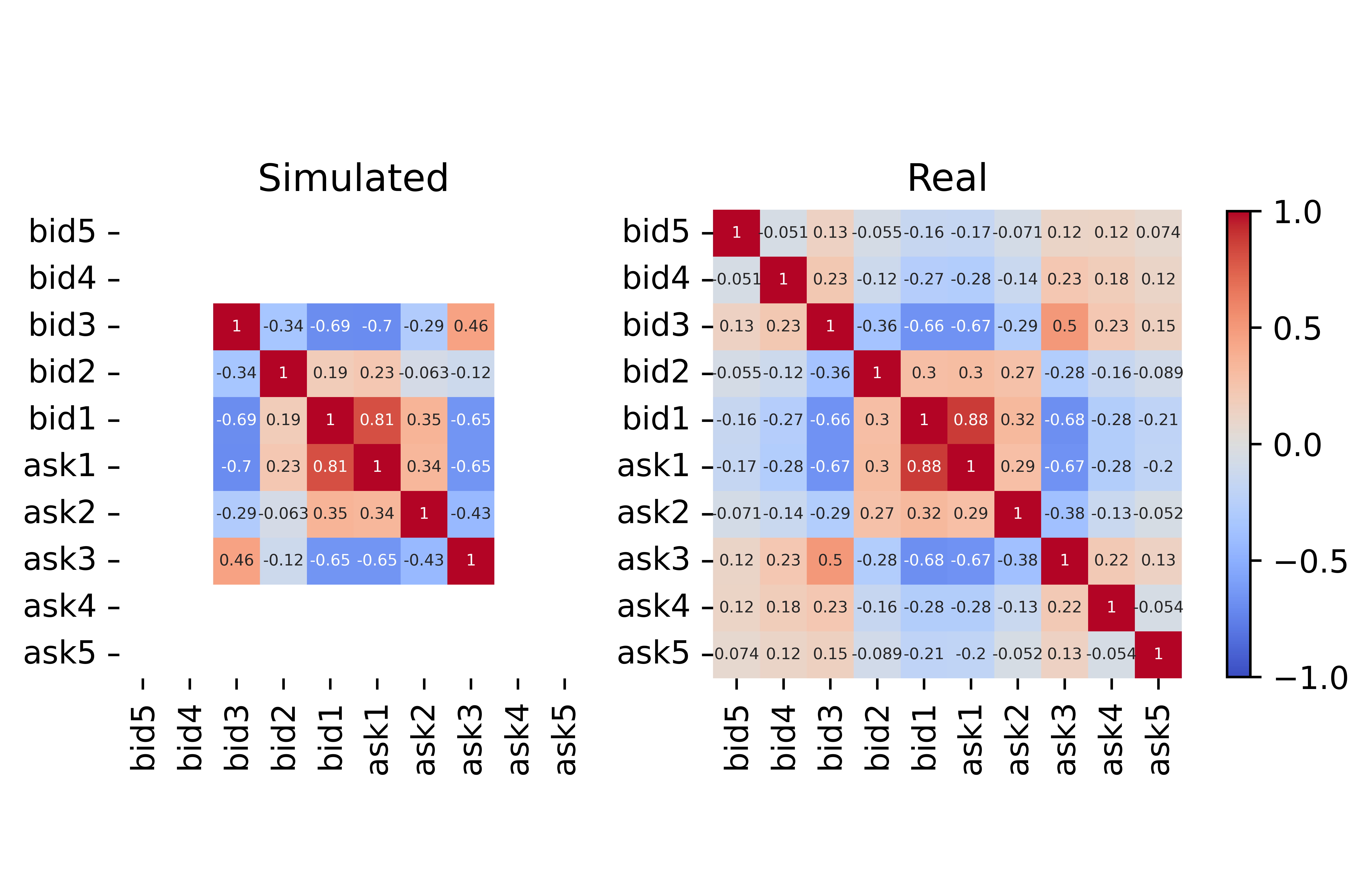}
		\label{fig:corrDiffsCGAN}
	\end{subfigure}
	\caption{Correlation for CGAN: Absolute Volumes (first row) Volume differences (second row)}
	\label{fig:corrVolCGAN}
\end{figure}

Figure \ref{fig:corrVolCGAN} corresponds to Figure \ref{fig:corrVol} in the main part. In the first plot in Figure \ref{fig:corrVolCGAN}, we observe that the CGAN method is effective in capturing the correlation structure for volume levels and only shows minor differences. The second plot in Figure \ref{fig:corrVolCGAN} reveals an overall good fit in the correlation of volume changes pointing to the effectiveness in conditional sampling of the CGAN method. For a few volume levels, we see some larger differences in correlations, e.g.\ ask2-bid2.   

\begin{figure}[h!]
	\begin{subfigure}{.5\textwidth}
		\centering
		\includegraphics[width=.8\linewidth]{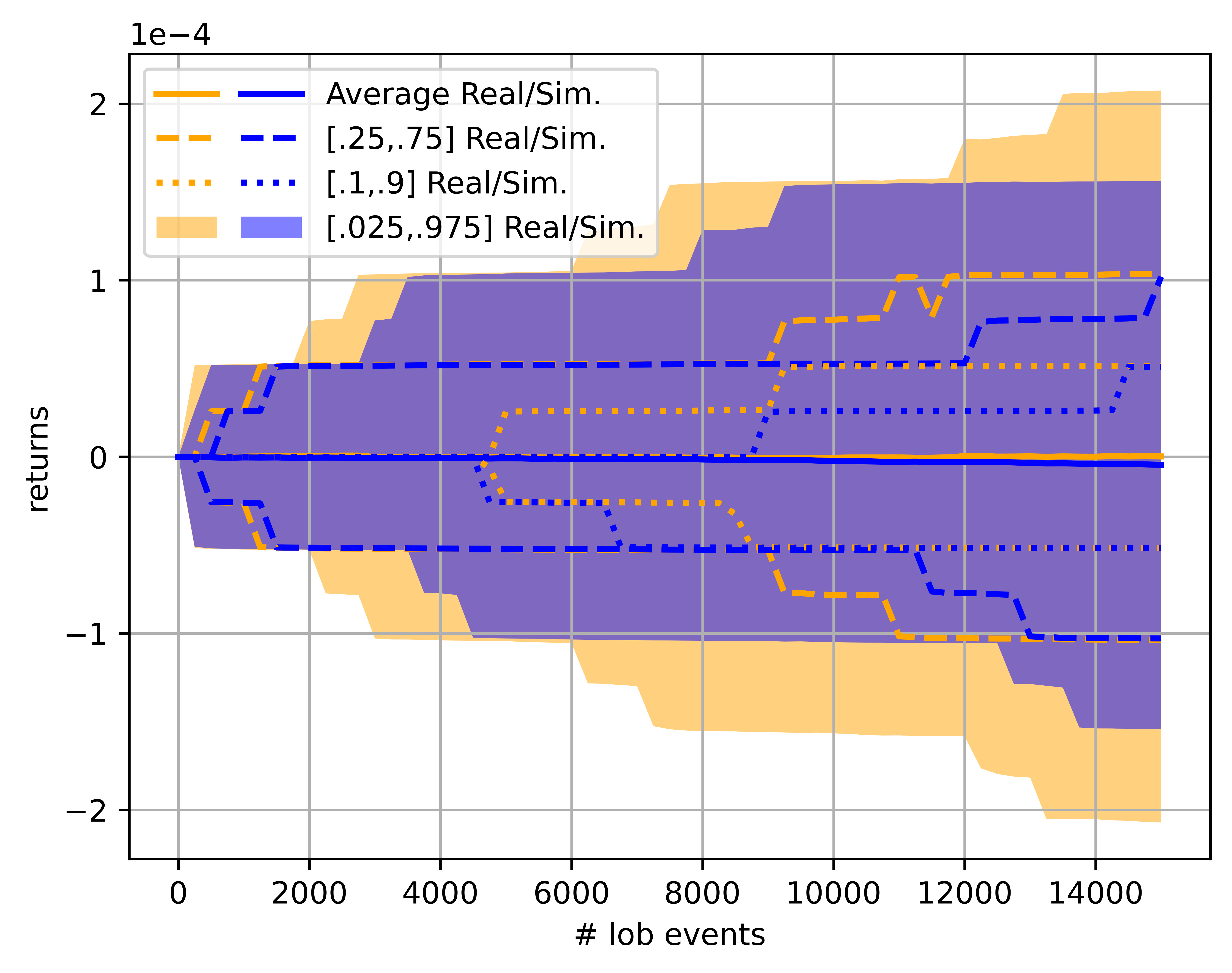}
		\label{fig:dynamicReturnMidCgan}
	\end{subfigure}%
	\begin{subfigure}{.5\textwidth}
		\centering
		\includegraphics[width=.8\linewidth]{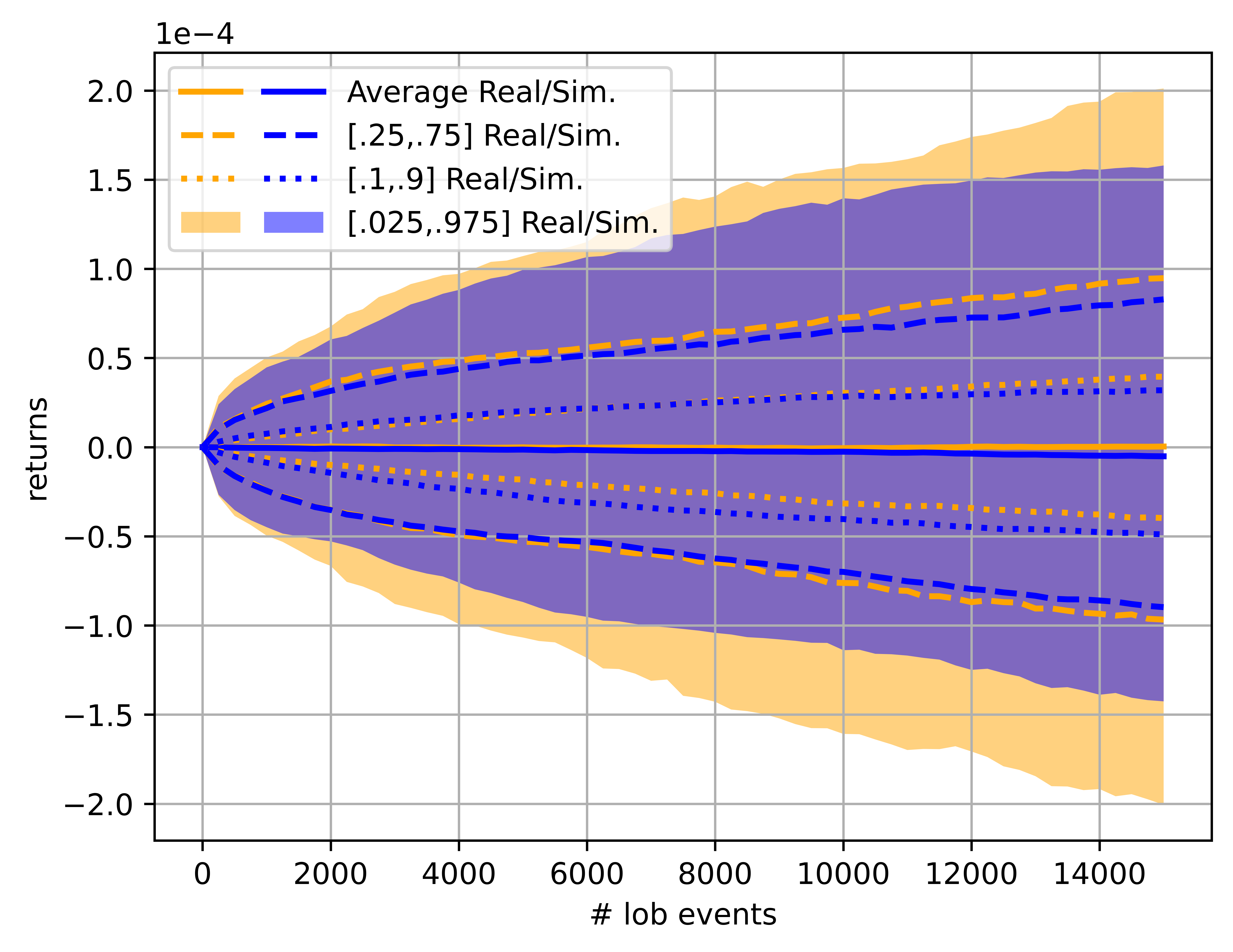}
		\label{fig:dynamicReturnWeightedCgan}
	\end{subfigure}
	\caption{Unconditional Return for CGAN: Mid-price (left); Weighted mid-price (right)}
	\label{fig:dynamicReturnCgan}
\end{figure}

Figure \ref{fig:dynamicReturnCgan} corresponds to Figure \ref{fig:dynamicReturn} in the main part. For both the weighted and the unweighted return series the central quantiles of the simulation fit well with the central quantiles of the real returns time series. As in the results for Algorithm \ref{alg:matchingSim}, the tails of the real returns extend further than the tails of the simulation. 

\begin{figure}[h!]
	\begin{subfigure}{.5\textwidth}
		\centering
		\includegraphics[width=.8\linewidth]{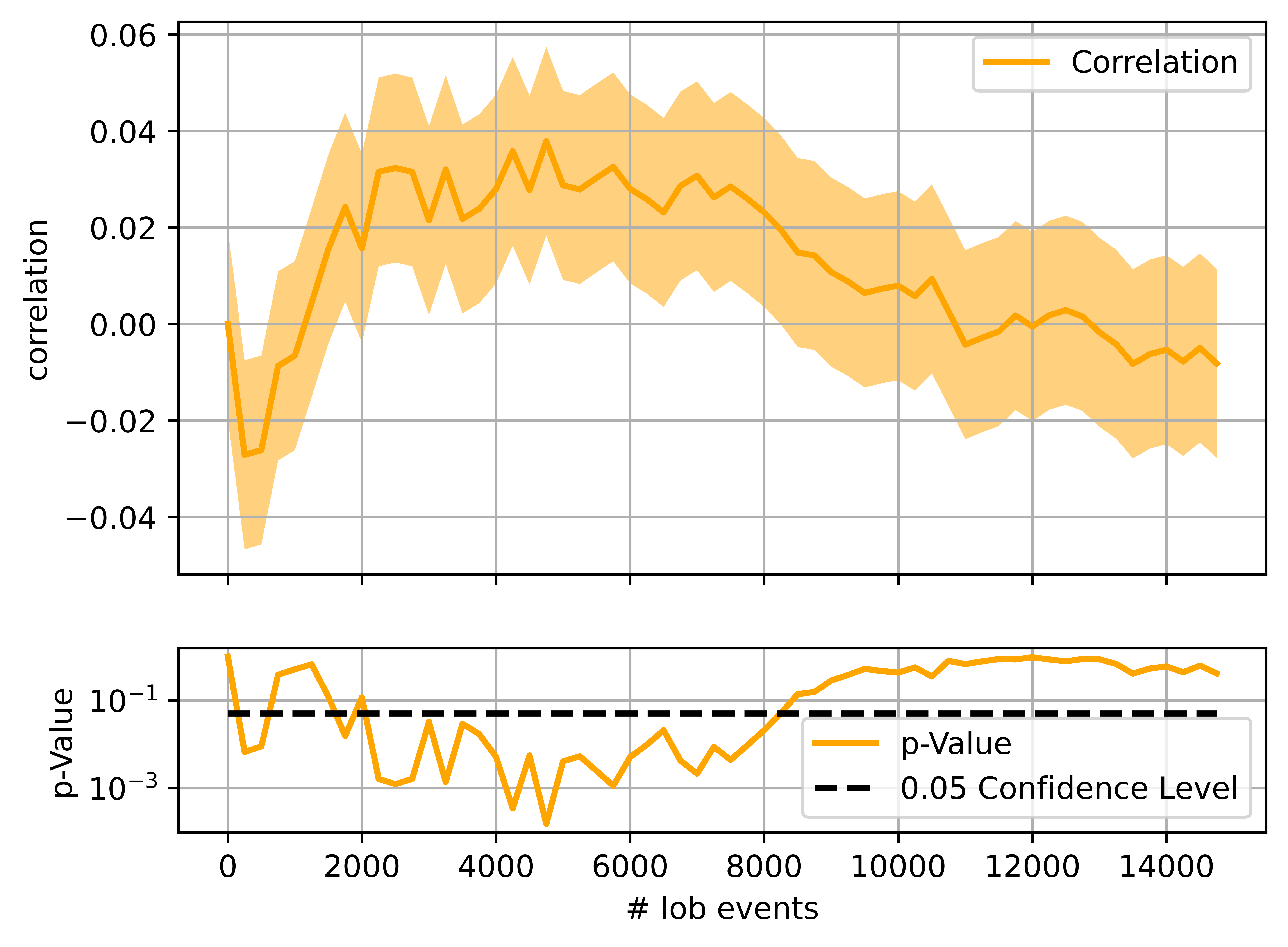}
		\label{fig:dynamicCorrReturnMidCgan}
	\end{subfigure}%
	\begin{subfigure}{.5\textwidth}
		\centering
		\includegraphics[width=.8\linewidth]{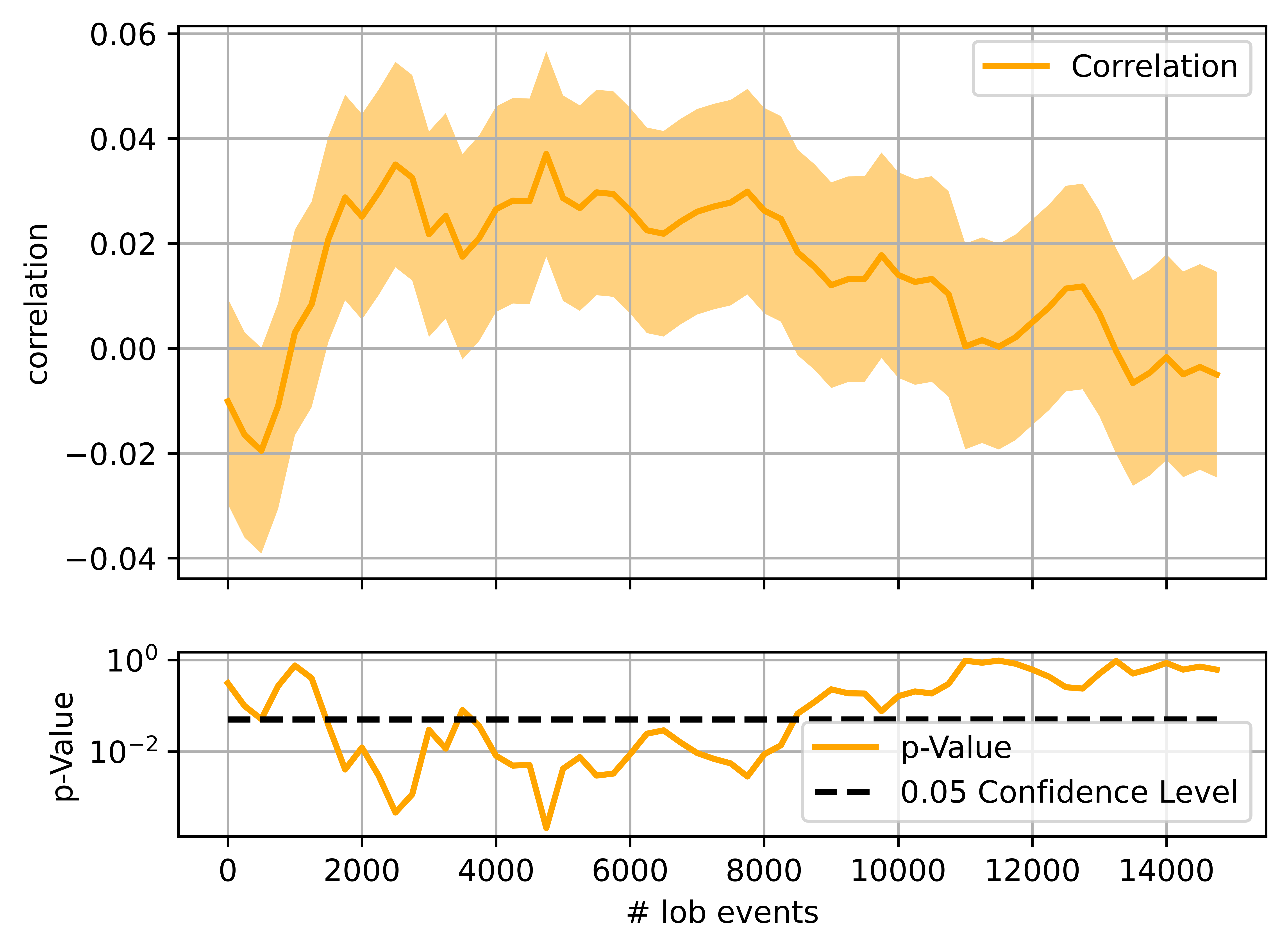}
		\label{fig:dynamicCorrReturnWeightedCgan}
	\end{subfigure}
	\caption{Return Correlation for CGAN: Mid-Price (left); Weighted mid-price (right)}
	\label{fig:dynamicCorrReturnCgan}
\end{figure}

Figure \ref{fig:dynamicCorrReturnCgan} corresponds to Figure \ref{fig:dynamicCorrReturn} in the main part. Figure \ref{fig:dynamicCorrReturnCgan} reveals a correlation between the real and the simulated return time series for both weighted and unweighted mid-price. Surprisingly, the correlation is initially  small for a short time and then becomes positive before vanishing again.

To conclude, we compare some other aspects of the CGAN approach with Algorithm \ref{alg:matchingSim}. First, we note that training CGAN requires us to solve a complicated optimization problem that is potentially unstable (cf.\ \cite{cont2023limit}). In contrast, Algorithm \ref{alg:matchingSim} does not require any optimization. Once trained, the parameters of the CGAN encode all learned market dynamics. First, this has the advantage that the system noise can be easily controlled in evaluation, which simplifies counterfactual comparisons. Second, the evaluation complexity of a CGAN data generator is constant in data set size. Computational costs for evaluation phase for algorithm \ref{alg:matchingSim} on the other hand grow on average logarithmically in the data set size (\cite{giegrich2023k}). The disadvantage of encoding the market dynamics in a neural network is that the simulator is effectively a black box and unrealistic order book transitions cannot be ruled out. Algorithm \ref{alg:matchingSim} is based on a simple rule based assignment and each sampled transition stems from a historical observation. Finally, the CGAN method as presented in \cite{cont2023limit} did not address evaluating trading strategies with limit orders, which we included in Algorithm \ref{alg:matchingSim}.

\end{document}